\documentclass[preprint]{aastex631}
 \hypersetup{linkcolor=orange,citecolor=blue,filecolor=cyan,urlcolor=blue}

\usepackage{amsmath}
 \usepackage{cancel}
\shorttitle{Linear \& Nonlinear Radial Transport}
\shortauthors{Osmane et al.}
\graphicspath{{./}{figures/}}

\begin{document}
%\linenumbers
\title{Linear, Quasi-Linear and Nonlinear Radial Transport in the Earth's Radiation Belts}

\correspondingauthor{Adnane Osmane}
\email{adnane.osmane@helsinki.fi}

\author[0000-0003-2555-5953]{Adnane Osmane}
\affiliation{University of Helsinki, Department of Physics, Helsinki, Finland, 00014}

%\author[0000-0002-8787-5170]{Yohei Kawazura}
%\affiliation{Tohoku University, Frontier Research %Institute for Interdisciplinary Sciences, Sendai, %980-8578
%Japan}

\author[0000-0002-4489-8073]{Emilia Kilpua}
\affiliation{University of Helsinki, Department of Physics, Helsinki, Finland}

\author[0000-0002-3715-4623]{Harriet George}
\affiliation{Laboratory of Atmospheric and Space Physics, University of Colorado Boulder, Boulder, USA}

\author[0000-0003-2353-8586]{Oliver Allanson}
\affiliation{University of Birmingham,
School of Engineering, Birmingham, UK}
\affiliation{University of Exeter, Department of Earth \& Environmental Sciences, Penryn,  UK} 
\affiliation{University of Exeter, Department of Mathematics \& Statistics, Exeter, UK}

\author[0000-0002-6445-5595]{Milla Kalliokoski}
\affiliation{Japan Aerospace Exploration Agency, Tokyo, Japan}

\begin{abstract} 
Observational studies of the Earth's radiation belts indicate that Alfv\'enic fluctuations in the frequency range of 2-25 mHz accelerate magnetically trapped electrons to relativistic energies. For decades, statistical models of the Earth's radiation belts have quantified the impact of Alfv\'enic waves in terms of quasi-linear diffusive models. However, quasi-linear models are inadequate to quantify Alfv\'enic radial transport occurring on timescales comparable to the azimuthal drift period of $0.1- 10$ MeV electrons. With recent advances in observational methodologies offering spatial and temporal coverage of the Earth's radiation belts on fast timescales, a theoretical framework that distinguishes between fast and diffusive radial transport can also be tested for the first time with in situ measurements. In this report, we present a drift kinetic description of radial transport for planetary radiation belts. We characterize linear processes that are too fast to be modelled by quasi-linear models and determine the conditions under which nonlinearities become dynamically significant. In the linear regime, wave-particle interactions are categorized in terms of resonant and non-resonant responses. We demonstrate that the phenomenon of zebra stripes is non-resonant and can originate from the injection of particles in the inner radiation belts. We derive a radial diffusion coefficient for a field model that satisfies Faraday's law and that contains two terms: one scaling as $L^{10}$ independent of the azimuthal number $m$, and a second one scaling as $m^2 L^6$. In the nonlinear regime, we show that azimuthally symmetric waves with properties consistent with in situ measurements can energize 10-100 keV electrons in less than a drift period. This coherent process provides new evidence that acceleration by Alfv\'enic waves in radiation belts cannot be fully contained within diffusive models.

\end{abstract}
\hypersetup{linkcolor=black}
  %\tableofcontents
\tableofcontents

\section{Introduction} \label{sec:intro}
\subsection{Motivation and background}
\noindent Radiation belts are torus-shaped plasma environments confined by planetary magnetic fields. Due to porous boundaries and energy-momentum deposition from the solar wind, the Earth's radiation belts are continuously driven away from a state of local thermodynamical equilibrium (LTE). With very low particle densities\footnote{The thermal component of the electrons has particle densities of the order of $n\leq 1$ cm$^{-3}$. The warmer electron populations of tens and hundreds of keV are much more dilute with densities several orders of magnitude lower.} and mean free times between collisions of the order of several months to a few years, the Earth's radiation belts are weakly collisional but respond rapidly to departure from LTE by sustaining a wide-range of plasma instabilities that mimic collisions and thermalise the plasma. The plasma instabilities result in a broad spectrum of fluctuations that accelerate particles to relativistic energies on timescales of a few hours to a few days. With electron's energies spanning almost seven orders of magnitude, and reaching as high as several MeV, the Earth's radiation belts are the closest natural laboratory in which charged particles are accelerated close to the speed of light \citep{Roederer}.\\

\noindent From a fundamental physics perspective, it is an observational fact that planetary radiation belts and a plethora of astrophysical plasma environments are efficient particle accelerators. The Earth's radiation belts constitute the most accessible environment to perform detailed \textit{in situ} studies relevant to a wide-range of fundamental physics' problems, such as cosmic rays' acceleration \citep{Cronin99}, upper and middle atmosphere climatology \citep{Turunen}, and even the microphysics of accretion disks \citep{Eliot99, Sironi15}. With electron to magnetic pressure ratio ($\beta_e=2\mu_0n_e k_BT_e/B^2\simeq 0.1-0.01$) and relativistic electron energies ($\gamma m_ec^2\simeq 10$ MeV) in accretion disks comparable to the Earth's radiation belts ($\beta_e \simeq 10^{-3}-10^{-1}$ \& $\gamma m_e c^2 \simeq 1-10$ MeV), kinetic plasma physics near black holes (but far from the event horizon), lies at our doorstep! From an applied physics perspective, and due to their high energies and confinement location around geostationary orbits, radiation belts' particles constitute a threat to satellites orbiting the Earth, and are therefore a research focus for communication and military industries. Driven by fundamental scientific questions and risk mitigation to communication infrastructures, radiation belts' research aims to quantify the acceleration and loss confinement processes of energetic electrons \citep{Cannon13, Horne18, Hands18}.\\ 

\noindent More than 60 years of research following the discovery of the Earth's radiation belts \citep{VanAllen}, plasma physicists have identified two dominant mechanisms responsible for the transport and acceleration of charged particles: 1) \textit{spatially localised} wave-particle interactions driven by small-scale kinetic fluctuations \citep{Thorne10}, and 2) \textit{large-scale} electromagnetic fluctuations induced by global magnetospheric currents and encompassed under the formalism of radial diffusion \citep{Lejosne20}. Both mechanisms can be understood in terms of adiabatic invariants' theory in nearly periodic Hamiltonian systems \citep{Cary09}. In the absence of collisions, the motion of magnetically trapped electrons can be decomposed fully in terms of three separate motions with very distinct timescales: 
\begin{enumerate}
    \item Larmor motion around the local magnetic field ($\Omega \simeq 1-10$ kHz),
    \item The bounce motion between magnetic mirror points ($\omega_b \simeq 0.1-1$ Hz),
    \item The azimuthal drift around the Earth's midplane ($\Omega_d \simeq 0.1-1$ mHz).
\end{enumerate}
In order to break one of the three periodic motions, a wave with a frequency comparable to one of the periodic motions has to interact with the particles. Since the Earth's radiation belts sustain broadband fluctuations with frequencies ranging between $10^{-4}$ Hz and  $10^4$ Hz \citep{Murphy20}, all three invariants can repeatedly be violated. Small-scale kinetic fluctuations accelerate electrons if one of the first two adiabatic invariants $\mu=E_{k\perp}/B$ and $\mathcal{J}=\int p_\parallel \ d s_\parallel$, defined in terms of the perpendicular kinetic energy $E_{k\perp}=|\mathbf{p_\perp}|^2/m$, the local magnetic field amplitude $B$, and the relativistic momentum along the local mean field $p_\parallel = \mathbf{p}\cdot \mathbf{B}/B=m\gamma v_\parallel$, are violated. On the other hand, the second dominant mechanism, radial diffusion, originates in large-scale Alfv\'enic waves in the Pc4 ($\omega \sim 8-25 $ mHz) and Pc5 ($\omega \sim 2-7 $ mHz) range that violate the third adiabatic invariant, i.e. the magnetic flux $\Phi=\int \mathbf{B}\cdot d \mathbf{A}$ \citep{Kulsrud, Roederer}.\\

\noindent In a dipole magnetic field the inverse of the magnetic flux can be expressed more simply as the normalised radial distance in the midplane $L=r/R_E$, in which $R_E$ is the Earth's radius \footnote{It should be kept in mind that when the background dipole magnetic field is deformed on long timescales compared to the drift period, that the third adiabatic invariant does not map into the normalised radial distance. The background magnetic field model used in this communication is dipolar and the third adiabatic invariant can be interpreted as the radial distance.}. Consequently, a collection of particles drift-resonant with Alfv\'enic fluctuations in the Pc5 range experience scattering along the radial distance. This scattering can be modelled statistically in terms of a Fokker-Planck equation and it's observational signature is a diffusive flattening of the distribution function along the radial distance $L^*$. With the first and second adiabatic invariant conserved, particles carried to closer to Earth gain energy through a betatron process \citep{Kulsrud} as they sample a larger magnetic field, whereas particles diffusing to higher radial distances sample a weaker magnetic field, loose energy, and experience greater likelihood for losses at the outer magnetopause boundary \citep{Turner12, George22b}. \\

\begin{figure}[ht!]
 %\label{fig_comparison}
     \centering
     %\begin{subfigure}
         \includegraphics[width=0.44\textwidth]{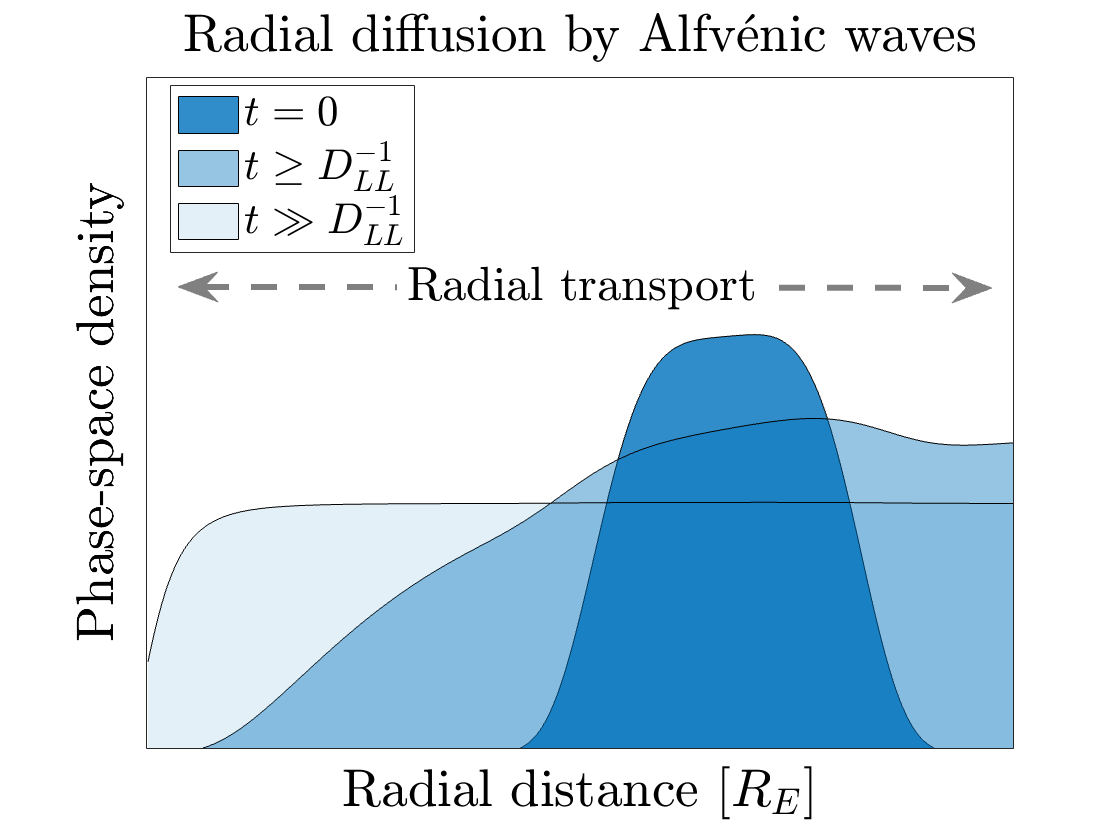}
              \includegraphics[width=0.44\textwidth]{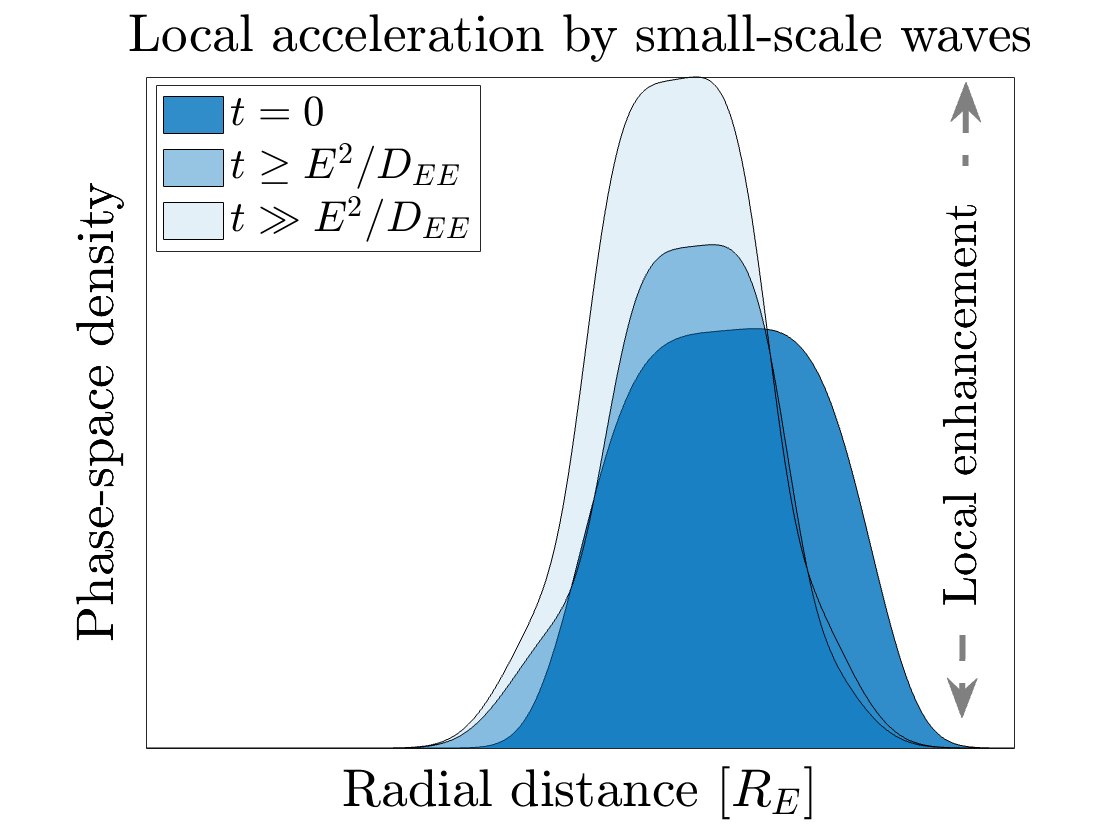}     
     %\end{subfigure}
\caption{\small{{Illustration of the conceptual frameworks for the acceleration of charged particles in planetary radiation belts. Following injection of particles at $t=0$ (darker shaded region) and the generation of plasma instabilities, the phase-space density will be deformed.  On the left panel, Alfvénic fluctuations drive radial diffusion and a flattening of the phase-space density along the equatorial radial distance \citep{Lejosne20}. Particles scattered to lower radial distance sample a larger magnetic field, and gain energy through a betatron process. In comparison, the signature of small-scale fluctuations consists in a localised enhancement along the radial distance \citep{Green04}, as shown on the right panel. The radial shift of the peak in the right panel illustrates that violation of the first and/or second adiabatic invariant results in a change in the third adiabatic invariant as well \citep{Ozturk07, Obrien14a, Desai21}. Both frameworks are expressed in terms of Fokker-Planck equations. Transport by Pc4 and Pc5 Alfv\'enic waves is encoded in a radial diffusion coefficient $D_{LL}$. Transport by small-scale interactions is encoded in an energy diffusion coefficient $D_{EE}$ \citep{Summers05, Shprits06}.}}}\label{fig_comparison}
%\caption{\small{{Illustration of the two conceptual frameworks for the acceleration of charged particles in planetary radiation belts. Following injection of particles at $t=0$ (darkest shaded regions), generated plasma instabilities deform phase-space density. The left panel shows Alfvénic fluctuations driving radial diffusion and a flattening of the phase-space density \citep{Lejosne20}. Particles scattered to lower radial distance gain energy through a betatron process. In comparison, the signature of small-scale fluctuations is seen as a localised enhancement \citep{Horne07}, as shown in the right panel. Both frameworks are expressed in terms of FKEs. Radial diffusion is encoded in a coefficient $D_{LL}$. Transport by small-scale interactions is encoded in an energy diffusion coefficient $D_{EE}$.}}}\label{fig:radbelts3}
\end{figure}

\noindent Similarly, violation of the first and second adiabatic invariants for a collection of particles is also modelled in terms of Fokker-Planck equations \citep{LichLieb}. Contrary to radial diffusion, scattering associated with the first two invariants results in a localised enhancement along the radial distance. From an observational perspective it has therefore been possible to infer which acceleration mechanism is dominant by computing from satellites data the distribution function in terms of the three adiabatic invariants, i.e. $f(\mu, \mathcal{J}, L^*)$ \citep{Green04}. As shown in Figure (\ref{fig_comparison}), if radial diffusion dominates, the distribution function results in a flattening along the radial distance, but if small-scale waves are primary drivers, localised enhancements along the radial distance should be observed. Contemporary observational and modelling studies of the radiation belts rely on this conceptual framework to determine which of the two mechanisms dominate \textit{on timescales of hours to several days} \citep{Chen07, Reeves13, Jaynes18}. 

%%%%%%%%%%%%%%%%%%%%%%%%% 
%Shall I put here examples of inconclusive observations, where it's difficult to determine what is going on? Ask Milla, Leon and Harriet? Check paper by Ian Mann. I would like to be intelligible to 
%the space science community but also to astrophysicists. 
%%%%%%%%%%%%%%%%%%%%%%%%%

\subsection{Benefits of quasi-linear models in the Earth's radiation belts}
\noindent  The theoretical framework to quantify and interpret the dynamical evolution of radiation belts on timescales of a few hours to several days rely exclusively on quasi-linear theories \citep{Kennel66, Falthammar65, Diamond10, Brizard22}. The overwhelming reliance on quasi-linear models in radiation belts' research is not fortuitous as it offers two benefits alternative computational and theoretical approaches lack:
\begin{enumerate}
    \item  \textbf{Computationally inexpensive reduced models}\\ 
The full particle motion requires a 7 dimensional description (three adiabatic invariants with three associated phases plus time). Since energetic electrons span four orders of magnitude in energy, and more than six orders of magnitude in time and space, reduced statistical models are necessary to account for geomagnetic storms occurring on timescales of at least a few hours. Quasi-linear models for small scale wave particle interactions \citep{Summers05, Shprits06} and radial diffusion \citep{Lejosne20} take the form of Fokker-Planck equations that are computationally inexpensive and can be easily implemented in global magnetospheric models.
    \item \textbf{Generalizability} \\ With sparse measurements of electric and magnetic fields responsible for violation of the three adiabatic invariants, quasi-linear models encode the wave-particle interactions in diffusion coefficients that have simple algebraic forms. For instance, radial diffusion coefficients are amenable to parametrisation in terms of ground magnetometers' measurements \citep{Brautigam} that are correlated with fluctuations that drive dynamically radiation belts. Current quasi-linear models can therefore be generalized to periods of unavailable \textit{in situ} measurements.
\end{enumerate} 

Quasi-linear model comparisons with data yields, in several events, accurate estimates of electron fluxes \citep{Reeves13, Thorne13, Jaynes15}. However, dominance of quasi-linear models also stems from the fact that building statistical models that are departing from quasi-linear assumptions is an outstanding theoretical challenge, since it falls into the class of multi-scale nonlinear problems \citep{Dupree66, Orszag67, Dupree72,  Scheko08, Diamond10, Davidson}\footnote{Studies of nonlinear multi-scale problems in kinetic plasma physics have a long history but only recently have we gained sufficient computational power to address them in plasma fusion and astrophysical environments \citep{Scheko2016, Adkins18, kawazura19, Meyrand2019}.}. Moreover, a multi-point satellite methodology that can quantify the evolution of energetic particle fluxes on timescale comparable than a drift period have only recently been developed with the availability of 140 keV-4 MeV electrons GPS fluxes calibrated with the Van Allen Probes \citep{Morley16, Kalliokoski22b}. GPS instruments combined with the Van Allen Probes offers, for the first time, an unprecedented large number of measurement points, and thus providing a broader spatial coverage of the radiation belts and a better temporal resolution in terms of drift-shells. Energetic electron fluxes inferred from GPS electron counts and calibrated against MagEIS and REPT instruments onboard the Van Allen Probe probes \citep{Morley17} can be used to quantify processes that are too fast to be quantified by radial diffusion. Thus, probing radiation belts' processes on timescales of the drift period is now observationally possible, and statistical models that quantify the impact of Pc4 and Pc5 waves on fast timescales comparable to the drift period are missing.\\

\noindent New tools for the radiation belts that can complement and supersede quasi-linear models would have to provide the benefits listed above in order to be incorporated in global models. In this communication we provide the theoretical framework to address the limitation of radial diffusion models and extend radial transport beyond a quasi-linear description. But before doing so, we describe the limits of quasi-linear theory and how it constrains interpretation of radiation belts' observational studies. \\

\subsection{On the need for a new theoretical framework of radial transport}
% Two weaknesses with QLT 2) it neglects nonlinear processes and mode-mode coupling 
%                         1) It's valid on timescales that are long/diffusive (we might miss localised structures caused by ULF)
%                         3) We do not have alternative models that encompass both QLT regime and beyond. This absence of alternative framework prevents the validation of QLT models. For instance, for strong driving and large-amplitudes are QL models valids for all timescales? 
 
\noindent Quasi-linear models in the radiation belts are mean field theories that assume that the \textit{average} interaction of electrons with small-amplitude waves will describe accurately the long timescale evolution of the particles and that nonlinearities arising due to mode-mode coupling or particle orbits can be neglected. Quasi-linear models in the radiation belts therefore contain the following inherent constraints\footnote{Current radial diffusion models also assume that the fluctuations are statistically homogeneous in space. This assumption is known from observations in the radiation belts to be incorrect \citep{Murphy20,Sandhu21}, but can nonetheless be modified under a quasi-linear framework so we have not included it as a limitation inherent to radial diffusion models.}: 
\begin{enumerate}
\item \textbf{Scale separation between fast and diffusive timescales}\\
In quasi-linear models the cumulative effect of many waves on the distribution functions is slow and diffusive \citep{Eijnden}. This slow timescale for diffusion is contrasted with the fast timescales associated with a single encounter/transit time of a wave with the particles. When the timescales for diffusion becomes comparable to the transit time for the wave-particle interactions the quasi-linear hierarchy breaks down \citep{Kennel66}.
\item \textbf{Absence of nonlinear processes}\\ 
 The fast response of the distribution function is assumed to be unperturbed and nonlinear processes such as particle trapping \citep{BGK, Art13, Osmane16} or mode-mode coupling \citep{Scheko2016, Adkins18} are ignored.
\end{enumerate} 
 On the basis of the first constraint, the slow diffusion expressed in terms of a Fokker-Planck equation cannot be used to describe particles acceleration on fast timescales comparable to a single interaction or transit time.  Nonetheless, current diffusion coefficients used for radial transport become sufficiently large during high geomagnetic activity \citep{Brautigam, Ozeke14, Sandhu21} to result in violation of the scale separation quasi-linear constraint. For instance, Figure 4 of \cite{Ozeke14} shows that the diffusion coefficient $D_{LL}$ can be of the order of $10^2-10^3$ days$^{-1}$ for Kp $>5$. Consequently, the diffusion time for a particle to be carried across one drift shell $\Delta L^{*}$ scales between $\tau_D\simeq 15$ minutes and a few minutes. Similarly, the impact of radial transport on losses cannot be quantified in terms of quasi-linear models if particles are depleted on timescales comparable or less than an azimuthal drift period. \cite{Olifer18} shows through observations that fast losses on timescales as short as half an hour can take place during intense magnetic storms. Such transport timescales are inconsistent with a quasi-linear theory relying on a scale separation between fast and slow timescales, with the fast timescales comparable to azimuthal drift orbits of the order of tens of minutes to a few hours.\\
 
 \noindent The second constraint can be justified on the basis that large-amplitude fluctuations are statistically rare occurrences: an electron will be scattered hundreds of times by small-amplitude fluctuations before encountering a large-amplitude wave. However, from a theoretical perspective, waves' amplitudes do not need to be very large for nonlinearities to become comparable to linear terms and for a quasi-linear theory to break down. This property of nonlinear system is well-known among astrophysical and fluid turbulence experts and underlies the assumption of critical balance in which the transit time becomes comparable to the nonlinear interaction time \citep{GS95}\footnote{In critical balance the linear transit timescale (time it takes for an Alfvén wave packet to transit across another Alfvén wave packet) becomes comparable to the nonlinear interaction time. In the nonlinear radial transport problem, the transit timescale (time it takes for a magnetically trapped particle to transit/sample an Alfv\'en wave) becomes comparable to the time it takes for nonlinear effects to be felt. This is quantified in Section \ref{nonlinear section}.}. \\

\noindent Observational evidence and theoretical studies of fast and nonlinear processes at the heart of the Earth's radiation belts have become substantial in the last 15 years but are typically associated with electron-scale whistlers and chorus \citep{Cattell08, Cully08, Bortnik08, Albert12, Mozer13, Malaspina14, Santolik14, Art13, Art15, Agapitov15, Osmane16, Osmane17, Tao20, Omura21} and ion-scale EMIC waves \citep{Hendry19, Grach22, Bortnik22}. With the exceptions of the numerical studies of \cite{Degeling08, Li18}, and extreme driving events such as the one reported by \cite{Kanekal16}, fast and nonlinear radial transport are rarely considered and have yet to be accounted for in global models. However, observational studies demonstrate the existence of large-amplitude fluctuations that can sustain radial transport. For instance, \cite{Hartinger13} demonstrated that transient foreshock perturbations during moderate geomagnetic periods lead to the generation of ultra low frequency (ULF) electric and magnetic fields as high as 10 mV/m and 10 nT, respectively. A statistical study by \cite{Simms18} and an information-theoretic analysis by \cite{Osmane22} characterised the statistical dependence of energetic electron fluxes in the Earth's radiation belts on ULF wave power measured on the ground and at geostationary orbit. Both studies demonstrated that ULF wave power is nonlinearly coupled to energetic electron fluxes\footnote{Counterintuitively, energetic electrons with 100 keV were shown to possess the largest statistical dependency with ULF waves that should only resonate with relativistic electrons $>$ 1 MeV. In Section \ref{fast_symmetric} we provide a non-resonant mechanism, unaccounted by quasi-linear radial diffusion, that can explain the results of \cite{Simms18} and \cite{Osmane22} as a result of ULF driven impulsive acceleration of 100-400 keV.}. And as nonlinear effects become significant, the scale separation constraint of quasi-linear models also breakdown. In this communication, we present a theoretical framework to distinguish quasi-linear diffusion from fast linear and nonlinear processes. \\

\subsection{Next generation of radial transport models for radiation belts}
\begin{figure}[ht!]
\epsscale{1.33}
\plotone{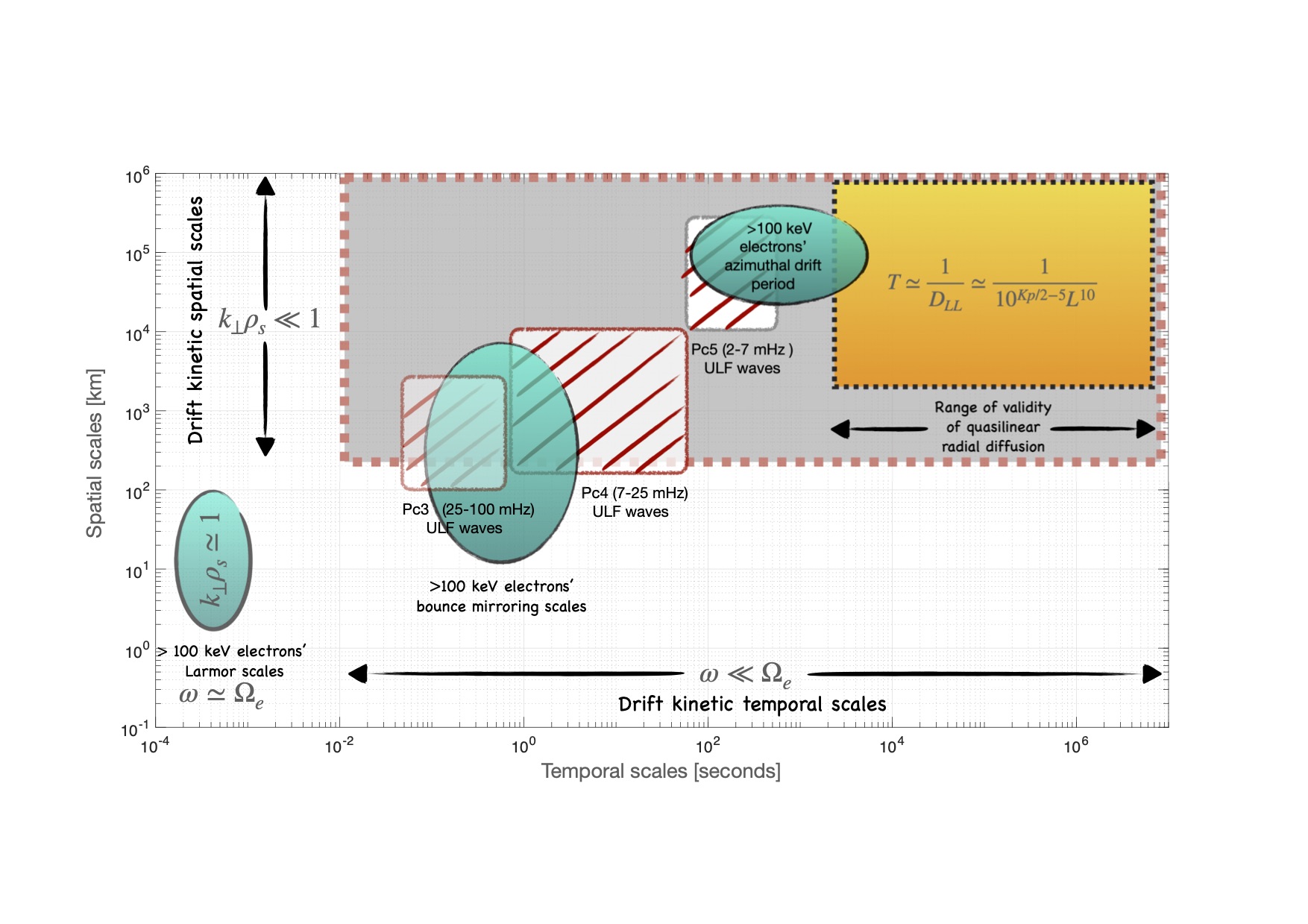}
\caption{Spatial and temporal scales of electromagnetic fields and particle motion in the Earth's radiation belts, and their relation to theoretical limits. The Larmor motion, bounce mirroring motion and azimuthal drift motions are represented as turquoise ellipses. ULF waves ranging from 2-100 mHz are shown in shaded rectangles. The regime of validity of quasi-linear radial diffusion is shown in yellow and the regime covered by drift kinetic, which encompasses quasi-linear radial diffusion is in gray. The left boundary of the quasi-linear regime is computed from the inverse of radial diffusion coefficient obtained from the \cite{Brautigam} for $L=8$ and Kp$=6$, which corresponds to strong geomagnetic conditions.  A $D_{LL}$ at $L=8$ and Kp$=6$ indicates radial transport over one L-shell on a timescale of 30 minutes. For a $>4 $ MeV electron, a drift period is of the order of 3 minutes and radial diffusion over one drift shell after 10 azimuthal drift periods is very fast, but perhaps possible through quasi-linear diffusion. For lower energy electrons, e.g., 400 keV, a complete azimuthal drift is of the order of 20 minutes, and a diffusion over one drift shell in less than two azimuthal drift is inconsistent with the quasi-linear assumption of small changes over fast timescales. It should therefore be kept in mind that the range of validity of quasilinear radial diffusion becomes smaller for less energetic particles. \label{schematic}}
\end{figure}

\noindent The physics of the Earth's radiation belts is nonlinear, high-dimensional and multi-scale and it is not computationally possible to resolve energetic particle motion ranging from milliseconds to hours during geomagnetic storms that can last from several hours to a few days. Consequently, reduced statistical models relying on quasi-linear theories have been developed to predict the dynamical evolution of energetic electrons in terms of physical drivers (i.e. in the solar wind and the magnetosphere). With growing satellite measurements and coverage, we now know that large-amplitude Alfv\'enic fluctuations and fast processes occurring on timescales beyond the reach of quasi-linear radial diffusion are commonly observed in the radiation belts \citep{Li93, Turner12, Hartinger13, Kanekal16, Olifer18}. The current modelling tools are therefore unable to quantify the impact of fast and/or nonlinear radial transport on the energetic electrons, and thus unable to distinguish it from small-scale wave particle interactions. Figure (\ref{schematic}) illustrates the spatial and temporal scales covered by radial diffusion in comparison to characteristic waves and particle motions. In order to characterise processes occurring on fast timescales we need to use a reduced statistical framework that accounts for variations during the drift motion. Drift kinetic models have been developed for decades, mostly for laboratory fusion plasma \citep{Goldston95, Parra08}, but is an ideal starting point to quantify the impact of Pc4 and Pc5 ULF waves on energetic electrons which belongs to the long wavelengths ($k\rho_e \ll 1$) and short frequency limit ($\omega/\Omega_e\ll 1$). 

\noindent

\subsection{Summary of main results}
\begin{itemize}
    \item The choice of the magnetic field model to quantify radial transport is essential for radial transport models and needs to respect Maxwell's equations. If Faraday's equation is violated, we show that Liouville's theorem is also not respected, and thus phase-space density is not conserved. This result also has implications for test-particle experiments in global magnetospheric simulations \citep{Weichao12}. If Faraday's equation is not respected in the simulation box, the construction of the distribution function from the particle trajectories can violate Liouville's theorem.
    \item The linear wave-particle response of the distribution function to a single Alfv\'enic ULF mode consists of three separate terms, two non-resonant processes and one resonant one: 1) a non-resonant modulation of the distribution function in terms of the ULF wave frequency $\omega$, 2) a non-resonant modulation of the distribution function in terms of particle's drift frequency  $\Omega_d$, known as drift echoes, and 3) a drift-resonant response in the instance where the frequency of the ULF wave corresponds the drift frequency of the particle, i.e. $\omega\simeq \Omega_d$. All three responses are a function of the radial gradient in the background distribution function, and the modulation in terms of the ULF wave frequency, sometimes interpreted as evidence of drift-resonance \citep{Claudepierre13}, can also be the product of a non-resonant interaction. 
     \item Zebra stripes' formation do not require drift-resonant interactions, and can be the signature of injected particles in the inner belts in the absence of ULF waves and radial gradients of the distribution function. We argue that the injection events reported by \cite{Zhao13} provide all the necessary ingredients for the formation of zebra stripes.
     \item We derive from the drift kinetic equation a quasi-linear radial diffusion coefficient that consists of two terms. The first term is independent of the wave azimuthal number $m$ and scales as $L^{10}$, and the second term is a function of the azimuthal wave number and scales as $L^6$. The diffusion coefficients accounts for electric and magnetic field fluctuations that respect Faraday's equations, and thus, the separation of the diffusion coefficient in terms of an electric and magnetic $D_{LL}$, as commonly used in the literature \citep{Fei06, Ozeke14, Sandhu21}, is made redundant. Our derived diffusion coefficient can be computed on the basis of the magnetic field wave power alone.
     \item We provide criteria to determine the limit where nonlinear radial transport processes become significant on timescales comparable to the drift period. We demonstrate that when nonlinear effects are accounted for, symmetric and compressive ULF waves can accelerate electrons with energies of the order of $10$ to a few hundreds keV by convecting them inward. This process is a nonlinear generalisation of the mechanism presented by \cite{Parker60} and does not require drift-resonance.  
\end{itemize}

%The electrostatic field, unless its due to some ionspheric wind, makes no fucking sense. So the lack of correlation between delta E and delta B has the insiduous effects of not only D_EE is correlated to D_BB, but that phase-space density is not conserved. I can probably make phase-space density work by using Faraday's law for appropriately chosen electric and magnetic field fluctuations directions. 

% If one wants to introduce empirical models of the electric and magnetic field fluctuations that are not homogeneous, than one can't use currently derived DLL, since they assume space and time homogeneity. 

%In his thesis Ali mentions that ULF wave power differs across different sectors. Is this consistent with homogeneous turbulence? 

%%THINGS TO DO%%%%
%Compute the flux, in the limit where phi is not necessarily equal to Omega_d t. 
%Under which limit do we have the nonlinear term in the phase comparable to the 
% Schulz and Lanzerotti argue that only m=1 contribute to radial diffusion. I suspect that this is because they do not keep delta f_m for m>1. And thus such a statement is incorrect. 
\section{Methodology}
\subsection{Drift kinetic} \label{sec:driftkinetics}
\noindent In a strongly magnetized plasma, charged particle motion can be split into a fast gyration around the local magnetic field and the motion of its guiding centre. The Larmor motion is analytically solvable when the electric and magnetic fields, $\mathbf{E}$ and $\mathbf{B}$, respectively, are assumed constant in time and uniform in space. However, this solution can also be extended to more general electromagnetic fields that are approximately constant on time scales comparable to the Larmor period $\Omega_s^{-1} = m_s/q_sB$ and spatial scales of the order of the Larmor radius $\rho = v/\Omega_s$, where $v$ is the characteristic speed of particles sampling the field, $q_s$ is the charge, and $m_s$ is the rest mass of a particle species ($s=p$ for protons and $e$ for electrons). \\

\noindent We consider a system with characteristic scale size $\mathnormal{l}$ and frequency $\omega\sim v/\mathnormal{l}$. The time and spatial scales of the system are estimated from derivatives of the electromagnetic fields: 
\begin{equation}
\nabla \mathbf{E} \sim \frac{\mathbf{E}}{\mathnormal{l}}, \hspace{2mm} \nabla \mathbf{B} \sim \frac{\mathbf{B}}{\mathnormal{l}}, \hspace{2mm} \frac{\partial \mathbf{E}}{\partial t} \sim \omega \mathbf{E},\hspace{2mm} \frac{\partial \mathbf{B}}{\partial t} \sim \omega \mathbf{B}.
\end{equation}
For a sufficiently strong background magnetic field, the small parameter $\varepsilon$ can be defined as:  
\begin{equation}
\label{rhostar}
\varepsilon=\frac{\rho}{\mathnormal{l}}=\frac{m v}{q B \mathnormal{l}} \ll 1, \hspace{2mm} \frac{\omega}{\Omega}= \frac{m\omega}{qB} \sim \varepsilon \ll 1.
\end{equation}
In this limit the particle does not sense significant variations in the electromagnetic field during characteristic Larmor time and spatial scales. By choosing appropriate coordinates, the fast gyration around the guiding centre can be ignored and a kinetic theory for a collection of particles in a magnetised plasma can be constructed \citep{Parra}.  Put differently, starting from the Lorentz equation or Hamilton's equations to compute the particle motion for slowly varying electromagnetic fields, one can build a statistical description of particles confined by large-scale inhomogeneous magnetic fields \citep{Goldston95, Parra08, Cary09, Hazeltine13}. In the Earth's radiation belts, such a description is therefore appropriate for energetic electrons with Larmor periods $\Omega_e^{-1}\sim 0.1-1$ ms, and interacting with electromagnetic fluctuations in the Pc4 ($\omega \sim 8-25 $ mHz) and Pc5 ($\omega \sim 2-7 $ mHz) ultra-low frequency (ULF) range \footnote{Terrestrial and planetary radiation belts also sustain high-frequency electromagnetic fluctuations with characteristic frequencies $\omega$ comparable to the Larmor frequency $\Omega_s$, e.g. the whistler-mode wave branch at Earth (ELF/VLF) (see \cite{sasha_review} for more detail). The drift-kinetic description relying on the small parameter ordering (\ref{rhostar}) can therefore not be generalised to wave-particle interactions with such modes and one needs to resort to a full Maxwell-Vlasov system \citep{Kulsrud}.}. \\

\noindent In this study, we use a kinetic theory of guiding centres known as drift kinetics to quantify the radial transport of energetic particles interacting with ULF fluctuations. Our starting point is the conservative drift kinetic equation derived recursively by \cite{Hazeltine73}\footnote{A pedagogical step by step derivation of \cite{Hazeltine73} results can be found in the lectures notes of \cite{Parra}. The notes are accessible on \url{http://www-thphys.physics.ox.ac.uk/people/FelixParra/CollisionlessPlasmaPhysics/CollisionlessPlasmaPhysics.html}.}:
%Should add a measure of electric field in the footnote to show that v>>E/B 
%%%%%%%%%%%%%%%%%%%%%%%%
\begin{equation}
\label{DKE}
\frac{\partial}{\partial t} (B \langle f \rangle) + \nabla\cdot(B \dot{\mathbf{r}}\langle f \rangle)+\frac{\partial}{\partial v_\parallel} (B\dot{v_\parallel} \langle f \rangle) +\frac{\partial}{\partial \mu} (B\dot{\mu} \langle f \rangle)=0, 
\end{equation}
%%%%%%%%%%%%%%%%%%%%%%%%
in terms of the gyro-averaged distribution function $\langle f \rangle$ defined as 
%%%%%%%%%%%%%%%%%%%%%%%%
\begin{equation}
\langle f \rangle=\frac{1}{2\pi}\int_0^{2\pi} f(\mathbf{r}, v_\parallel, \mu, \theta_g, t) d\theta_g,
\end{equation}
%%%%%%%%%%%%%%%%%%%%%%%%
the guiding-centre position vector $\mathbf{r}$, parallel velocity $v_\parallel$, and gyrophase $\theta_g$,  first adiabatic invariant $\mu$,
\begin{equation}
\label{First_adiabatic_invariant}
\mu=\frac{1}{2} \frac{m_ec^2 (\gamma^2-1)}{B}\sin^2(\alpha).
\end{equation}
Equation (\ref{First_adiabatic_invariant}) for $\mu$ is written in terms of the parallel velocity $v_\parallel$, pitch-angle $\alpha=\tan^{-1} (v_\perp/v_\parallel)$ and relativistic Lorentz factor $\gamma=(1-v^2/c^2)^{-1/2}$ to account for the relativistic correction that appear for particles with kinetic energies $E_c=m_ec^2 (\gamma-1)$ comparable to the electron rest mass $m_e c^2=511$ keV \footnote{In the Earth's radiation belts particles are injected at energies of the order of 1-100 keV, but are accelerated to energies comparable to the rest mass and as high as a few MeV \citep{Turner17}. It is therefore crucial to keep track of the relativistic effects. In our particular problem limited to equatorially trapped particles, the relativistic effects appear in the first adiabatic invariant but an extension to non-equatorially trapped particles will require a relativistic representation of the drift kinetic equation in terms of the parallel momentum $p_\parallel=m_e \gamma v_\parallel$.} .\\

\noindent The appearance of the magnetic field amplitude $B$ in Equation (\ref{DKE}) originates from the Jacobian when one transforms variables from $(\mathbf{r}, \mathbf{v})$ to $(\mathbf{r}, \mu, v_\parallel, \theta_g)$. In the absence of collisions, conservation of phase-space density for a collection of guiding centre particles requires that the following equation be respected:
\begin{equation}
\label{Liouville}
\frac{\partial}{\partial t} (B) + \nabla\cdot(B \dot{\mathbf{r}})+\frac{\partial}{\partial v_\parallel} (B\dot{v_\parallel}) +\frac{\partial}{\partial \mu} (B\dot{\mu})=0, 
\end{equation}
Equation (\ref{Liouville}) is a statement of Liouville's theorem, and is a function of the electromagnetic field model and of the guiding centre's particle trajectory. In open systems the impact of electromagnetic fluctuations will naturally lead to transport to the boundaries, and thus to irreversible losses. Terrestrial and planetary radiation belts are not closed systems and the inner and outer boundaries allow for particles' injection and losses \citep{Millan07, Aryan20, Walton22}. However, the wave-particle interactions with ULF waves, in the absence of boundary effects, have to conserve phase-space density. Equation (\ref{Liouville}) is therefore a different statement, independent of the presence of porous boundaries, and determines whether phase-space density, and thus the number of particles, are conserved in a closed phase-space volume. The choice of a fields' model that violate phase-space density is unphysical and necessarily results in erroneous quasi-linear diffusion coefficients.  For instance if a field model that does not conserve phase-space density is chosen, and boundary effects are added, the resulting losses would either be amplified or underestimated. Liouville's theorem can therefore be used as a constraint for the electromagnetic fields, as shown in Section (\ref{sec:review}). \\

\noindent The particle guiding-centre description in the $(\mathbf{{r}}, v_\parallel, \mu)$ phase-space, for a given problem, is a function of the strength of the electric field when compared with the magnetic force.  If the characteristic speed of the particle is comparable to the $E\times B$ drift, additional sources for perpendicular drifts can be ignored. For instance, in the collisionless MHD approximation, the perpendicular velocity of ion and electron fluids are to first order comparable to the $E\times B$ drift and MHD fluid equations can be derived from the kinetic equation with the perpendicular velocity approximated by the $E\times B$ \citep{Hazeltine18}. However if additional drifts are comparable in size to the $E\times B$ drift, or if the characteristic speed of a particle population is much greater than the $E\times B$ drift, perpendicular velocities of ions and electrons are going to decouple, and  additional drifts have to be taken into account. \\

\noindent \cite{Hazeltine73} suggests two regimes to account for the ordering of the $E\times B$ in a given problem: the high flow regime, with strong perpendicular electric fields $|\mathbf{E_\perp}| \simeq v B$ , and the low flow regime, with small electric fields, making the $\mathbf{E \times B}$ drift small compared to the characteristic speed of the particle. Thus, in the high flow regime, the perpendicular electric field can be comparable to the magnetic force, and the $\mathbf{E}\times\mathbf{B}$ drift is the dominant drift. In the low flow ordering, the perpendicular electric field cannot balance the magnetic force, and since the $\mathbf{E}\times\mathbf{B}$ drift is not dominant, additional magnetic drifts, such as the curvature drift and the magnetic gradient drift $-\mu \nabla B$ have to be included. \\

\noindent For an application to energetic electrons in the Earth's radiation belts possessing kinetic energy ranging between hundreds of keV and a few MeV, and interacting with ULF waves, the low flow regime is the correct limit since it accounts for the dominance of the magnetic gradient drift over the $\mathbf{E}\times\mathbf{B}$ drift. Dominated by the magnetic gradient drift, energetic electrons in the Earth's radiation belts perform one complete azimuthal loop on timescales ranging from few minutes, for MeV electrons, to a few hours for 50 to a few hundreds of keV electrons. In comparison, the additional drifts present in Equation (\ref{drifts_LF}) are weaker on such timescales. However, we keep track of additional drifts since they are cumulatively responsible for irreversible transport of particles across drift shells on long timescales of several hours to a few days \citep{Lejosne20}. \\

In the low flow regime, the position is to first order in the small parameter $\varepsilon$ evolving according to\footnote{Terms of order $\varepsilon^2\simeq (\rho/l)^2$ are neglected.}: 
\begin{eqnarray}
\label{drifts_LF}
\mathbf{\dot{r}}=\left(v_\parallel + \frac{ \mu}{q_s} \mathbf{b\cdot\nabla\times b}\right)\mathbf{b}-\frac{\mathbf{E}\times \mathbf{b}}{B}+\frac{v_\parallel^2}{\Omega_s}\mathbf{b}\times\mathbf{(b\cdot\nabla) b}+\frac{\mu}{q_s B} \mathbf{b}\times\nabla B, 
\end{eqnarray}
in terms of the local magnetic field direction $\mathbf{b}=\mathbf{B}/B$. The five terms are, respectively,  the velocity parallel to the magnetic field, the Ba\~nos parallel drift, the $E$ cross $B$ drift, the curvature drift and the magnetic gradient drift. Coupled with particle's position, the evolution of the parallel velocity is given by
\begin{eqnarray}
\label{drifts_LFv}
{\dot{v}_\parallel}=\left[\frac{q_s}{m_s}\mathbf{E}-\left(\frac{\mu+\tilde{\mu}}{m_s}\right)\nabla B\right]\cdot\mathbf{b}+\frac{v_\parallel}{\Omega_s}\left[\mathbf{b}\times(\mathbf{b}\cdot\nabla)\mathbf{b}\right]\cdot\left(\frac{q_s}{m_s}\mathbf{E}-\frac{\mu}{m_s}\nabla B\right)-v_\parallel\frac{ \mu}{q_s}\mathbf{b}\cdot\nabla\left[\mathbf{b}\cdot\nabla\times\mathbf{b}\right],
\end{eqnarray}
in terms of the correction to the first adiabatic invariant 
\begin{equation}
\tilde{\mu}=-(v_\parallel \mu/ q_s B)\mathbf{b}\cdot\nabla\times\mathbf{b}.
\end{equation} 
\noindent The evolution equation for the first adiabatic invariant is given by

\begin{eqnarray}
\label{drifts_LFmu}
\dot{\mu}=-m_sv_\parallel\mathbf{b}\cdot\nabla\tilde{\mu}-(q_s\mathbf{b}\cdot{\mathbf{E}}-\mu\mathbf{b}\cdot{\nabla}B)\frac{\partial \tilde{\mu}}{\partial v_\parallel}.
\end{eqnarray}
Combining Equations (\ref{DKE}), (\ref{drifts_LF}), (\ref{drifts_LFv}) ,(\ref{drifts_LFmu}) with a model of electromagnetic fields consistent with Liouville's theorem (Equation \ref{Liouville}), one can quantify the evolution of the distribution function for a collection of energetic particles in planetary magnetosphere on timescales much shorter than quasi-linear times and therefore comparable to the azimuthal drift periods of magnetically confined particles. The drift kinetic approach therefore provides the foundation for a variety of models (linear, quasi-linear, nonlinear, with or without porous boundaries) to account for ULF radial transport of particles. \\

\noindent \textit{A priori} the set of drift-kinetic equations are nonlinear and therefore not easily tractable analytically. However, the equations can be simplified when  energetic particles confined to the equator of the Earth's magnetosphere are studied. Equatorially trapped particles have pitch-angles $\alpha=\tan^{-1} v_\perp/v_\parallel \simeq \pi/2$ and thus $v_\parallel=0$. Moreover, the absence of ULF parallel electric field results in $\dot{\mu}=0$, $\dot{v}_\parallel=0$, and the evolution of the conservative kinetic equation for the distribution function $f(\mathbf{r}, v_\parallel=0, \mu=\mu_c)$, for a fixed magnetic moment $\mu_c$, takes the simple form: 
\begin{equation}
\label{DKE_finalform}
\frac{\partial}{\partial t}(B\langle f \rangle)+\nabla\cdot(B\mathbf{\dot{r}}\langle f\rangle)=0.
\end{equation}
In the remaining part of this communication, we will use kinetic Equation (\ref{DKE_finalform}) to describe equatorially trapped particles and leave the generalisation to non-equatorial particles ($\alpha \neq \pi/2$) for future work\footnote{Since ULF waves propagate off the equatorial plane \citep{Sarris22}, additional drifts have to be accounted for non-equatorial particles.}. But before solving the kinetic equation we need to complement it with an electromagnetic fields' model. \\% In the next section we present a brief review of the electromagnetic field models that have been used in the past 60 years to theoretically model radial transport.  \\ %If a particle is in the equatorial plane, does it end up out of the plane as it experiences diffusion? I guess, only if the second adiabatic invariant is conserved. If violated, it will result in a diffusion in L and pitch-angle (?). 

% Equation of motion, Liouville's theorem. keep the final form for alpha=pi/2 

\subsection{Review of electromagnetic fields used for radial diffusion models} \label{sec:review}
\noindent In this section, we review the electromagnetic fields that have been chosen to model ULF radial transport. We focus solely on electromagnetic models that can be written analytically and that have been used to model coefficients for Fokker-Planck equations. Our aim in this section is also to demonstrate that an arbitrary choice of electromagnetic fields can violate conservation of phase-space density given by Equation (\ref{Liouville}). \\
 
\subsubsection{Mead field}\label{MEAD}
\noindent The Mead field \citep{Mead} consists in the superposition of two perturbations: an azimuthally symmetric fluctuation with amplitude $S(t)$ and an azimuthally asymmetric fluctuation $A(t) r \cos(\varphi)$ superposed to a background magnetic dipole field of amplitude $B_E R_E^3/ r^{3}$. The Mead model has the benefit to be mathematically simple yet to contain all the necessary ingredients, through the presence of an asymmetric perturbation, for the violation of the third adiabatic invariant experienced by a collection of magnetically trapped particles. The Mead field was therefore a natural choice for early models of radial diffusion \citep{Falthammar65, Schulz69, Schulz74} and has been used as the field model for empirical \citep{Brautigam, Cunningham16, Sarma20} and theoretical studies \citep{Lejosne19, Osmane21a} of quasi-linear radial diffusion in the past decades. \\

\noindent In our analysis, we will argue that the choice of the Mead field is preferable for analytical studies. As stated in Section \ref{sec:driftkinetics} we will focus here exclusively on equatorially trapped particles, but note that a generalisation to non-equatorial particles can also be done. We also generalise the Mead field to anti-symmetric perturbations with azimuthal wave numbers $m\neq 1$ This generalisation of the Mead field will have little incidence for the linear and quasilinear radial transport equation since the perturbed distribution function due to various $m$ modes are independent from one another another. In the nonlinear regime of radial transport in turn, as shown in Section (\ref{nonlinear section}), mode coupling of various $m$ modes can interact with one another. \\

\noindent Thus, the magnetic field for equatorial particles can be written in cylindrical coordinates ($r, \varphi, z$), with $r$ the radial distance, and $\varphi$ the azimuthal angle, and $z$ the cylindrical axis direction:
\begin{equation}
\mathbf{B} =-\left(\frac{B_E R^3_E}{r^3}- S(t) -\sum_m A_m(t) r e^{im\varphi}\right) \hat{z} 
\end{equation}
in terms of the magnetic field dipole moment $B_E$ and the Earth's radius $R_E$. The original simplified Mead field can be recovered by setting $m=1$ and taking the real part in the Fourier sum decomposition. This generalisation of the Mead to some arbitrary number of $m$ modes is based on observational measurements demonstrating that the Earth's radiation belts can sustain a broad spectrum in $m$ of ULF waves \citep{Sarris14, Barani19} and that the $m=1$ model is inaccurate during large driving  conditions quantified by a geomagnetic Kp index greater than 4 \citep{Lejosne13}.\\

\noindent Using Faraday's law, the inductive electric field can be written as: 
\begin{equation}
\mathbf{\delta E} =\left(\frac{1}{7} r^2 \sum_m \frac{i\dot{A}_m}{m} e^{im\varphi}\right)\hat{r} -\left(\frac{r \dot{S}}{2}+\frac{8 r^2}{21}\sum_m \dot{A}_m e^{im\varphi}\right)\hat{\varphi}.
\end{equation}
The above Mead field results in two drifts, the $E \times B$ drift,
\begin{equation} 
-\frac{\mathbf{\delta E}\times \mathbf{b} }{B} =\left( \frac{r \dot{S}}{2B}+\frac{8 r^2}{21 B} \sum_m \dot{A}_m e^{im\varphi}\right)\hat{r}+\left(  \frac{1}{7} \frac{r^2}{B} \sum_m \frac{i\dot{A}_m}{m} e^{im\varphi}\right) \hat{\varphi}
\end{equation}
and the magnetic gradient drift\footnote{On the other hand non-equatorial trapped particles ($\alpha \neq \pi/2$) will experience the Ba\~nos and curvature drift.} written for the electron charge $e=-q$: 
\begin{equation}
\frac{\mu}{q\gamma} \frac{\nabla B \times \mathbf{b}}{B}=\left(\frac{3\mu B_0}{q B\gamma r}+\frac{\mu}{q\gamma B} \sum_m A_m e^{i m\varphi}\right)\hat{\varphi} -\left(\frac{\mu}{q\gamma B}\sum_m i m A_m e^{im\varphi}\right) \hat{r}, 
\end{equation}
written in terms of the background magnetic dipole magnitude $B_0=B_ER^3/r^3$ and the magnitude $B=B_0- S(t) -\sum_m A_m(t) r e^{im\varphi}$. \\

\noindent Conservation of phase-space density for a collection of particles trapped in a magnetic dipolar  field and interacting with ULF fluctuations can be written as: 
%%%%%%%%
\begin{eqnarray}
\label{eq9}
\frac{\partial B} {\partial t} +\nabla\cdot(B \mathbf{\dot{r}})+\cancelto{0}{\frac{\partial (B\dot{v}_\parallel)}{\partial v_\parallel}}+\cancelto{0}{\frac{\partial (B\dot{\mu})}{\partial \mu}}&=&  \frac{\partial B} {\partial t} + \nabla\cdot\left(\mathbf{\delta E} \times \hat{z}-\frac{\mu}{q\gamma} \nabla B \times \hat{z}\right) \nonumber\\
&=&\hat{z} \cdot\underbrace{\left( \frac{\partial \mathbf{B}} {\partial t} +\nabla \times \mathbf{\delta E}\right)}_{\text{=0, by Faraday's law.}}-\frac{\mu}{q\gamma}\underbrace{{\nabla\cdot \left(\nabla B \times \hat{z}\right)}}_{\text{=0, identically.}}  \\
&=& 0 \nonumber
\end{eqnarray}
%%%%%%%%
in which the first term on the right-hand side of Equation (\ref{eq9}) is the projection of Faraday's law along the background magnetic field direction $\mathbf{b}$. Since we are focussing solely on equatorially trapped particles for the Mead field we can switch to cylindrical coordinates $(r, \varphi, \theta=z)$.  Thus, phase-space density is always conserved for particles confined in a magnetic dipole if Faraday's law projected unto the mean field is respected. A corollary is that the choice of time-varying electric fields that does not satisfy Faraday's law does not satisfy Maxwell's equation and also have the additional undesirable consequence that it does not conserve phase-space density. Since the electric field in the Mead model satisfies Faraday's equation, the Mead field conserves phase-space density. The choice of the Mead field is therefore appropriate to develop a kinetic theory of radial diffusion. 

\subsubsection{Asymmetric background field}
\noindent \cite{Elkington03} argued that enhanced radial diffusion could take place by accounting for an asymmetric background magnetic field attributed to periods of high solar wind pressure and solar wind speeds. In their model, \cite{Elkington03} chose a background dipole magnetic field with a superposed perturbation $\Delta B$: 
%%%%%%%%
\begin{equation}
\label{B_EK}
B^{EK}(r, \varphi) = \frac{B_E R^3}{r^3}+\Delta B(r) \cos(\varphi)
\end{equation} 
%%%%%%%%
Here the azimuthal angle is chosen to be zero at noon and we denote the model as $B^{EK}$ to distinguish it from the Mead field. In addition to the background field, ULF wave perturbations in the electric and magnetic field are chosen to be the sum of azimuthal Fourier components: 
%%%%%%%%
\begin{equation}
\label{EK_E}
\mathbf{\delta E} = \sum_m \delta \mathbf{E}_m (r, t) e^{im\varphi}
\end{equation}
%%%%%%%%
\begin{equation}
\label{EK_B}
\mathbf{\delta B} = \sum_m \delta \mathbf{B}_m (r,t) e^{im\varphi}
\end{equation}
%%%%%%%%
The above perturbations have no particular polarisation, with unspecified toroidal $(\delta E_{r,m})$ and poloidal  $(\delta E_{\varphi,m})$ electric fields components, and the relation between the magnetic and electric components are ignored.  In order for these fields to conserve phase-space density, two constraints have to independently hold: the first one applies to the stationary background magnetic field given by (\ref{B_EK}),
%%%%%%%%
\begin{equation}
\nabla\cdot\left(\nabla B^{EK} \times \hat{z}\right)=0,
\end{equation}
%%%%%%%%
and is respected for a perturbation $\Delta B (r)$ with an existing first derivative along the radial direction. The second one is Faraday's law for the time varying electric and magnetic perturbations (\ref{EK_E})-(\ref{EK_B})
which results in the following three constraints for the electric and magnetic field amplitudes: 
%%%%%%%%
\begin{equation}
\label{C1_EK}
\frac{\partial }{\partial z}  \delta E_{m, \varphi}= \frac{\partial }{\partial t}\delta B_{m, r}
\end{equation}
%%%%%%%%
\begin{equation}
\label{C2_EK}
\frac{\partial }{\partial z}  \delta E_{m, r}= -\frac{\partial }{\partial t}\delta B_{m, \varphi}
\end{equation}
%%%%%%%%
\begin{equation}
\label{C3_EK}
\frac{1}{r}\frac{\partial }{\partial r}  (r\delta E_{m, \varphi})-im \delta E_{m, r}=- \frac{\partial }{\partial t}\delta B_{m, z}
\end{equation}
%%%%%%%%
For the sake of simplicity we assume that the magnetic field perturbations have no poloidal ($B_\varphi=0$) or toroidal ($B_r=0$) component, and thus only require the constraint (\ref{C3_EK}) to be enforced. In terms of a Fourier decomposition in time ($\delta B_{m, z} \sim e^{-i \omega t}$), Equation (\ref{C3_EK}) can thus be written as:
%%%%%%%%
\begin{equation}
\label{Faraday_constraint}
\frac{1}{r}\frac{\partial }{\partial r}  (r\delta E_{m, \varphi})-im \delta E_{m, r}=i \omega \delta B_{m, z}.
\end{equation}
%%%%%%%%
This last equation constrains the choice of a poloidal or toroidal electric fields. For a purely toroidal electric field ($\delta E_{m, r} \neq 0$, $\delta E_{m, \varphi} = 0$) the complex coefficients have the following constraint: $\delta E_{m,r} =-\omega \delta B_{m,z}/ m$. For a purely poloidal electric field ($\delta E_{m, \varphi} \neq 0$, $\delta E_{m, r} = 0$) that has no radial dependence the following equality must be held: $\delta E_{m,\varphi}/r =\omega \delta B_{m,z}/ m$. We therefore conclude that the asymmetric model used to compute the radial diffusion coefficients in \cite{Fei06} does not conserve phase-space density and that the diffusion coefficients derived on the basis of this field model yields unphysical results. The violation of Faraday's law in the model \cite{Fei06} has already been noted by \cite{Lejosne19} and shown to enhance the diffusion coefficient by a factor of 2. By treating this problem kinetically, we have also shown that it violates Liouville's theorem.\\

\noindent Equation (\ref{Faraday_constraint}) also provides a constraint on the electrostatic model ($\mathbf{\nabla \times \delta E}=0$) of \cite{Falthammar65}. For the case of a purely poloidal component, Faraday's equation requires  $\frac{1}{r}{\partial }(r\delta E_{m, \varphi})/{\partial r}  =0$, and thus $\delta E_{m, \varphi} \sim  1/r$. The assumption of a poloidal field independent of the radial distance used in \cite{Falthammar65} therefore also violates Liouville's theorem and yields unphysical radial transport coefficient. \\

\noindent We note that both the \cite{Fei06} electromagnetic model and \cite{Falthammar65} electrostatic models can nonetheless be corrected by accounting for Faraday's law. This correction can be done by enforcing Equation (\ref{Faraday_constraint}) when computing the diffusion coefficient with or without the asymmetry introduced by \cite{Elkington03}. On the basis of this section and the previous one, we choose to use the Mead model since it conserves phase-space density for equatorially trapped particles and already contains all the key ingredients to model radial transport in the Earth's radiation belts \footnote{A reader might then wonder why not simply use the field in \cite{Fei06} after enforcing the constraint given by  Equation (\ref{Faraday_constraint}). The short answer is that the main benefit in using the asymmetric field results in a modification of the diffusion coefficient of the order of $\Delta B^2/B^2 \ll 1$. This modification is therefore negligible.}. \\

\section{Linear, quasi-linear and nonlinear limits of radial transport}
\subsection{Multiscale dynamics \& separation between slow and fast variables}\label{sec:multiscale} 
\noindent In this section we develop a mean-field theory from the drift-kinetic equation (\ref{DKE}) for charges confined in a magnetic dipole and interacting with ULF fluctuations given by the Mead field (\ref{MEAD}). We will solely focus on particles confined in the equatorial plane ($\alpha=\pi/2$) and leave the more involved case of particles bouncing off at mirror points at higher and lower latitudes to future studies. In order to build a mean-field theory we separate slow changes in the third adiabatic invariant $L^*$ and background quantities and fast changes in the associated invariant phase and fluctuation timescales parts of the distribution function\footnote{A scale separation between fast and slow motion is the basis of quasilinear theories in astrophysical plasmas \citep{Kulsrud, Schekochihin_notes, Diamond10}. This approach is identical to the one performed in \cite{Kennel66} for a quasilinear theory of magnetised charged particles interacting with plasma waves of frequencies comparable to the Larmor frequency. The resulting diffusion models written in the form of Fokker-Planck equations would not be possible without such a scale separation and constrains the timescales upon which the quasilinear theory can be used.}: 
%%%%%%%%
\begin{equation}
\label{def_average2}
f(r, \varphi, t) = f_0(r, \varepsilon^a \varphi, \varepsilon t) + \delta f (r, \varphi, t)
\end{equation}
%%%%%%%%
in which $r$ is the radial distance at the equator, $\varphi$ is the azimuthal angle $\varphi \in [0, 2\pi]$, and the small parameter $\varepsilon$ characterises the scale separation between large-scale and small-scale parts of the distribution. We note that it is possible to build a background distribution function with azimuthal dependence. For instance, in the presence of an azimuthal dependent source or loss term that evolves slowly in time compared to the azimuthal drift period of the particles. Such an azimtuhal dependence can then be accounted for in terms of $\varepsilon^a \varphi$, for $a>0$, and resulting in $\partial f_0/\partial \varphi = \varepsilon ^a f_0$. But for simplicity, and comparison with previous radial transport model, we will assume that the background distribution function has no dependence on the azimuthal angle, i.e., 
%%%%%%%%
\begin{equation}
f_0=f_0(r,\varepsilon t)
\end{equation}
%%%%%%%%
Formally, this equilibrium distribution can be defined as the average of the exact distribution function over the range of azimuthal angle and over timescales that are intermediate between the fast and the slow ones:
%%%%%%%%
\begin{equation}
\label{def_average}
f_0=f_0(r,t) =\langle f(r, \varphi, t) \rangle =\frac{1}{2\pi\Delta t} \int_{t-\Delta t/2}^{t+\Delta t/2} dt' \int_0^{2\pi} d\varphi f(r, \varphi, t')
\end{equation}
%%%%%%%%
for $\omega^{-1}\ll \Delta t \ll t_{eq}$, where $\omega\sim \frac{1}{A_m}\frac{d A_m}{dt} \sim  \frac{1}{S}\frac{d S}{dt}$ denotes the frequency of ULF fluctuations, and $t_{eq} \sim  \frac{1}{f_0}\frac{\partial f_0}{\partial t}$, the timescale for an equilibrium in the distribution to form. This definition of $f_0$ constrains the time and spatial scales upon which the background distribution function can be computed. It is shown in Section \ref{sec:linear} that particles with azimuthal drift frequencies $\Omega_d$, as defined by Equation (\ref{Azimuthal}), comparable to ULF wave frequency with azimuthal mode number $m$ experience resonance. Thus, since resonance requires $\omega\simeq m\Omega_d$, Equation (\ref{def_average}) also constrains the evolution of the background  background distribution function $f_0$ on timescales much larger to $1/m\Omega_d$. For the mode $m=1$, the implication on the quasi-linear theory is that the diffusion cannot take place on timescales comparable to the azimuthal drift periods.  \\ 

\noindent For equatorial particles with a conserved first adiabatic invariant $\mu$ interacting with a Mead field, the kinetic Equation (\ref{DKE_finalform}) takes the form: 
%%%%%%%%
\begin{eqnarray}
\label{starting_equation}
g B_0 \frac{\partial f}{\partial t} + \frac{3\mu B_0}{q\gamma r^2} \frac{\partial f}{\partial \varphi} &+& \sum_me^{im\varphi}\left[\frac{\mu A_m}{q\gamma r}+i \frac{r\dot{A}_m}{7m}\right]\frac{\partial f}{\partial \varphi}=-\left[\frac{r \dot{S}}{2} +\sum_me^{im\varphi}\left(\frac{8r^2\dot{A}_m}{21}-im \frac{\mu A_m}{q\gamma }\right)\right]\frac{\partial f}{\partial r}  
\end{eqnarray}
%%%%%%%%
with the function $g(r, \varphi, t)= 1- S(t)/B_0-\sum_m e^{im\varphi}r A_m(t)/B_0$.  We now define the drift frequency for equatorially trapped particles
\begin{equation}
\label{Azimuthal}
\Omega_d=3\mu/q\gamma r^2
\end{equation}
in terms of the first adiabatic invariant $\mu$, and decompose the perturbed fluctuations along the azimuthal angle in Fourier space\footnote{The generalisation of the Mead field in section (\ref{MEAD}) was already expressed in terms of Fourier modes for the anti-symmetric perturbations.}:
%%%%%%%%
\begin{equation}
\label{basis}
f(r, \varphi, t)=f_0(r, t)+\sum_me^{i m\varphi} \delta f_m(r,t).
\end{equation}
Replacing the decomposition (\ref{basis}) in Equation (\ref{starting_equation}) for $m=0$ (the azimuthal average), and averaging over time according to (\ref{def_average}) results, as shown in Appendix \ref{AppendixA}, in the quasi-linear equation:
%Recompute whole QLT by keeping things as they stand and replace the partial derivative of delta f_m with Eq 20. 
\begin{eqnarray}
\frac{\partial f_0}{\partial t}&=&-\sum_m \left[ \frac{i m\mu}{qB_0\gamma r}\frac{\partial}{\partial r} \left(r\langle A^*_{m}\delta f_m\rangle\right) -\frac{r}{B_0}\frac{\partial}{\partial t}\langle {A}^*_{m}\delta f_m\rangle +\frac{8}{21}\frac{1}{rB_0}\frac{\partial }{\partial r}\langle r^3\dot{A}^*_{m}\delta f_m\rangle \right]
\label{kinetic4}
\end{eqnarray}
%%%%%%%%
The right-hand side of (\ref{kinetic4}) describes the slow evolution of the background distribution due to the effect of fluctuations. As its often the case in space and astrophysical plasmas we need a closed equation for the evolution of the background. The correlation $\langle \delta f_m A^*_{m} \rangle$\footnote{Since the magnetic field amplitude is real, we can write the Fourier coefficient $A_{-m}=A^*_{m}$.} can be computed if we can write an equation for the perturbation $ \delta f_m$, replace it in Equation (\ref{kinetic4}), and take the average defined by  (\ref{def_average}). The detail of this calculation can be found in the Appendix \ref{AppendixB}, and results in the following nonlinear equation for the perturbation:
%%%%%%%%
\begin{eqnarray}
\label{QLT2} 
\frac{\partial \delta f_m}{\partial t}+\underbrace{i m\Omega_d \delta f_m}_{\text{particle streaming}} =\underbrace{\frac{A_m r}{B_0}\frac{\partial f_0}{\partial t}-\left( \frac{8r^2\dot{A}_m}{21 B_0}-im \frac{\mu A_m}{qB_0\gamma } \right)\frac{\partial f_0}{\partial r}}_{\text{Linear wave-particle interaction}}\underbrace{-\sum_{m'} \mathcal{Q}[S, A_{m-m'}; \delta f_{m'}].}_{\text{Nonlinear wave-particle interaction}}
\end{eqnarray}
%%%%%%%%
The three terms that control the evolution of the perturbed distribution in (\ref{QLT2}) represent free ballistic motion, or streaming, linear wave-particle interaction, and nonlinear wave-particle interaction. The term $\mathcal{Q}$, given by Equation \ref{nonlinear_term}, is negligible in the limit $m\Omega_d \delta f_m \gg \mathcal{Q}$, otherwise it has to be accounted for and will result in mode-mode coupling even if the ULF wave amplitudes are considered small, i.e., $\delta B\simeq r A_m \ll B_0$ and $S(t)\ll B_0$. In the next sections we solve these equations in the linear and quasi-linear regimes and describe the conditions in which nonlinear processes become significant. \\

\subsection{Linear theory and radial transport on fast timescales}\label{sec:linear}
\noindent In the linear theory we consider small perturbations of the equilibrium that evolve on fast time scales comparable to the drift period. All nonlinear terms can then be ignored and the background distribution is assumed as constant in time, i.e. $f_0(t)=$ const. The linear equation is therefore given by:  
%%%%%%%%
\begin{eqnarray}
\label{QLT3}
\frac{\partial \delta f_m}{\partial t}+{i m\Omega_d \delta f_m} =-\left( \frac{8r^2\dot{A}_m}{21 B_0}-im \frac{\mu A_m}{qB_0\gamma } \right)\frac{\partial f_0}{\partial r}.
\end{eqnarray}
%%%%%%%%
Equation (\ref{QLT3}) is linear and can be solved by Duhamel's principle for the initial condition $\delta f_m(r, t=0)$ as: 
%%%%%%%%
\begin{eqnarray}
\label{Linear_EQ1}
\delta f_m(r,t) =\underbrace{\delta f_m(r, 0)e^{-i m\Omega_d t}}_{\text{Ballistic response}} \underbrace{-\frac{\partial f_0 (r)}{\partial r}  \int_{0}^t dt' \ e^{i m\Omega_d (t'-t)} \left( \frac{8r^2\dot{A}_m(t')}{21 B_0}-im \frac{\mu A_m(t')}{qB_0\gamma } \right)}_{\text{Linear wave-particle response}}.
\end{eqnarray}
%%%%%%%%
The first term on the right-hand side of Equation (\ref{Linear_EQ1}) is a ballistic mode that we will see as responsible for the formation of transient structures in the phase-space $(r,\varphi)$. The second term on the right-hand side is the linear wave-particle response of the distribution function to the ULF wave. This problem is almost identical to the self-consistent electrostatic problem solved by \cite{Landau}, in which perturbations of the background distribution results in growing or decaying fluctuations. However, the radial transport problem in the linear regime, contained in Equation (\ref{Linear_EQ1}), is simpler than the one solved by \cite{Landau}, since the resonant energetic electrons with densities of the order of $0.1$ \% or less are passive tracers and self-consistent effects can be to very good degree of accuracy ignored\footnote{A basic dimensional analysis shows that ULF waves can be a significant source for energetic electrons' acceleration, but that energetic electrons densities are too low to act as an energy sink for ULF waves. With magnetic ULF amplitudes $\delta B \geq 0.01$ nT, and in some instances reaching as high as a few nT \citep{Hartinger13}, and energetic electrons of 100 to 1000 keV with densities $n_e \leq 10^{-3}$ cm$^{-3}$, the ratio of kinetic energy density to ULF magnetic field energy density scales as $2 \mu_0 n_e m_e c^2 (\gamma-1)/\delta B^2 \leq 10^{-6}$.}. One therefore has freedom to model the ULF fluctuations in a manner consistent with \textit{in situ} observations, as long as Faraday's law is respected. For a ULF fluctuation given as a single Fourier mode ($A_m(t)\sim e^{-i\omega t}$) or some stochastic noise, one can solve the linear system analytically. To the best of our knowledge, the analytical solution of Equation (\ref{QLT3}), i.e., Equation (\ref{Linear_EQ1}), has not appeared in peer-reviewed studies of terrestrial radial transport before so we proceed hereafter with a detailed analysis. \\%\footnote{Check Dungey stuff}.\\

%FIGURE 1 %%%%%%%%%%%%%%%%%%%%%%
\begin{figure}[ht!]
\plotone{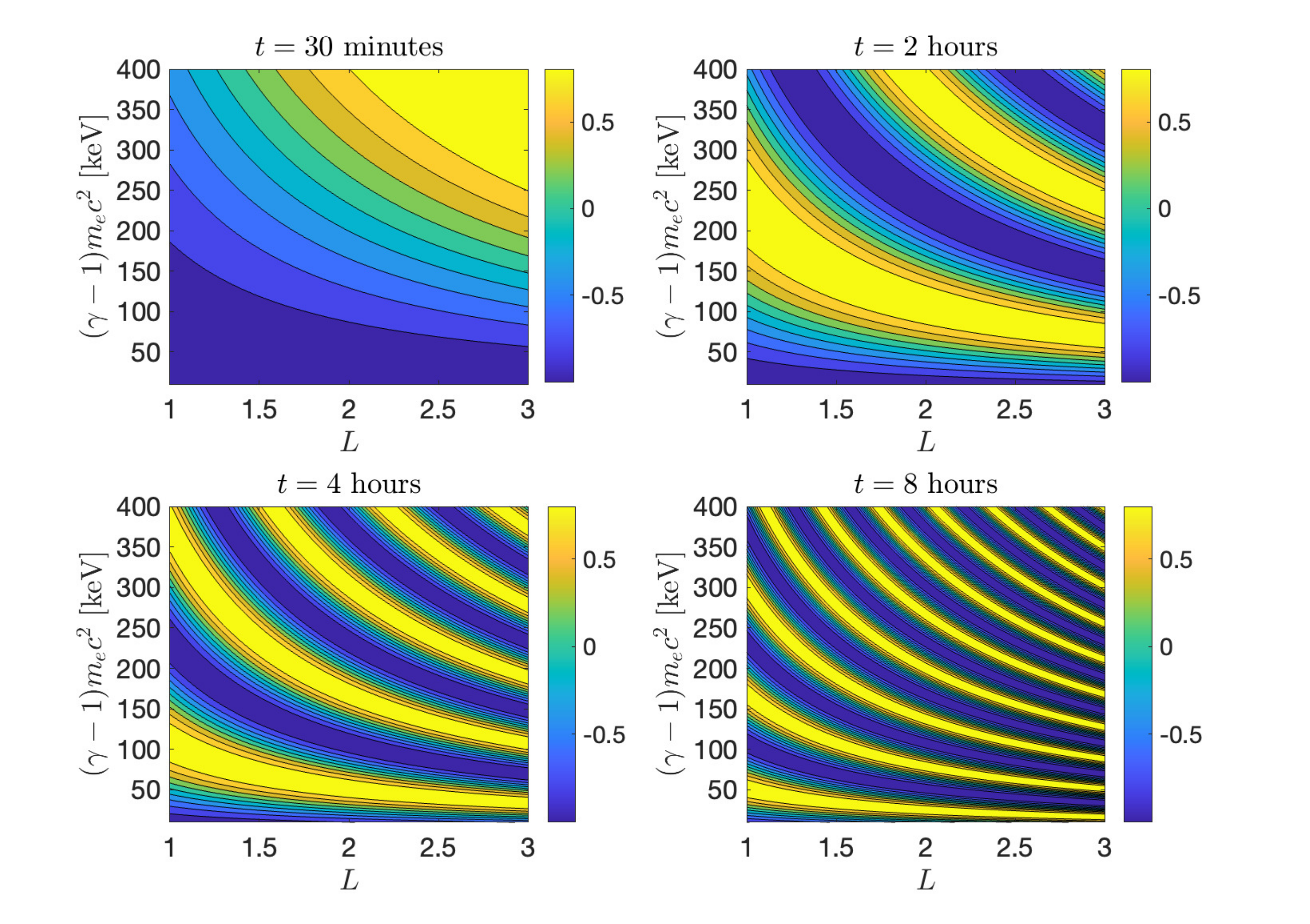}
\caption{Ballistic motion for particles trapped in a dipolar field result in zebra stripes formation. The top left, top right, bottom left and bottom right are solutions at $\varphi_0=0$ for $t=$ 30 minutes, 1 hour, 2 hours and 8 hours, respectively. The initial distribution in uniform in $L$ and kinetic energy $E_c$.   \label{fig_zebra_stripes1}}
\end{figure}
%FIGURE 1 %%%%%%%%%%%%%%%%%%%%%%

%FIGURE 2 %%%%%%%%%%%%%%%%%%%%%%
\begin{figure}[ht!]
\plotone{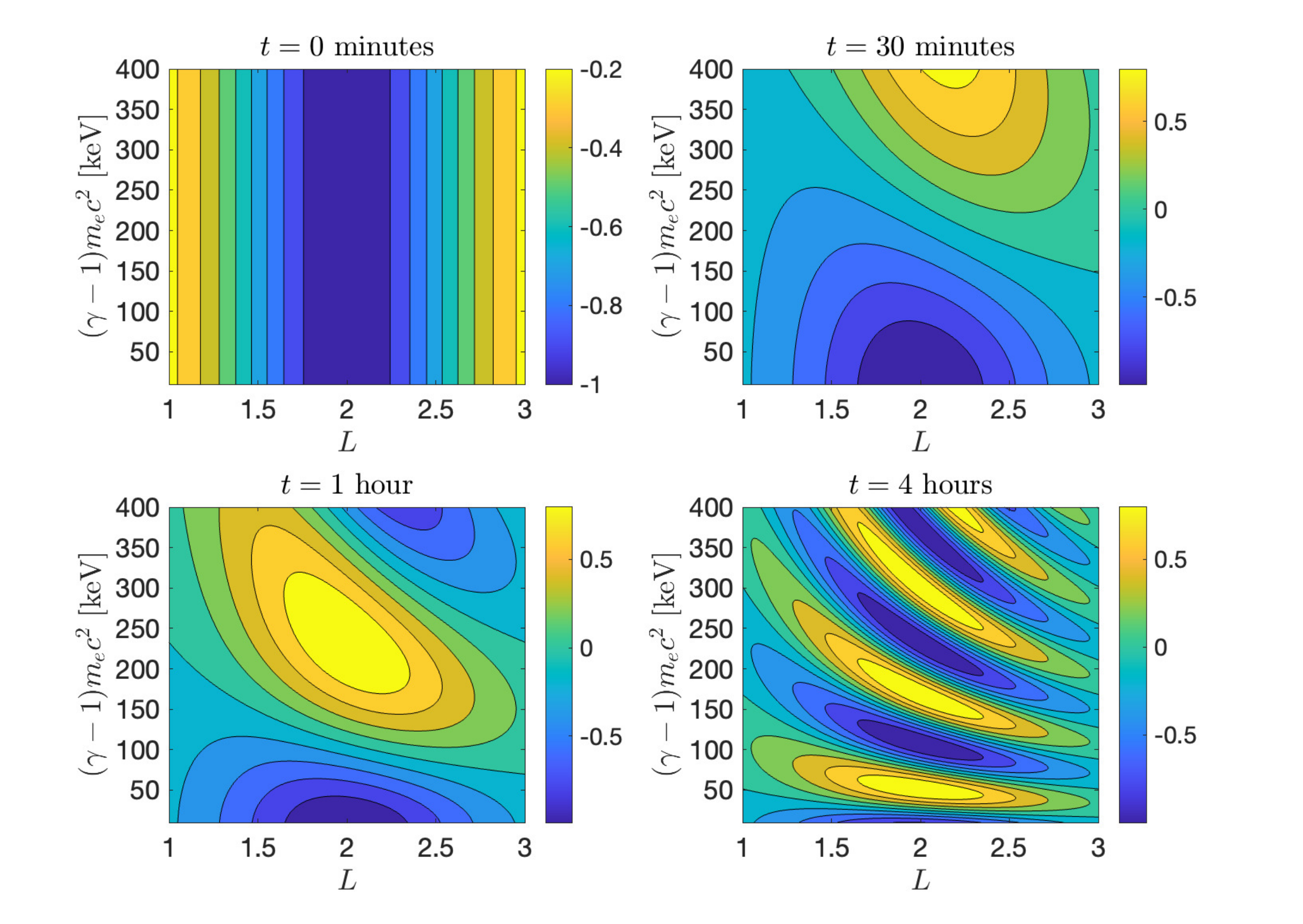}
\caption{Zebra stripes formation after the injection of particles  centred at $L=2$ with a spread in radial distance of $\Delta L= 0.75$. \label{fig_zebra_stripes2}}
\end{figure}
%FIGURE 2 %%%%%%%%%%%%%%%%%%%%%%

\subsubsection{Ballistic solution and the formation of zebra stripes}\label{zebra_stripes_section}
\noindent Inserting the ballistic solution in the perturbed distribution, i.e., the term $\delta f_m(r,0) e^{-i m\Omega_d t}$ in  (\ref{Linear_EQ1}), in Equation (\ref{basis}), the total distribution in the linear limit is given by: 
\begin{equation}
\label{basis2}
f(r, \varphi, t)=f_0(r)+\sum_m \delta f_m(r, 0)e^{-i m\Omega_d t}e^{i m\varphi}.
\end{equation}
%%%%%%%%
We can consider the ballistic solution separately of the linear wave-particle response since the former is independent on the radial gradient of $f_0$ and the latter is not. The ballistic response is therefore the only possible observable response when radial gradients in the distribution function are very small. \\

\noindent If we consider a single Fourier mode $m=1$, we note that an initial perturbation $\delta f(r,t=0, \varphi)$ will develop fine structures in the $(r,\varphi)$ space as $t\longrightarrow \infty$. The formation of fine structures in space occurs because an initial perturbation $\delta f_m(r, 0)$ will experience a differential shearing along the radial position $r$. We can use the solution (\ref{basis2}) to quantify the parametric dependence of the structures arising from ballistic motion in a magnetic dipolar field $B=B_E/L^3$. For an initial phase $\varphi_0$, the perturbed distribution, $\delta f$ is constant along the curve
\begin{eqnarray}
\label{zebra_curve}
\Delta \varphi=\varphi(t)-\varphi_0&=&\Omega_d t \nonumber \\ 
&=&\frac{3}{2} \frac{m_e c^2 (\gamma^2-1)}{q\gamma B_E R_E^2} L t \nonumber \\
&=&\frac{3}{2} \frac{m_e c^2 }{{qB_E R_E^2}} \frac{E_c/m_ec^2+2}{E_c+m_ec^2} E_c Lt \nonumber \\ 
&\sim& \frac{E_c+2m_ec^2}{E_c+m_ec^2} E_c L t \nonumber\\
&\simeq& E_c L t
\end{eqnarray} 
in which we replaced the Lorentz factor $\gamma$ by the kinetic energy $E_c=m_ec^2 (\gamma-1) $. For a fixed time $t\neq 0$, Equation (\ref{zebra_curve}) indicates that energetic particles will have phase-space structures with a kinetic energy that is inversely proportional to the radial distance, i.e., $E_c\sim 1/L$. \\

\noindent The time evolution of the perturbed distribution function is shown in Figure (\ref{fig_zebra_stripes1}) for time snapshots of 10 minutes, 1 hour, 2 hours and 8 hours. Energetic particles ranging between $50-400$ keV experience a full azimuthal drift on the order of a few hours. Figure (\ref{fig_zebra_stripes1}) shows that phase-space structures can form on timescales of the order of a single drift-period, that is on timescales that are far too rapid to be accounted for by radial diffusive effects. After several drift periods, phase-space structures in $(E_c, L)$ become thinner even though their numbers grow. This behaviour of the ballistic solution is consistent with the phenomenon of \textit{zebra stripes} commonly observed in the inner part of the Earth's radiation belts \citep{Imhof65, Datlowe85}. Zebra stripes are transient structured peaks and valleys observed on spectrograms of inner radiation belts' electrons with energies ranging between tens to hundreds of keV. The zebra stripes that are measured \textit{in situ} are also characterised by energy peaks and dips that vary as the inverse of the radial distance, i.e., $E_c\sim 1/L$. They are also associated with substorms onsets and correlated with various geomagnetic indices, such as Kp and Dst, but are also able to form during quiet geomagnetic conditions \citep{Sauvaud13, Lejosne16zebra, Lejosne20a, Lejosne20b}. Mechanisms explaining formation of zebra stripes must therefore reproduce the $E_c\sim 1/L$ dependence and explain the processes responsible for their transient nature and appearance under a wide range of geomagnetic conditions. \\

\noindent Mechanisms suggested for the formation of zebra stripes can be categorised into two types. In the first type, particles sample an electric field that varies on timescales consistent with their drift motion [see, e.g., \citet{Lejosne22} and references therein for the most recent advances on the subject]. Consequently, a collection of trapped particle can experience drift resonance with the field, and result in zebra stripes structures as resonant particles are scattered to different drift-shells. In the second type, illustrated by the study of \cite{Sasha_zebra}, zebra stripes also sample an electric field but are non-resonant. The formation of zebra stripes for this mechanism is akin to a phase-mixing process. Magnetically trapped particle's drifts are faster for more energetic particles. When fluxes are projected in energy and radial distance, the shearing of the distribution leads to a $E_c\sim 1/L$ dependence.\\ 

\noindent However, our analysis of the ballistic motion also demonstrates that phase-space structures consistent with \textit{in situ} observations of zebra stripes can form in the absence of both drift-resonance and electric field perturbations. The formation occurs on time-scales comparable to the drift period of energetic particles and is equivalent to the phase-mixing scenario presented by \cite{Sasha_zebra} in that it does not require drift-resonance. However, the ballistic solution we derived assumes a perturbation of the distribution function $\delta f_m(t=0, r)$ at some arbitrary time. This perturbation of the distribution function can either be due to particles being lost $\delta f_m(t=0, r) <0$, e.g. to the boundaries, or particles being injected $\delta f_m(t=0, r)>0$. While more quiescent than the outer belts, the inner belts experience injection events of energetic electrons even during moderate geomagnetic storms \citep{Zhao13}\footnote{Albedo neutron decay is also a constant source of energetic particles' injection in the inner belts \citep{Li2017} but the density might be too low for observational measurements of zebra stripes formation in energetic electrons or protons.}. \\

\noindent In order to inject electrons in the inner belts a radial transport mechanism, such as a convective electric field, is required. But once injected in the inner belts the ballistic term shows that zebra stripes can form in the absence of any ULF perturbations. In Figure (\ref{fig_zebra_stripes2}), we show the formation of the zebra stripes following localised loss of energetic electrons centred at $L=2$ and spread with a standard deviation along radial distances of $\Delta L= 0.75$. Localised injection and losses also result in stripes on timescales comparable to the drift period but shearing of the distribution function results in structures spreading across radial distances beyond the injection or loss location. \\

\begin{figure}[ht!]
\epsscale{1.25}
\plotone{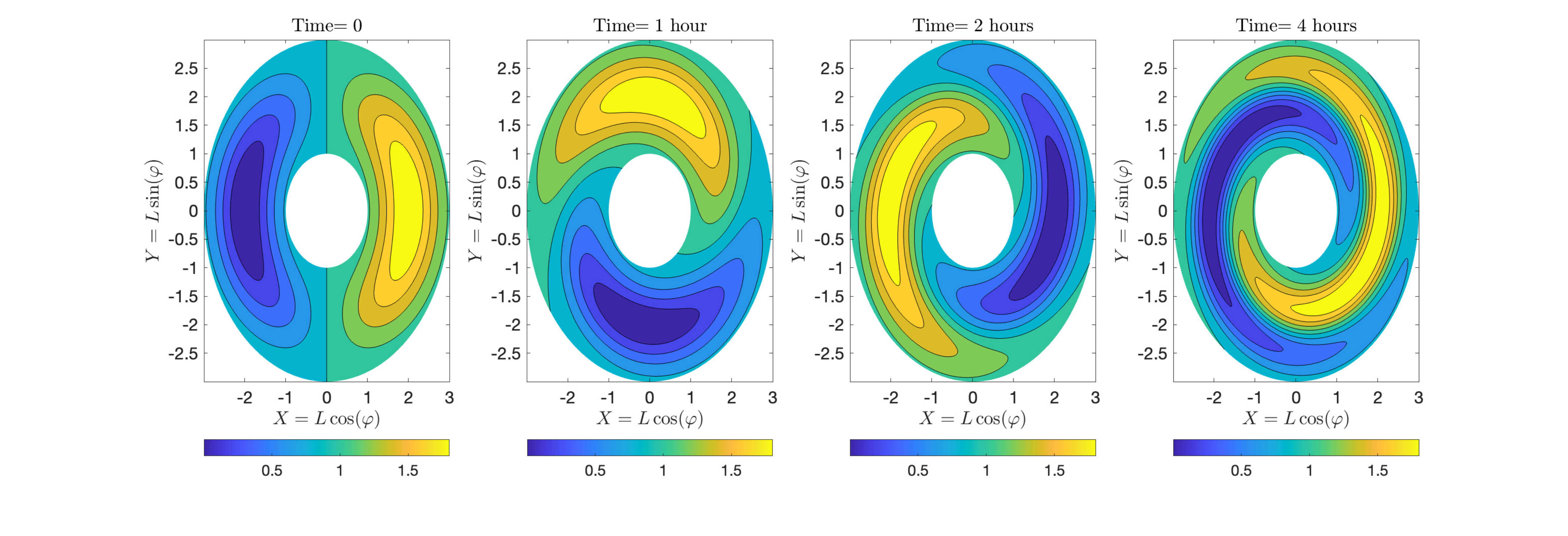}
\caption{Zebra stripes formation for 100 keV equatorially trapped electrons in terms of $(L-\varphi)$. The initial distribution correspond to a Gaussian distributed beam centered at $L=2$ for $\varphi\in[0, \pi]$ and a Gaussian distributed drop centred at $L=2$ for $\varphi\in[\pi, 2\pi]$. After four hours the zebra stripes remain visible. \label{fig_zebra_stripes3}}
\end{figure}

\begin{figure}[ht!]
\epsscale{1.25}
\plotone{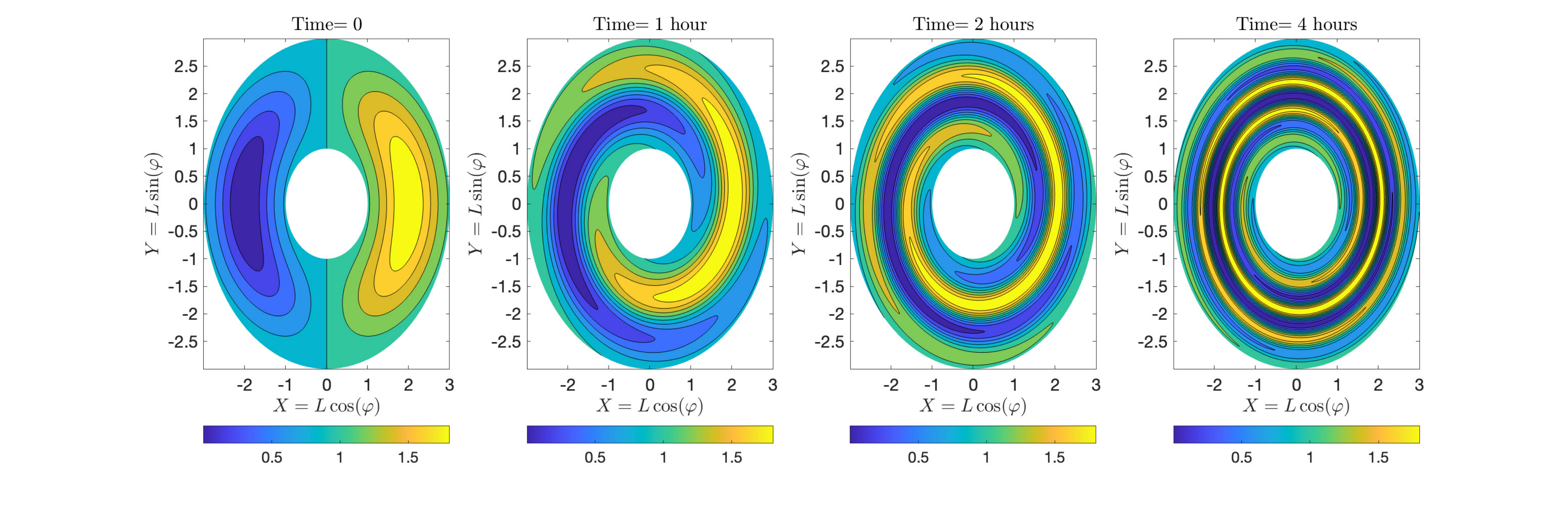}
\caption{Zebra stripes formation for 400 keV equatorially trapped electrons in terms of $(L-\varphi)$. The initial distribution correspond to a Gaussian distributed beam centered at $L=2$ for $\varphi\in[0, \pi]$ and a Gaussian distributed drop centred at $L=2$ for $\varphi\in[\pi, 2\pi]$. After four hours the zebra stripes have phase-mixed. \label{fig_zebra_stripes4}}
\end{figure}

\noindent The transient nature of zebra stripes can also be evidenced when projecting the ballistic solution in the equatorial plane.  Figure (\ref{fig_zebra_stripes3}) shows the temporal evolution of 100 keV electrons's injection (at $\varphi \in [0, \pi]$) and losses  (at $\varphi \in [\pi, 2\pi]$). The drift period of 100 keV electrons between $L=1$ and $L=3$ ranges between 2.6 hours and 8 hours. After a single drift period the distribution function preserve their initial shape and have yet to phase-mix. In comparison, Figure (\ref{fig_zebra_stripes4}) shows the temporal evolution of 400 keV electrons's injection (at $\varphi \in [0, \pi]$) and losses  (at $\varphi \in [\pi, 2\pi]$). The drift period of 400 keV electrons between $L=1$ and $L=3$ ranges between 45 minutes and 2.3 hours. For more energetic particles, since the drift period is shorter, shearing of the initial distribution phase-mixes the distribution on faster timescales. After 4 hours, the zebra stripes of $400$ keV have very fine-scale structures in the equatorial plane. \\

\noindent Injection or losses of particles can therefore result in the formation of zebra stripes without the need for drift-resonance or the presence of an electric field. The injection and losses are encoded in the ballistic solution but since shearing of the distribution function occurs on timescales of a few drift periods, the most energetic electrons develop quickly fine-scale structures in the distribution function that might not be resolved by spacecraft instruments. Nonetheless, the ballistic solution does not preclude the possibility for zebra stripes formation as a response to a ULF electric field for resonant or nonresonant particles. In the next section, we compute the linear solution to include the impact of ULF waves on the distribution function and differentiate between resonant and nonresonant responses. \\

%FIGURE 3 %%%%%%%%%%%%%%%%%%%%%%
%FIGURE 3 %%%%%%%%%%%%%%%%%%%%%%

%FIGURE 4 %%%%%%%%%%%%%%%%%%%%%%
%FIGURE 4 %%%%%%%%%%%%%%%%%%%%%%

\subsubsection{Solution to the linear wave-particle interaction}
\noindent In the previous section we described the time evolution of the ballistic term in the distribution function and argued that it should dominate the particle's response when radial gradients in the distribution function are small. However, in the absence of phase-space injection and/or loss terms, and thus in instances where the ballistic term is zero, i.e., $\delta f(t=0, r)=0$, and $\partial f_0 /\partial r \neq 0$, the linear wave-particle response should dominate. \\

\noindent In this section we describe the linear wave-particle solution found in Equation (\ref{Linear_EQ1}). For the sake of simplicity, we assume a single Fourier mode for the ULF wave: 
\begin{equation}
\label{Fourier_mode_linear}
A_m(t)=a_m e^{-i\omega_m t+ \gamma_m t}
\end{equation}
with initial amplitude $a_m$,  frequency $\omega$, and growth/damping rate $\gamma_m$. We can generalise this solution to a spectrum of Fourier modes, but since the solution is linear, each are independent of one another. The linear solution is valid in the limit where the growth rate is sufficiently small, for the fluctuations to remain sufficiently small in amplitude and nonlinear effects negligeable \citep{Davidson}\footnote{See Section \ref{nonlinear section} in which we quantify the conditions for the linear regime to breakdown. It will not come as a surprise to readers' familiar with solar wind turbulent problems that nonlinear effects can become dynamically significant for small amplitude electromagnetic fluctuations. In the magnetohydrodynamic limit this condition is associated with a state of critical balance \citep{GS95} at fluid scales, but has also been generalised to kinetic problems in space plasmas \citep{Scheko2016, Meyrand2019}. For the problem of radial transport the nonlinear regime is reached even in the limit where ULF wave amplitudes and the perturbed distribution function are small, i.e., $\delta B/B_0 \ll 1$, and $\delta f/f_0 \ll 1$, respectively.}. We insert Equation (\ref{Fourier_mode_linear}) into Equation (\ref{Linear_EQ1}) to find the following linear wave-particle response $\delta f^L_m(r,t)$: 
%%%%%%%%%
\begin{eqnarray}
\label{linear_solution}
\delta f^L_m&=&-a_me^{-im\Omega_d t}\left[ \frac{8r^2}{21 B_0}(\gamma_m-i\omega_m)-\frac{i m \mu}{qB_0\gamma } \right] 
\frac{\partial f_0}{\partial r}\int_0^t dt'  e^{im\Omega_d t'-i\omega_m t' +\gamma_m t'} \nonumber\\
&=& -a_me^{-im\Omega_d t}\left[ \frac{8r^2}{21 B_0}(\omega_m+i\gamma_m)+\frac{m \mu}{qB_0\gamma } \right] 
\frac{\partial f_0}{\partial r} \left(\frac{e^{im\Omega_d t-i\omega_m t +\gamma_m t}-1}{\omega_m-m\Omega_d+i\gamma_m}\right).
\end{eqnarray}
%%%%%%%%%
Equation (\ref{linear_solution}) contains a resonant part indicating that particles with a drift frequency $\Omega_d$ can be scattered across drift shell efficiently with ULF waves of frequencies $\omega_m$. We can decompose equation (\ref{linear_solution}) in terms of a linear wave-particle resonant part that can grow in time, and two oscillating parts as follows: 
%%%%%%%%%
\begin{eqnarray}
\label{linear_solution2}
\delta f^L_m&=&-\frac{a_m r^2}{21B_0} \frac{\partial f_0}{\partial r}\left[\underbrace{8 e^{-i\omega_m t +\gamma_m t}}_{\text{ULF modulation with growth/decay}}-\underbrace{{8 }e^{-im\Omega_d t}}_{\text{Ballistic motion}}+\underbrace{15 m \Omega_d e^{-im\Omega_d t}\left(\frac{e^{im\Omega_d t-i\omega_m t +\gamma_m t}-1}{\omega_m-m\Omega_d+i\gamma_m}\right)}_{\text{Wave-particle drift-resonance}}\right].
\end{eqnarray}
The first term on the right-hand side of Equation (\ref{linear_solution2}) is a non-resonant term oscillating mode that can grow or damp with the ULF wave at a rate $\gamma_m$ and modulates the distribution function at a frequency $\omega_m$. The second term on the right-hand side of (\ref{linear_solution2}) is a non-resonant ballistic term that indicates that a ULF fluctuation of arbitrary frequency $\omega_m$ can sustain fluctuation in the distribution function at frequencies $\Omega_d$ without drift-resonances involved. In the past 60 years, time series of particles fluxes observed to have temporal frequencies comparable to the drift period have been termed drift echoes \citep{Lanzerotti67}. For instance, Figure 6 of \cite{Kokobun77} shows simultaneous association of transverse ULF wave mode with oscillations in energetic ion fluxes and energetic electron fluxes of 79, 158 and 266 keV. The low energy fluxes are modulated by the ULF wave, and the phases of modulations are energy dependent. The oscillations reported by \cite{Kokobun77} are occurring on timescales comparable to the drift periods of energetic populations and are therefore produced too quickly to be sustained by a quasi-linear radial diffusive process. This second term, responsible for drift echoes, is another source responsible for the formation of zebra stripes (\ref{zebra_stripes_section}) for non-resonant particles and corresponds to the mechanism explained in \cite{Sasha_zebra}. The difference between this second term and the zebra stripe source derived in (\ref{zebra_stripes_section}) is that the latter requires the phase-space loss ($\delta f(t=0)<0$) or injection of particles ($\delta f(t=0)>0$) and no electric fields \footnote{It should be noted that in the terrestrial radiation belts injection and losses are separated between adiabatic and non-adiabatic ones. Reversible losses are associated with adiabatic perturbations, whereas irreversible losses are associated with non-adiabatic effects, for instance the scattering of particles inside the atmosphere \citep{Millan07}. The ballistic amplitude $\delta f(t=0)$ can account for both reversible and irreversible losses.}, whereas the former requires the perturbation of the distribution function from a ULF fluctuation with amplitude $a_m$ \textit{and} a gradient in the distribution function, i.e. $\partial f_0/\partial r \neq 0$. \\ 

\noindent The third term on the right-hand side of (\ref{linear_solution2}) represents the linear wave-particle resonance between ULF fluctuations of frequencies $\omega_m$ and particles with drift frequency $\Omega_d$. It can be shown that this last term can grow in time for the limit $m\Omega_d\sim \omega_m$ and for what we here call intermediate times: 
\begin{equation}
\label{intermediate_times}
\frac{1}{m\Omega_d}\ll t \ll \frac{1}{\gamma_m}.
\end{equation}
The intermediate time range defined by Equation (\ref{intermediate_times}) means that the ULF wave has had time to oscillate, but the perturbation has not yet been damped away significantly, or grown appreciably, for linear effects to breakdown \citep{Schekochihin_notes}. In this limit, the resonant term in Equation (\ref{linear_solution2}) dominates over the other two, and the perturbation in the linear response of the distribution function takes the following form: 
\begin{eqnarray}
\label{linear_solution3}
\delta f^L_m \simeq m\Omega_d e^{-i\omega_mt} \frac{\left( e^{\gamma_m t}- e^{im\Omega_d t-i\omega_m t +\gamma_m t}\right)}{m\Omega_d-\omega_m-i\gamma_m}\simeq m\Omega_d e^{-i\omega_mt} \frac{1- e^{i(m\Omega_d t-\omega_m) t}}{m\Omega_d-\omega_m}
\end{eqnarray}
Equation (\ref{linear_solution3}) is valid at intermediate times given by (\ref{intermediate_times}) and assumes that $m\Omega_d-\omega_m \gg \gamma_m$. In the limit where $m\Omega_d-\omega_m \ll 1/t$, the exponential term in Equation (\ref{linear_solution3}) can be expanded as a Taylor series, and the dominant term for the perturbed distribution function gives: 
\begin{equation}
\delta f_m^L \simeq m\Omega_d t e^{-i\omega_mt},
\end{equation}
and thus demonstrates that fluctuations grow linearly in time due to resonant interactions. Equation (\ref{linear_solution3}) is an instance of a \textit{Case-van Kampen mode}, initially derived for a Vlasov-Poisson plasma \citep{Kampen, Case}, but rederived here in the context of radial transport. In the limit where $t\longrightarrow \infty$ but $\gamma_m t \ll 1$ necessary to respect (\ref{intermediate_times}), the right-hand side of Equation (\ref{linear_solution3}) tends to a delta function\footnote{These non-eigenmodes are not only of theoretical interest. Non-eigenmodes have to be tracked in order to quantify entropy production in kinetic systems. See for instance Section 5.6 of \cite{Schekochihin_notes} for an introduction in terms of a Vlasov-Poisson system and \cite{Zhdankin22} for an application that can be used for wave-particle interactions in the radiation belts.}. \\

\noindent The resonant linear response presented in this section occurs on timescales comparable or larger than the drift period but smaller than $1/\gamma_m$, while phase-mixing and zebra stripes are taking place on timescales comparable to drift periods. For finite damping ULF rate $\gamma_m<0$, the resonant part decays $e^{\gamma_t \longrightarrow 0}$ on timescales $|\gamma_m| t \gg 1$, and the ballistic response proportional to $e^{-i\Omega_d t}$ in Equation (\ref{linear_solution2}) dominates. This criterion can be used to distinguish non-resonant to resonant drift particle interactions from spacecraft data since both require a radial gradient in the MLT averaged distribution function $f_0$. The requirement for a non-zero radial gradient in  $f_0$ of a given energetic population is an experimental constraint on the observation of phase-space structures, as reported by \cite{Hartinger20} and \cite{Sarris21}, and is discussed further in Section \ref{subsection_fast}. \\

\begin{figure}
    \centering
    \plotone{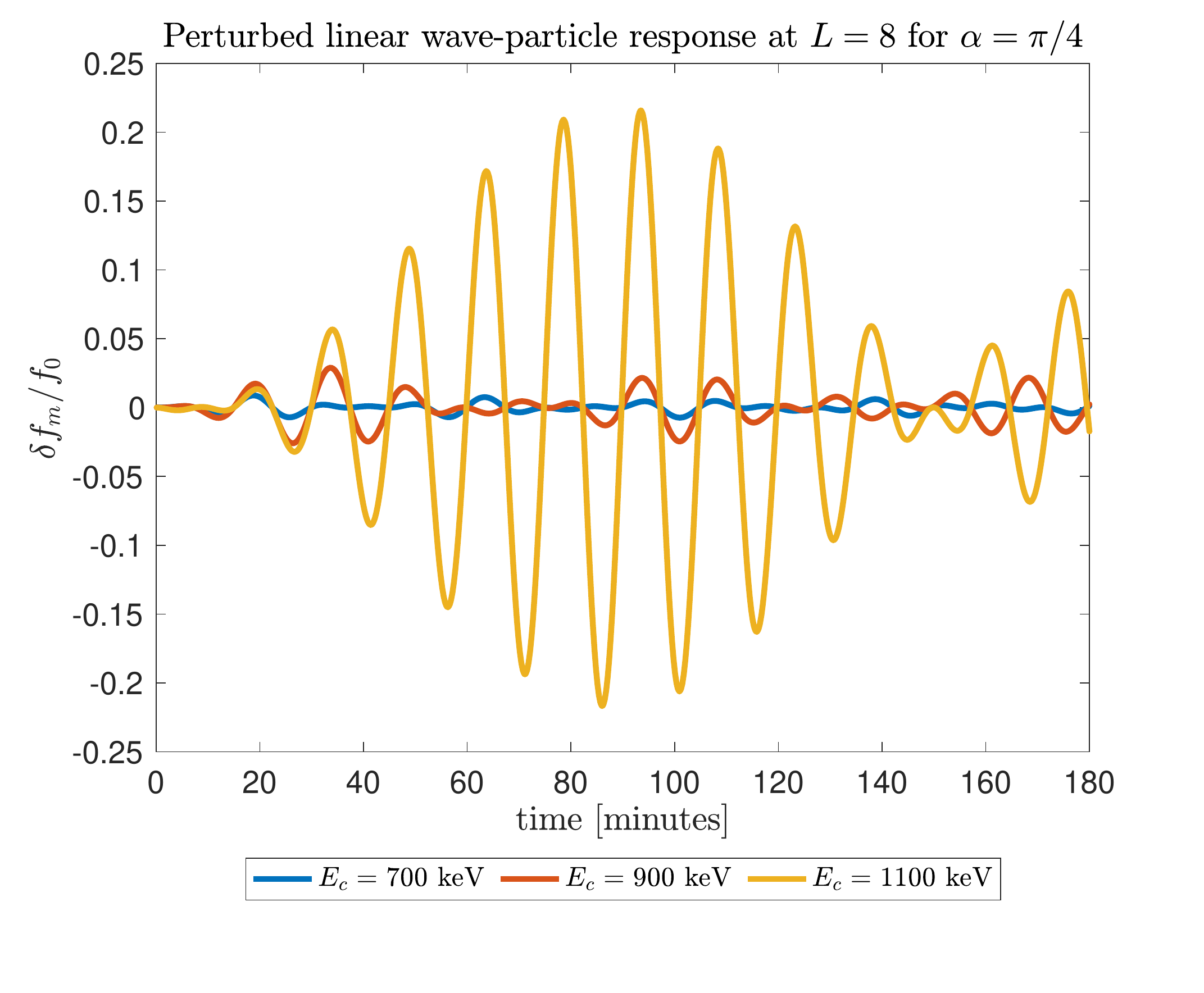}
    \caption{Example of resonant and nonresonant response in the electron distribution function. The ULF wave has a frequency $\omega=7$ mHz and a mode number $m=1$. The particles are located at $L=8$ with pitch-angle $\alpha=45$ degrees. Particles with kinetic energies of the order of $\simeq 1.2$ MeV ($2\pi/\Omega_d\simeq 15$ minutes) are resonant, but particles with energies less than 1 MeV ($2\pi/\Omega_d\geq 17$ minutes) are not. The resonant particles experience fluctuations almost one order of magnitude greater than nonresonant particles with comparable kinetic energy.}
    \label{fig:linear_resonance_example}
\end{figure}
%%%%%%%%%

\noindent An additional criterion to distinguish resonant from non-resonant particle's response can also be achieved observationally for instruments recording energy-dependent fluxes. Drift-resonance is energy dependent, and the signature of resonance for resonant energies should be markedly different than for non-resonant particles, even though Equation (\ref{linear_solution3}) shows that both experience oscillations with frequencies comparable to the ULF wave frequency $\omega_m$. Figure (\ref{fig:linear_resonance_example}) shows the perturbed distribution function of 1.1 MeV electrons at $L=8$ in comparison to the particle's response for energies at 700 and 900 keV. Thus, a shift in energy can take particles out of resonances and result in perturbed distribution function that are more than 5 times smaller in amplitude.\\

\begin{figure}
    \centering
    \epsscale{1.4}
    \plottwo{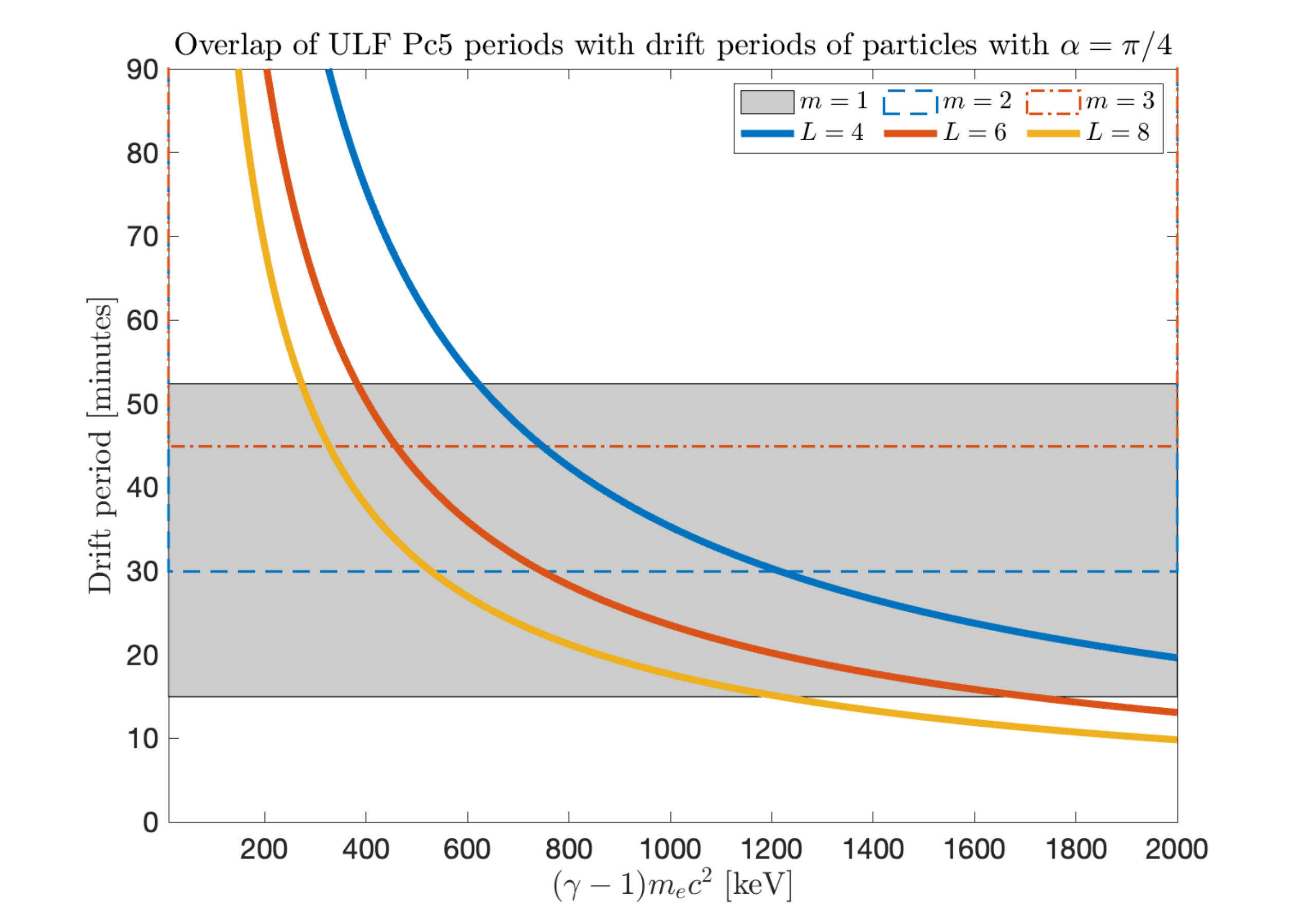}{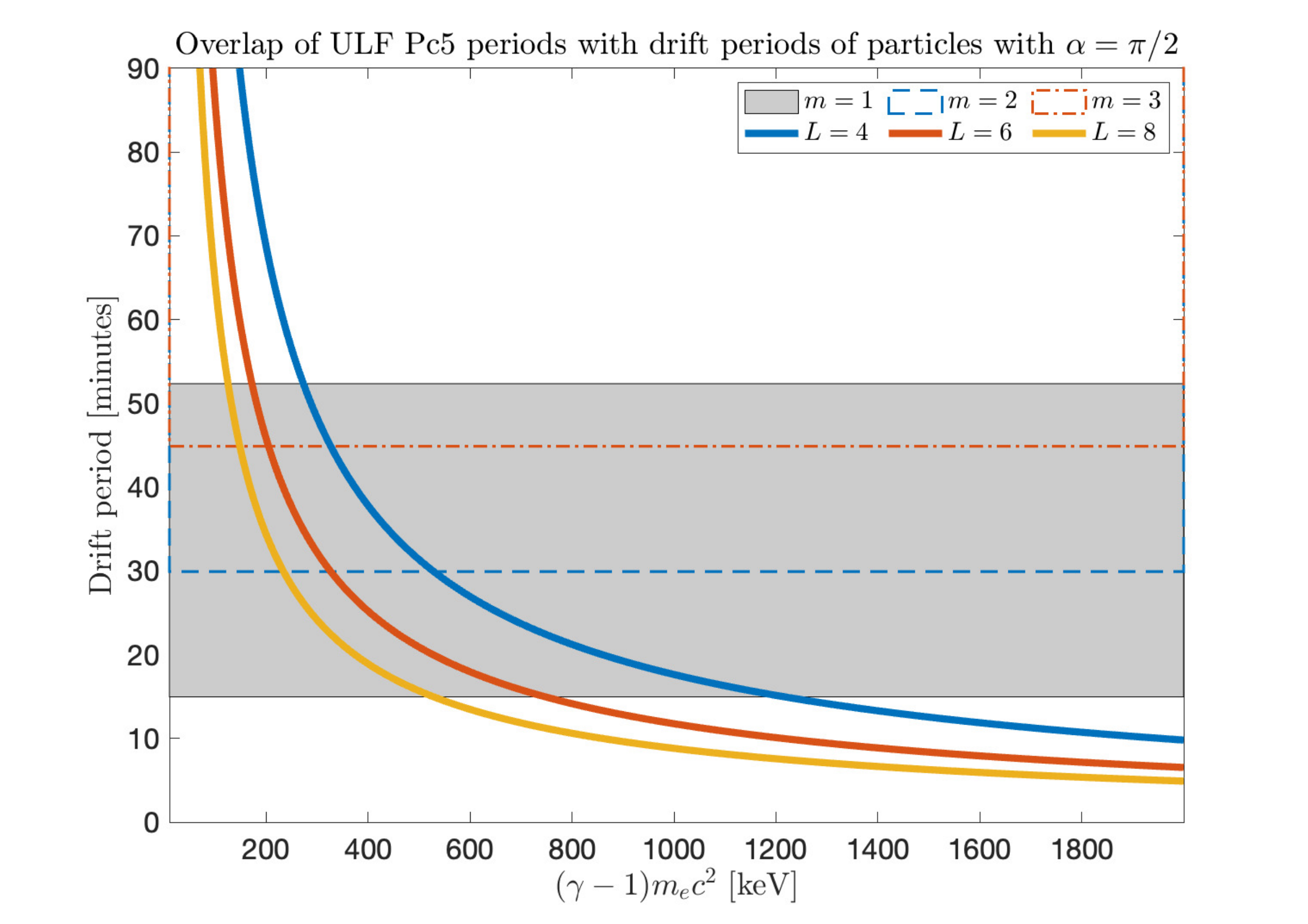}
    \caption{Azimuthal drift period ($2\pi/\Omega_d$) dependence in terms of the kinetic energy $E_c=[50-2000]$ keV and normalised radial distance $L=r/R_E= [4, 6, 8]$. The top panel is for $\alpha=45$ and the bottom one for $\alpha=90$. The grey shaded area is when the drift frequency matches the Pc5 ULF fluctuations with $\omega=[2, 7]$ mHz and resonant interactions is possible. The areas bounded in dashed and dotted lines show the resonant boundary for $m=2$ and $m=3$ modes, respectively.}
    \label{fig:Resonance_condition_Pc5}
\end{figure}

\begin{figure}
    \centering
    \epsscale{1.4}
    \plottwo{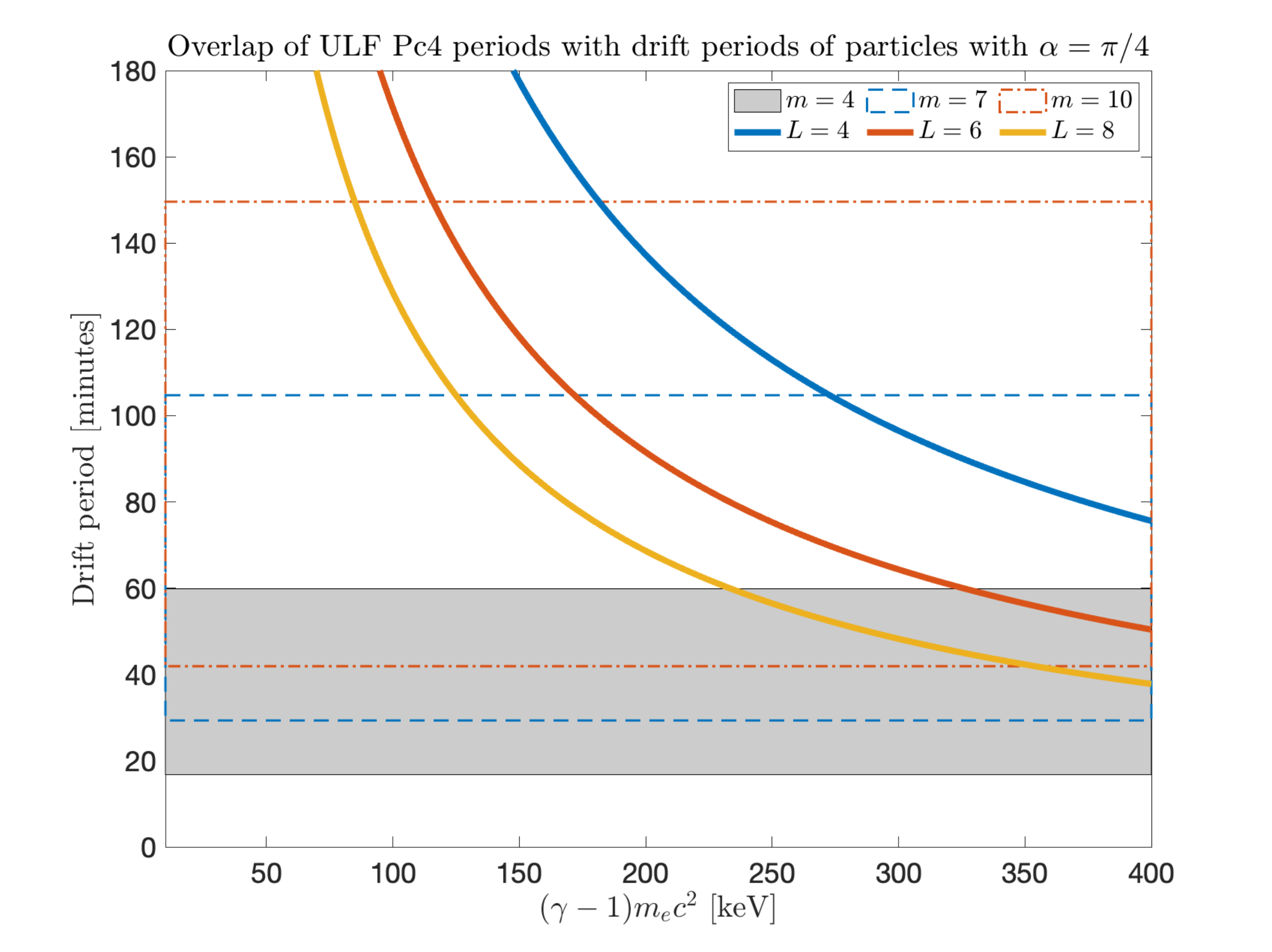}{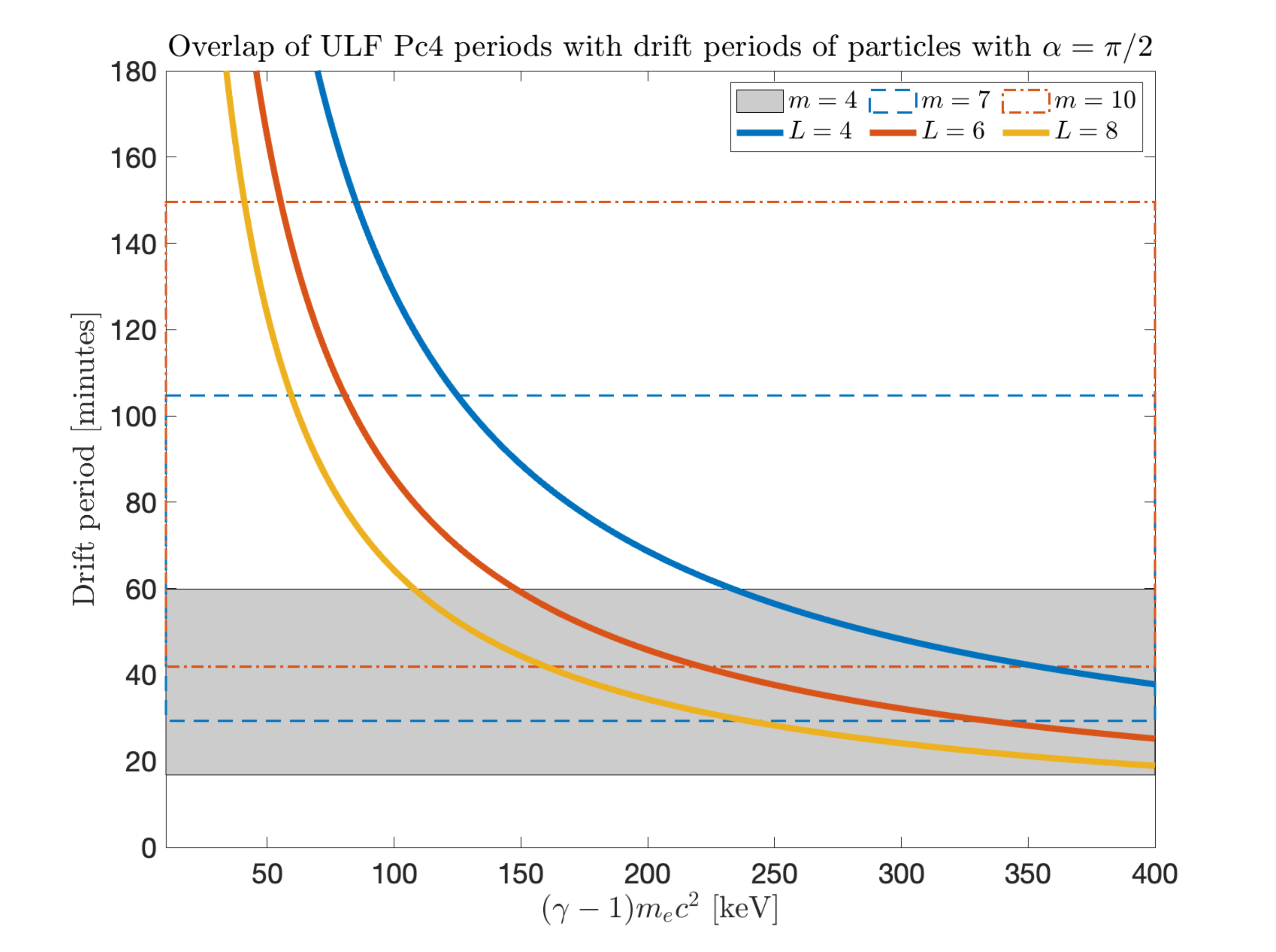}
    \caption{Azimuthal drift period ($2\pi/\Omega_d$) dependence in terms of the kinetic energy $E_c=[50-400]$ keV and normalised radial distance $L=r/R_E= [4, 6, 8]$. The top panel is for $\alpha=45$ and the bottom one for $\alpha=90$. The grey shaded area is when the drift frequency matches the $m=4$ Pc4 ULF fluctuations with $\omega=[7, 25]$ mHz and resonant interactions is possible. The areas bounded in dashed and dotted lines show the resonant boundary for $m=7$ and $m=10$ Pc4 modes, respectively.}
    \label{fig:Resonance_condition_Pc4}
\end{figure}

\noindent Drift-resonance is therefore an efficient mechanism for ULF waves to exchange energy with energetic electrons. In Figure (\ref{fig:Resonance_condition_Pc5}) we plotted the drift period as a function of kinetic energy and parametrized in terms of the radial distance $L$. The top panel is made for 45 degrees pitch-angles and the bottom panel for 90 degrees pitch-angles. The shaded and dashed rectangles bound the resonant frequency $\omega_m/m$ for Pc5 ULF frequencies with azimuthal mode numbers $m=1, 2$ and $m=3$. From Figure (\ref{fig:Resonance_condition_Pc5}) we note that energetic electrons with kinetic energies larger than $200$ keV and up to a few MeV have access to drift orbit resonance across broad drift-shells. Figure (\ref{fig:Resonance_condition_Pc4}) is the same as Figure (\ref{fig:Resonance_condition_Pc4}) but the bounded rectangles are drawn for Pc4 waves with azimuthal wave numbers $m=4, 7$ and. 10 In the case of Pc4 waves, they can sustain drift resonance for energetic electrons with kinetic energy less than 400 keV, but require larger azimuthal wave numbers \citep{Barani19}. Even though drift-resonance is strongly energy dependent, Figures (\ref{fig:Resonance_condition_Pc5}) and (\ref{fig:Resonance_condition_Pc4}) show that they can be accessible to a broad range of energy and pitch-angles across the radiation belts. \\

\noindent We therefore conclude this section by pointing out that the linear perturbation of the distribution function due to ULF electromagnetic fluctuations, particle injections ($\delta f(r, t=0)>0$) or losses ($\delta f(r, t=0)<0$), all result in phase-space drift structures on non-diffusive timescales comparable to the drift periods. Some of the phase-space structures for the lower energetic electrons ($E_c<m_e c^2$), assuming particle injection or gradient in the background distribution, can appear as zebra stripes in the inner radiation belts. Even though Equation (\ref{linear_solution2}) shows that the resonant part of $\delta f$ also experiences phase-mixing, drift echoes and zebra stripes nonetheless form for \textit{non-resonant} drift frequencies $m\Omega_d \neq \omega_m$, and thus, stringent resonant conditions $m\Omega_d \simeq \omega_m$ do not constitute \textit{sine qua non} constraints for the formation of drift echoes and zebra stripes. \\

\subsection{Quasi-linear theory of radial diffusion}\label{section_diffusion}
\noindent In the previous section we have described the fast linear response of the perturbed distribution function to an electromagnetic ULF wave. We assumed that the background distribution $f_0$ was time independent, which is equivalent as saying that it did not experience significant variations on fast time scales. In this section, we compute the evolution of the background distribution function according to quasi-linear assumptions \citep{Kennel66, Diamond10, Schekochihin_notes, Allanson22}. In quasi-linear theories one assumes that perturbations start modifying the equilibrium before they reach nonlinear amplitudes. In other words, the nonlinear term $\mathcal{Q}$ in Equation (\ref{QLT2}) can be ignored when the characteristic time for nonlinear effects is longer than the time for the equilibrium to be reached. We also neglect the linear term $\frac{A_m r}{B_0}\frac{\partial f_0}{\partial t}$ on the right-hand side of (\ref{QLT2}) since it provides a correction of order $|\delta B|^2/B_0^2\ll 1$ in the quasilinear limit, as shown in Appendix \ref{AppendixC}. Thus, for our purpose, we assume that the evolution of the perturbation is determined by Equation (\ref{QLT3}). Similarly to the previous section, this linear equation can be solved by Duhamel's principle, for the initial condition $\delta f_m(r, t=0)=0$: 
%%%%%%%%
\begin{eqnarray}
\label{QLT4}
\delta f_m(r,t) =-e^{-i m\Omega_d t} \int_{-\infty}^t dt' \ e^{+i m\Omega_d t'} \left( \frac{8r^2 \dot{A}_m(t')}{21 B_0}-im \frac{\mu A_m(t')}{qB_0\gamma } \right)\frac{\partial f_0}{\partial r}.
\end{eqnarray}
%%%%%%%%
The linear solution given by Equation (\ref{QLT4}) can then be combined with the following quasilinear equation to described the time evolution of $f_0$:
%%%%%%%%
\begin{eqnarray}
\frac{\partial f_0}{\partial t}&=&-\sum_m \left[ \frac{i m\mu}{qB_0\gamma r}\frac{\partial}{\partial r} \left(r\langle A^*_{m}\delta f_m\rangle\right) +\frac{8}{21}\frac{1}{rB_0}\frac{\partial }{\partial r}\langle r^3\dot{A}^*_{m}\delta f_m\rangle -\frac{r}{B_0}\langle {A}^*_{m}\frac{\partial}{\partial t}\delta f_m\rangle-\frac{r}{B_0}\langle {\dot{A}}^*_{m}\delta f_m\rangle \right]
\label{kinetic4b}
\end{eqnarray}
%%%%%%%%
We note that the first two terms on the right-hand side of Equation (\ref{kinetic4b}) will result in a diffusion term, and the last two expressions in advection terms. Replacing the linear solution of $\delta f_m$ into (\ref{kinetic4b}) to compute the correlation terms $\langle A_m^*(t)\delta f_m(t)\rangle$ and $\langle \dot{A}_m^*(t)\delta f_m(t)\rangle$, results in the following two integrals:
%%%%%%%%
\begin{eqnarray}
\label{integral_one}
\langle A_m^*(t)\delta f_m(t)\rangle=- \int_{-\infty}^t dt' \ e^{+i m\Omega_d (t'-t)} \left( \frac{8r^2 \langle A_m^*(t)\dot{A}_m(t')\rangle}{21 B_0}-im \frac{\mu \langle A_m^*(t)A_m(t')\rangle}{qB_0\gamma } \right)\frac{\partial f_0}{\partial r}.
\end{eqnarray}
%%%%%%%%
\begin{eqnarray}
\label{integral_two}
\langle \dot{A}_m^*(t)\delta f_m(t)\rangle=- \int_{-\infty}^t dt' \ e^{+i m\Omega_d (t'-t)} \left( \frac{8r^2 \langle \dot{A}_m^*(t)\dot{A}_m(t')\rangle}{21 B_0}-im \frac{\mu \langle \dot{A}_m^*(t)A_m(t')\rangle}{qB_0\gamma } \right)\frac{\partial f_0}{\partial r}.
\end{eqnarray}
%%%%%%%%

\noindent To compute the autocorrelations analytically we need to make some assumptions about the nature of the ULF amplitude $A_m(t)$. To account for finite and zero correlation times we choose to model the fluctuations as different realisations of an Ornstein-Uhlenbeck process \citep{Papoulis} given by the following time evolution equation\footnote{Energetic electrons in the Earth's radiation belts are passive tracers and the self-consistent response onto the field can therefore be ignored. This freedom allows one to model the ULF wave amplitudes in a manner empirically consistent with \textit{in situ} measurements.}: 
%%%%%%%%
\begin{equation}
\frac{\partial A_m}{\partial t} = -A_m/\tau_c+ \sqrt{2 D} \chi(t),
\end{equation}
%%%%%%%%
where $\tau_c$ is a correlation time, $\sqrt{2D}$ is a measure of the root mean square value of $A_m$ and $\chi(t)$ is a unit Gaussian white noise, $\langle \chi(t) \chi(t')\rangle=\delta (t-t')$. The solution for $A_m$, assuming $A_m(t=0)=0$, is given by
%%%%%%%%
\begin{equation}
\label{OUP}
A_m(t) = \sqrt{2D}e^{-t/\tau_c} \int_{-\infty}^t dt' \ e^{t'/\tau_c} \chi(t').
\end{equation}
%%%%%%%%
Using Equation (\ref{OUP}) we can compute the following quantities for a finite correlation time $\tau_c\neq 0$: 
%%%%%%%%
\begin{eqnarray}
C_1(t, t')=\langle A_m(t) A_m(t') \rangle ={\tau_cD}e^{-|t-t'|/\tau_c}
\end{eqnarray}
%%%%%%%%
\begin{eqnarray}
C_2(t, t')=\langle \dot{A}_m(t) \dot{A}_m(t') \rangle =\frac{D}{\tau_c} e^{-|t-t'|/\tau_c}+2D \delta (t-t')
\end{eqnarray}
%%%%%%%%
\begin{eqnarray}
C_3(t, t')=\langle A_m(t) \dot{A}_m(t') \rangle =-{D} e^{-|t-t'|/\tau_c}
\end{eqnarray}
%%%%%%%%
The above correlators are only a function of the time difference $t-t'$, and not the particular times $t$ and $t'$, indicating that the Ornstein-Uhlenbeck process is stationary, or time-homogeneous. \\ 

\noindent Returning to the integrals (\ref{integral_one}) and (\ref{integral_two}) it should be stressed that the gradient in the background distribution functions in the integrals is a function of time, i.e., $f_0=f_0(r,t)$. The last step before solving the integral is to assume that a short decorrelation time $\tau_c$ exists, such that the correlators $C_i(t-t') \ll C_i(0)$ if $t-t' > \tau_c$. We can thus replace $f(r, t')=f(r, t-\tau)$ by $f(r, t)$ on the basis that $C_i(\tau=t-t')$ changes appreciably before any significant variation in the background distribution \citep{Eijnden}. This quasi-linear assumption indicates that the ULF wave amplitude cannot alter the background distribution function on timescales comparable to the ULF wave and drift period. The diffusion coefficient that follows in the next lines can therefore not lead to changes on timescales comparable to the azimuthal drift period and justifies the ensemble-average defined by Equation (\ref{def_average}). \\

\noindent For the sake of simplicity, we now assume zero correlation time\footnote{By keeping $\tau_c$ finite but small ($\Omega_d \tau_c\ll 1$), the diffusion coefficient in the quasilinear limit is rescaled by a factor $\frac{1}{1+\Omega_d^2 \tau_c^2}$, thereby introducing an energy dependence to the radial transport, as shown in \cite{Osmane21a}.}, which means $e^{-|t-t'|/\tau_c}\longrightarrow \tau_c\delta(t-t')$ with $D=|A_m|^2/\tau_c$. Using the above expressions, we compute the following correlators: 
%%%%%%%%
\begin{eqnarray}
\langle A_m(t) \delta f_m(t)\rangle= \left( \frac{8r^2 D\tau_c}{21 B_0}+im \frac{\mu D \tau_c}{\gamma_d qB_0\gamma } \right)  \frac{\partial f_0}{\partial r} 
\end{eqnarray}
%%%%%%%%
%%%%%%%%
\begin{eqnarray}
\langle \dot{A}_m(t) \delta f_m(t)\rangle= -\left( \frac{8r^2 D}{21 B_0}+im \frac{\mu D \tau_c}{qB_0\gamma } \right)  \frac{\partial f_0}{\partial r} 
\end{eqnarray}
The quasilinear diffusion equation therefore takes the general form: 
%%%%%%%%
\begin{eqnarray}
\label{QLT_Final}%Recompute the whole QLT by keeping things as they stand and replace the partial derivative of delta f_m with Eq 20. 
\frac{\partial f_0}{\partial t}&=&\sum_m \left[ \left(\frac{m \mu r}{\gamma q B_E R_E}\right)^2\frac{\partial }{\partial r} \left(\frac{D \tau_c^2 r^4}{R_E^4} \frac{\partial f_0}{\partial r}\right)+\left(\frac{8}{21}\right)^2\frac{r^2}{R_E^2} \frac{\partial}{\partial r}\left(\frac{D r^8}{B_E^2R_E^4} \frac{\partial f_0}{\partial r}\right)- \frac{1}{3}m^2\Omega_d^2\tau_c^2\frac{D r^9}{B_E^2R_E^6} \frac{\partial f_0}{\partial r}\right].
\end{eqnarray}
We normalise time and the radial distance in the quasi-linear Equation (\ref{QLT_Final}) as $\tau =t/\tau_c$ and $L=r/R_E$, and write  $|\delta B_m|^2=r^2|A_m|^2$ to find: 
%%%%%%%%
\begin{eqnarray}
\label{QLT_Final2}\
\frac{\partial f_0}{\partial \tau}&=&\sum_m \left[ L^2\frac{\partial}{\partial L} \left(\frac{1}{9}m^2\Omega_d^2\tau_c^2L^6\frac{|\delta B_m|^2}{B_E^2} \frac{\partial f_0}{\partial L}\right)+L^2\frac{\partial}{\partial L} \left(\frac{8^2}{21^2}L^8\frac{|\delta B_m|^2}{B_E^2} \frac{\partial f_0}{\partial L}\right)\right] 
\nonumber \\ 
&=& L^2\frac{\partial}{\partial L} \left(\frac{D_{LL}}{L^2} \frac{\partial f_0}{\partial L}\right),
\end{eqnarray}
%%%%%%%%
in which the diffusion coefficient $D_{LL}$ normalised by $\tau_c$ is given by 
%%%%%%%%
\begin{equation}
\label{DLL_new}
D_{LL}=\sum_m\left( \frac{1}{9}m^2\Omega_d^2\tau_c^2+\frac{64}{441}\right)\frac{|\delta B_m|^2}{B_E^2}L^8.
\end{equation}
Equation (\ref{QLT_Final2}) conserves particles confined within a bounded volume since the total rate of change of particles is given by $dN/dt\simeq \int_{L_{min}}^{L_{max}} \ B \ L \ dL \ \partial f/\partial t \simeq \frac{D_{LL}}{L^2}\frac{\partial f_0}{\partial L}\big{|}_{L_{min}}^{L_{max}}$. Moreover, since this diffusion coefficient has been derived for an electromagnetic field model that respects Faraday's law it can be expressed in terms of the wave power in the magnetic field alone and does not require the separation in terms of an electric $D_{LL}^E$ and magnetic  $D_{LL}^B$ diffusion coefficients commonly used in radial transport studies \citep{Ozeke14, Sandhu21}. \\

\noindent The diffusion coefficient is dependent on the first adiabatic invariant $\mu$ contained in the azimuthal drift frequency $\Omega_d$. We note that for large $m\gg 1$ azimuthal wave number, the diffusion coefficient is energy dependent and has a radial distance dependence that goes as $L^6$, even though the short-correlation time assumption would constrain $\Omega_d \tau_c\ll1$. For $m\simeq 1$, and $\Omega_d \tau_c\ll1$, the diffusion coefficient is independent of energy and has an $L^{10}$ scaling: 
\begin{equation}
D_{LL} =
    \begin{cases}
            \frac{m^2\mu^2}{q^2\gamma^2}\tau_c^2 \frac{|\delta B_m|^2}{B_E^2} L^4\sim L^6,&      \text{if } 0.77 m^2\Omega_d^2\tau_c^2\gg 1,\\
            \frac{8^2}{21^2} \frac{|\delta B_m|^2}{B_E^2} L^8\sim L^{10}, &         \text{if } 0.77m^2\Omega_d^2\tau_c^2\ll 1.
    \end{cases}
\end{equation}
This distinction between $D_{LL}$ for high and low azimuthal wave numbers is important for modelling of the Earth's radiation belts because solar wind perturbations can result in both broad and narrow ULF azimuthal wave number spectrums \citep{Murphy20}. For instance, interplanetary shocks can cause a broad spectrum in azimuthal wave numbers with $\{m \in \mathbb{Z}^+ : m<20\}$ \citep{Sarris14, Barani19}. In such an instance, the model predicts an energy dependent $D_{LL}$ that can scales as $L^6$ if the wave power of the high $m$ modes is comparable to the wave power in the low $m$ modes. On the other hand a narrow ULF wave spectrum along the azimuthal wave number $m=1$ should result in a diffusion that is independent of energy and with a radial scaling dependence more sensitive to the radial distance. In other words, the parametric dependence of $D_{LL}$  is a function of how broad the ULF wave spectrum is in $m$. If the magnetospheric plasma is dominated by an $m=1$ mode, with several orders of magnitude less power in $m>1$ modes, a quasilinear modelling of the diffusion coefficient with an $L^{10}$ dependence should be chosen. If the choice of a quasilinear model with an $L^6$ dependence and an energy dependence in $D_{LL}$ provides better accuracy, it would nonetheless be inconsistent with the above radial diffusion coefficients derived for a Mead field.  \\ 
 
\subsection{Beyond a quasi-linear theory of radial transport: nonlinear regime}\label{nonlinear section}
\noindent In the preceding sections we have described the linear response of the perturbed distribution function $\delta f$ and written a Fokker-Planck equation for the quasi-linear evolution of the ensemble-averaged distribution function $f_0(L, t)$. Even in the quasi-linear limit the perturbed distribution function is assumed to be linear while the evolution of the background distribution is nonlinear in the sense that it depends on the correlator $\langle \delta B_m \delta f_m \rangle$. However, the perturbed response given by Equation (\ref{QLT2}) contains a nonlinear term and this section aims to determine when linear assumptions of radial transport breakdown and nonlinear processes become dynamically important.\\

\noindent We distinguish two type of nonlinear regimes. In the first type, nonlinear structures associated with ULF waves are produced but isolated in the sense that they cannot interact with one another. Such structures have been covered in the case of ULF radial transport by \cite{Li18, Wang18} and their observational signatures consist in the appearance of fluxes trapped in the potential well of electric fields. This regime of isolated trapped structures is equivalent to the formation of Bernstein-Green-Kruskal (BGK) mode for a Vlasov-Poisson system \citep{BGK} and requires a sufficiently large-amplitude fluctuation to confine particles in their respective phase-space.\\ 

\noindent In the second type, nonlinearities arise because multiple ULF modes are present and resulting fluctuations in the distribution function interact with one another. This second type of nonlinearity, unlike the first one, can be facilitated by the presence of large-amplitude fluctuations but does not require them. This regime is equivalent to the one presented by \cite{Dupree72} for a Vlasov-Poisson system and associated with the formation of phase-space granulations. These phase-space granulations can consist in linear fluctuations arising due to ballistic trajectories, such as drift echoes, or nonlinear trapped fluctuations equivalent to BGK modes. Theoretical and observational studies have indicated that such a nonlinear regime of non-isolated structures might be common in weakly collisional plasmas \citep{Scheko08, Scheko2016, Meyrand2019, Servidio2017, Kunz18}, prevent Landau damping from dissipating fluctuations \citep{Wu19}, and can result in a phase-space turbulent cascade akin to what is observed in fluid and MHD turbulent systems \citep{GS95}.  \\

\noindent While we acknowledge that ULF wave amplitude in the Earth's radiation belts can be sufficiently large to sustain trapped structures derived by \cite{Li18, Wang18}, the trapping along magnetic local time does not result in irreversible energy gain by the trapped populations. We focus hereafter on the second nonlinear regime which relies on the presence on more than one ULF Pc4 and Pc5 mode. We show hereafter that the second nonlinear regime can result in the transport of particles along magnetic drift shells, and thus irreversible energising of populations that would otherwise be unable to experience drift-resonance. We also demonstrate that the inclusion of nonlinear effects associated with the symmetric ULF fluctuation, which in the linear and quasi-linear regime had no impact, can suddenly become drivers of acceleration and losses. 

\subsubsection{Criteria to determine when nonlinear radial transport becomes significant}
\noindent The nonlinear terms contained in $\mathcal{Q}$ is given by Equation (\ref{nonlinear_term}) can be understood as coupling terms in which a mode with azimuthal wave number $p=m-m'$ couples with a mode $q=m'$ to pump or sink energy from a mode number $m$. For instance, a collection of particles interacting with azimuthal wave numbers $m=3$ and encoded in $\delta f_{m=3}$ and azimuthal wave number $m'=1$ encoded in $\delta f_{m'=1}$ can couple to another through a ULF mode with $p=2$ with $A_{p=2}$. This nonlinear wave-particle coupling can lead to acceleration of nonresonant energetic particles with slow azimuthal drift periods compared to Pc4 and Pc5 ULF frequencies, i.e. $m\Omega_d \ll \omega$. \\

\noindent However, satisfying the condition $p+q=m$ is not enough to make nonlinear effects relevant dynamically for radial transport. The nonlinear coupling terms becomes significant when it becomes comparable to the linear transit term of a particle experiencing an azimuthal drift which is given by the second expression on the left-hand side of Equation (\ref{linear_solution2}). For instance, if we account for the nonlinear term associated with the symmetric ULF amplitude, $S(t)$, with mode number $p=0$, with the particle response to a mode $q=m$, $\frac{S}{B_0} \frac{\partial \delta f_m}{\partial t}\simeq \omega \frac{S}{B_0} \delta f_m $, we find the following two criteria
%%%%%%%%%%%%%%%%%%%
\begin{equation}
\label{criteria_1}
I_1=\frac{\text{symmetric nonlinear term \# 1}}{\text{linear transit time}}\simeq\frac{\omega_m}{m\Omega_d} \frac{S}{B_0} \simeq 1,
\end{equation}
%%%%%%%%%%%%%%%%%%%
%%%%%%%%%%%%%%%%%%%
\begin{equation}
\label{criteria_2}
I_2=\frac{\text{symmetric nonlinear term \#2}}{\text{linear transit time}}\simeq\frac{L}{2}\frac{\omega}{m\Omega_d} \frac{S}{ B_0}\frac{\partial}{\partial L} \log \delta f_m\simeq 1,
\end{equation}
%%%%%%%%%%%%%%%%%%%
in which the frequency $\omega_m$ is associated with time variations of the perturbed distribution function $\delta f_m$ and the frequency $\omega$ with the symmetric ULF wave amplitude $S(t)$. We note that the linear ballistic response of the perturbed distribution function given by Equation (\ref{linear_solution2}) resulted in time variations with frequencies $\omega_m=m\Omega_d$, thus nonlinear effects can be felt whenever the symmetric ULF amplitude becomes comparable to the local magnetic field. However, criteria $I_1$ can also be satisfied in the limit where the symmetric ULF fluctuations are small in amplitude, i.e. $S(t)/B_0 \ll 1$, if $\omega_m \gg m\Omega_d$. For criteria $I_2$, nonlinear effects become significant for large gradients in the perturbed distribution ($\partial \log(\delta f_m)/\partial L \gg 1$) even in the limit $S\ll B_0$.\\  %These two criteria are associated with Nonlinear processes become dynamically important when their impact can be felt on timescales comparable to the drift period of the trapped particles. \\ 

\noindent We can account for nonlinearities associated with the anti-symmetric perturbation $\frac{r A_{m-m'}}{B_0}\frac{\partial \delta f_{m'}}{\partial t}\simeq \omega_{m'} \frac{rA_{m-m'}}{B_0} \delta f_{m'}$  by defining two additional criteria $I_3$ and $I_4$ in terms of the nonlinear terms 
%%%%%%%%%%%%%%%%%%%
\begin{equation}
\label{criteria_3}
I_3=\frac{\text{anti-symmetric nonlinear term \# 1}}{\text{linear transit time}}\simeq\frac{\omega_{m'}}{m\Omega_d} \frac{rA_{m-m'}}{B_0} \frac{\delta f_{m'}}{{\delta f_m}} \simeq 1,
\end{equation}
%%%%%%%%%%%%%%%%%%%
\begin{equation}
\label{criteria_4}
I_4=\frac{\text{anti-symmetric nonlinear term \# 2}}{\text{linear transit time}}\simeq L\frac{8}{21}\frac{\omega_{m-m'}}{m\Omega_d} \frac{rA_{m-m'}}{B_0} \frac{\delta f_{m'}}{{\delta f_m}}\frac{\partial}{\partial L} \log \delta f_{m'} \simeq 1.
\end{equation}
%%%%%%%%%%%%%%%%%%%
We note once more that since the linear response of the perturbed part $\delta f_{m'}$ can result in time variations with frequencies $\omega\simeq m' \Omega_d$, nonlinear effects can be sensed whenever $\frac{m'}{m} \frac{\delta B}{B_0} \frac{\delta f_{m'}}{\delta f_m} \simeq 1$. If $m=1$, $m'>m$, or in the presence of large gradients, small amplitude ULF fluctuations $r A_{m-m'}=\delta B_{m-m'} \ll B_0$ can nonetheless result in dynamically relevant nonlinear effects.  \\ 

\subsubsection{Nonlinear impact of symmetric perturbations on fast timescales}\label{fast_symmetric}
\noindent In the previous section we have defined four criteria to argue that nonlinear effects can become significant even for small amplitude ULF fluctuations. In this section we focus on the nonlinearity arising from the symmetric ULF perturbation $S(t)$. The nonlinear Equation ({\ref{QLT2}}) for the perturbed distribution function $\delta f_m$ can be solved analytically on fast timescales comparable to the drift period of particles. The linear solution to Equation  ({\ref{QLT2}}) is independent of the symmetric perturbation $S(t)$. But the nonlinear term $\mathcal{Q}$ contains a coupling term between the symmetric perturbation and $\delta f_m$. This nonlinear response of the particles with a mode $m$ is due to the coupling between the  $m=0$ ULF mode contained in the symmetric perturbation and itself. If we assume that the nonlinear coupling due to $S(t)$ is greater than the one due to the anti-symmetric ULF waves $S\gg r A_m $, Equation ({\ref{QLT2}}) becomes:  
%%%%%%%%
\begin{eqnarray}
\label{fast_1} 
\frac{\partial \delta f_m}{\partial t}+{i m\Omega_d \delta f_m} ={-\left( \frac{8r^2\dot{A}_m}{21 B_0}-im \frac{\mu A_m}{qB_0\gamma } \right)\frac{\partial f_0}{\partial r}}{+\frac{S}{B_0}\frac{\partial \delta f_{m}}{\partial t}} -\frac{r\dot{S}}{2B_0}\frac{\partial \delta f_{m}}{\partial r}.
\end{eqnarray}
%%%%%%%%
In order to isolate the impact of the symmetric perturbation arising due to nonlinear coupling we split the perturbed distribution in terms of a linear part $\delta f_m^L$ given by Equation (\ref{linear_solution2}) and a nonlinear part $\delta f_m^{NL}$ that can be extracted from the following equation:
%%%%%%%%
\begin{eqnarray}
\label{fast_2}
\frac{\partial \delta f_m^{NL}}{\partial t}+{i m\Omega_d \delta f_m^{NL}} ={\frac{S}{B_0}\frac{\partial \delta f_{m}^{L}}{\partial t}} -\frac{r\dot{S}}{2B_0}\frac{\partial \delta f_{m}^{L}}{\partial r}.
\end{eqnarray}
%%%%%%%%
\noindent In Equation (\ref{fast_2}) we assume that the nonlinear perturbation remains smaller than the linear response, $| \delta f_m^{NL}| <  | \delta f_m^{L}|$, and thus we can solve the nonlinear equation perturbatively to drop the coupling terms proportional to $\delta f_m^{NL} S(t)$. Equation (\ref{fast_2}) is linear in $\delta f_m^{Nl}$ and can now be solved if we prescribe a solution for the linear response $\delta f_m^L$. \\

\noindent For the sake of simplicity, and in order to highlight that ULF radial transport can have an impact on non resonant particles on fast timescales comparable or less than the drift period, we assume that the linear perturbation $\delta f_m^{L}$  is given by an injection or a loss of $100$ keV electrons consistent with the linear solution, and set $A_m=0$\footnote{The linear response is taken as the ballistic one $\delta f_m=\delta f_m(t=0, r) e^{-im\Omega_d t}$. Inclusion of the linear wave-particle response for $A_m\neq 0$ leads to the same physical process and is left for future more detailed studies of higher order radial transport.}. Particles with 100 keV confined in the equatorial plane at normalised radial distances $L\leq8$ have azimuthal drift periods of the order of 90 to 120 minutes. Thus, frequencies of the order of $\omega \simeq 1$ mHz would require azimuthal wave numbers of $m \geq 10$ \citep{Barani19}. \\

\noindent We assume an injection of $50$ keV given by a Gaussian centred at a radial distance $L_c$ and with a radial spread $\Delta L$ 
%%%%%%%%
\begin{eqnarray}
\delta f_m^L(L, t)&=&\delta f_m(0,L) e^{-im\Omega_d t} \nonumber\\
&=& e^{-\frac{(L-L_c)^2}{\Delta L^2}}e^{-im\Omega_d t}.
\end{eqnarray}
%%%%%%%%
\noindent The symmetric perturbation is modeled as a compression of the magnetic field with a decay time $\tau_c^S$, 
\begin{eqnarray}
S(t) = \delta b \ e^{-t/\tau_c^S}.
\end{eqnarray}
%%%%%%%%
The perturbed solution for the distribution function $\delta f_m^{L}+\delta f_m^{NL}$ following the Gaussian shaped injection and decaying symmetric ULF mode is given by
%%%%%%%%
\begin{eqnarray}
\label{sol_fast}
\delta f_m^{L}+\delta f_m^{NL}=\delta f_m(0,L)e^{-im\Omega_d t}\left[1-\frac{\delta b}{B_0}\bigg{(}1-e^{-t/\tau_c^s}\bigg{)}\left( i m\Omega_d \tau_c^S+\frac{L(L-L_c)}{\Delta L^2} \right)\right].
 \end{eqnarray}
%\delta f_m^{NL}=\frac{\delta b}{B_0} \delta f_m(0,L) \left[m\Omega_d\tau_c^S \sin(m\Omega_d t)\left(2-e^{-t/\tau_c^S}\left(2+\frac{t}{\tau_c^S}\right)\right)-\cos(m\Omega_d t)\left(\frac{L(L-L_c)}{\Delta L^2}-e^{-t/\tau_c^S}\right)\right]

\begin{figure}[ht!]
\plotone{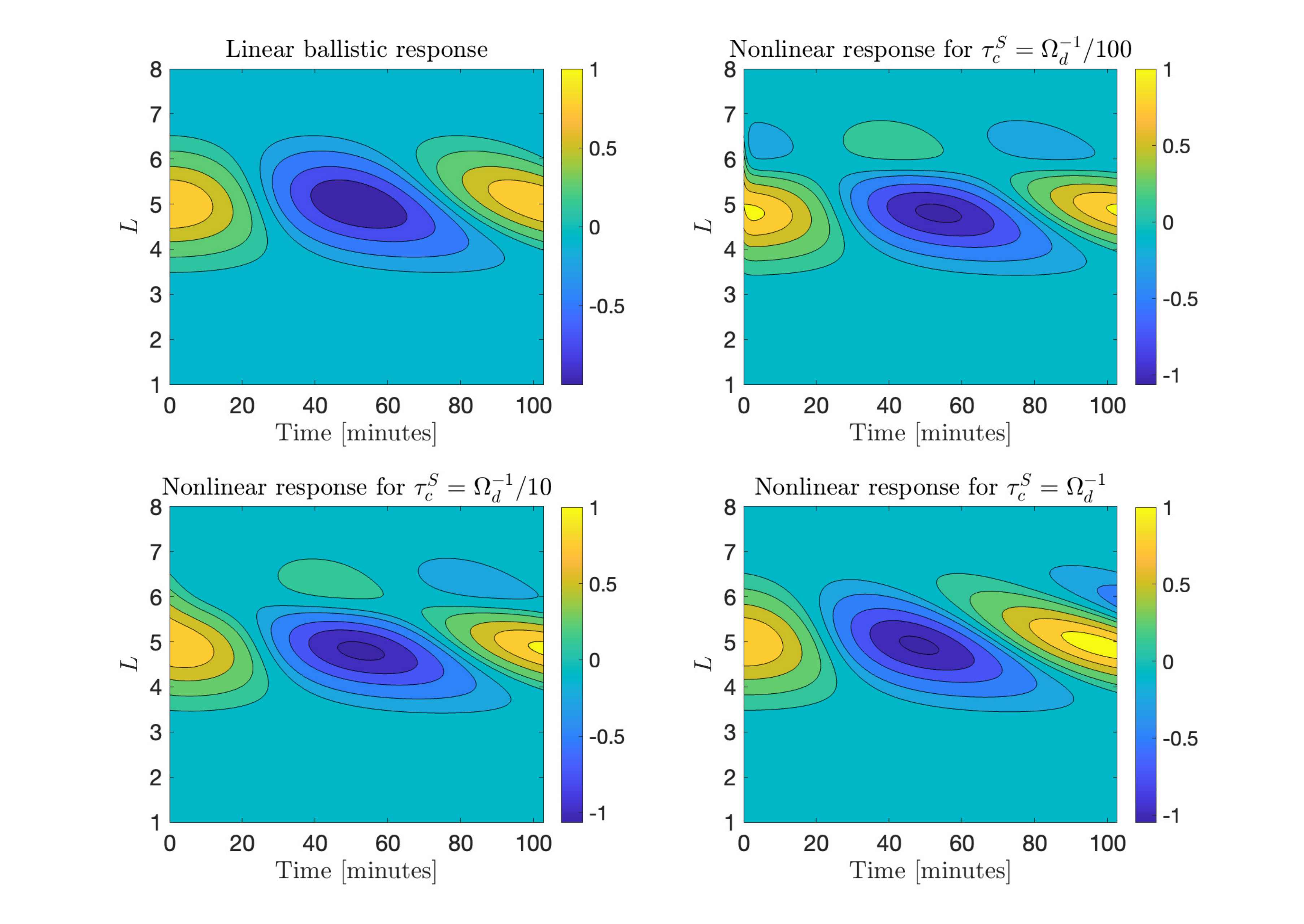}
\caption{Impact on the perturbed distribution function of $100$ keV injected electrons at $L=5$ by symmetric perturbation on timescales less than one drift period for symmetric perturbation of amplitude $\delta b=0.12 B_0$ at $L=5$. The color scale denote the perturbed distribution amplitude.}
\label{Figure_4}
\end{figure}

%%%%%%%%
\noindent The nonlinear response given by (\ref{sol_fast}) for the $m=1$ mode is shown in Figure (\ref{Figure_4}) for a symmetric ULF wave amplitude of $\delta b = 0.12 B_0$. The top left panel corresponds to the linear response. After the injection of the particles at $L_c=5$, the distribution function oscillates in time and gets sheared along $L$. However, when we introduce a symmetric perturbation with a decay time that is smaller than the drift period (with $\tau_c^s< \Omega_d$), the distribution function splits at the injection point. This non-adiabatic behavior is shown in the top right and bottom left panels of Figure (\ref{Figure_4}). In comparison, an adiabatic decay of the ULF mode with $\tau_c^S \geq \Omega_d$ has no impact on the distribution function, as shown in the bottom right panel of Figure (\ref{Figure_4}). \\

\begin{figure}[ht!]
\plotone{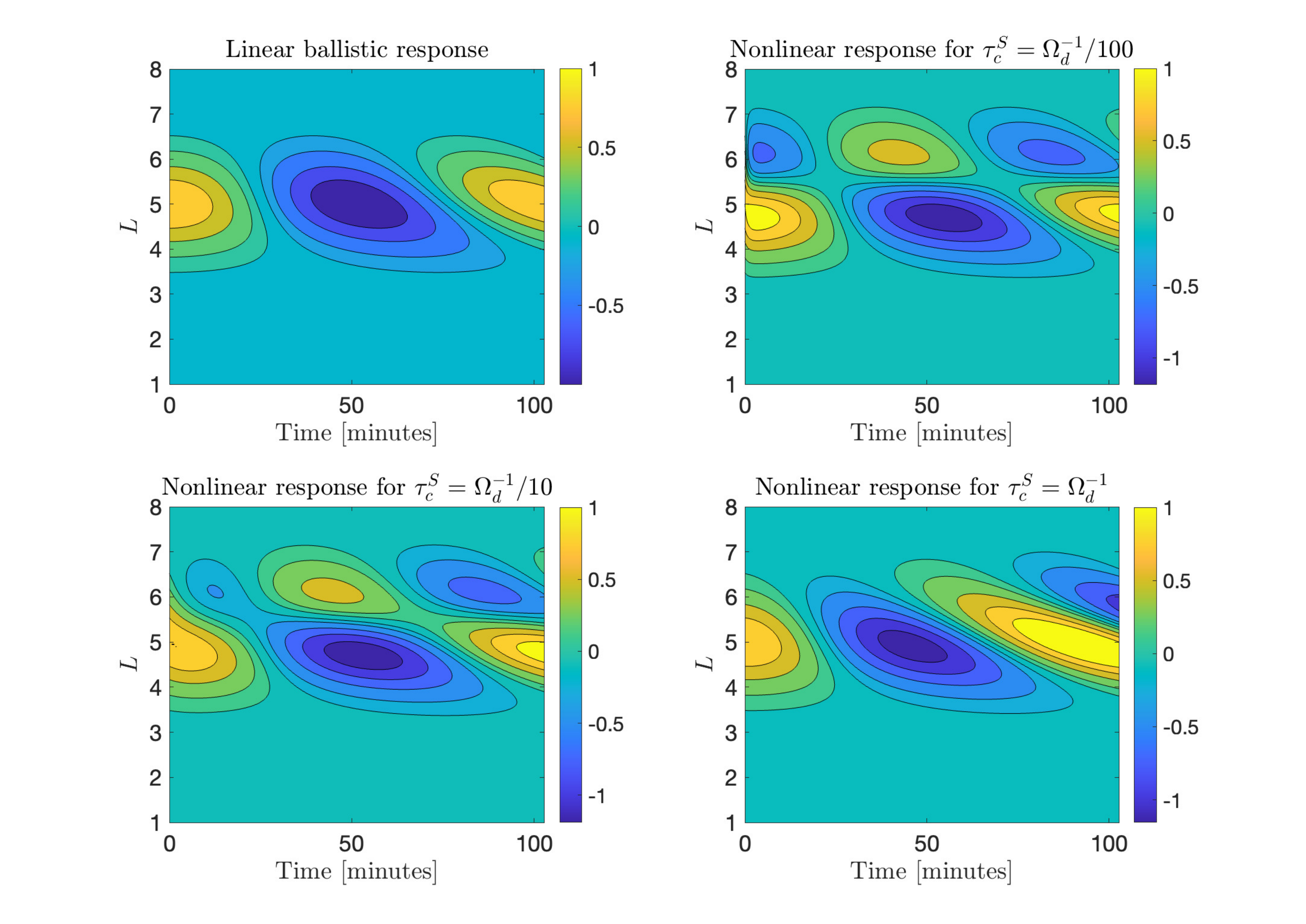}
\caption{Same as Figure \ref{Figure_4} but with symmetric perturbation of amplitude $\delta b=0.25 B_0$ at $L=5$.}
\label{Figure_5}
\end{figure}

\begin{figure}[ht!]
\plotone{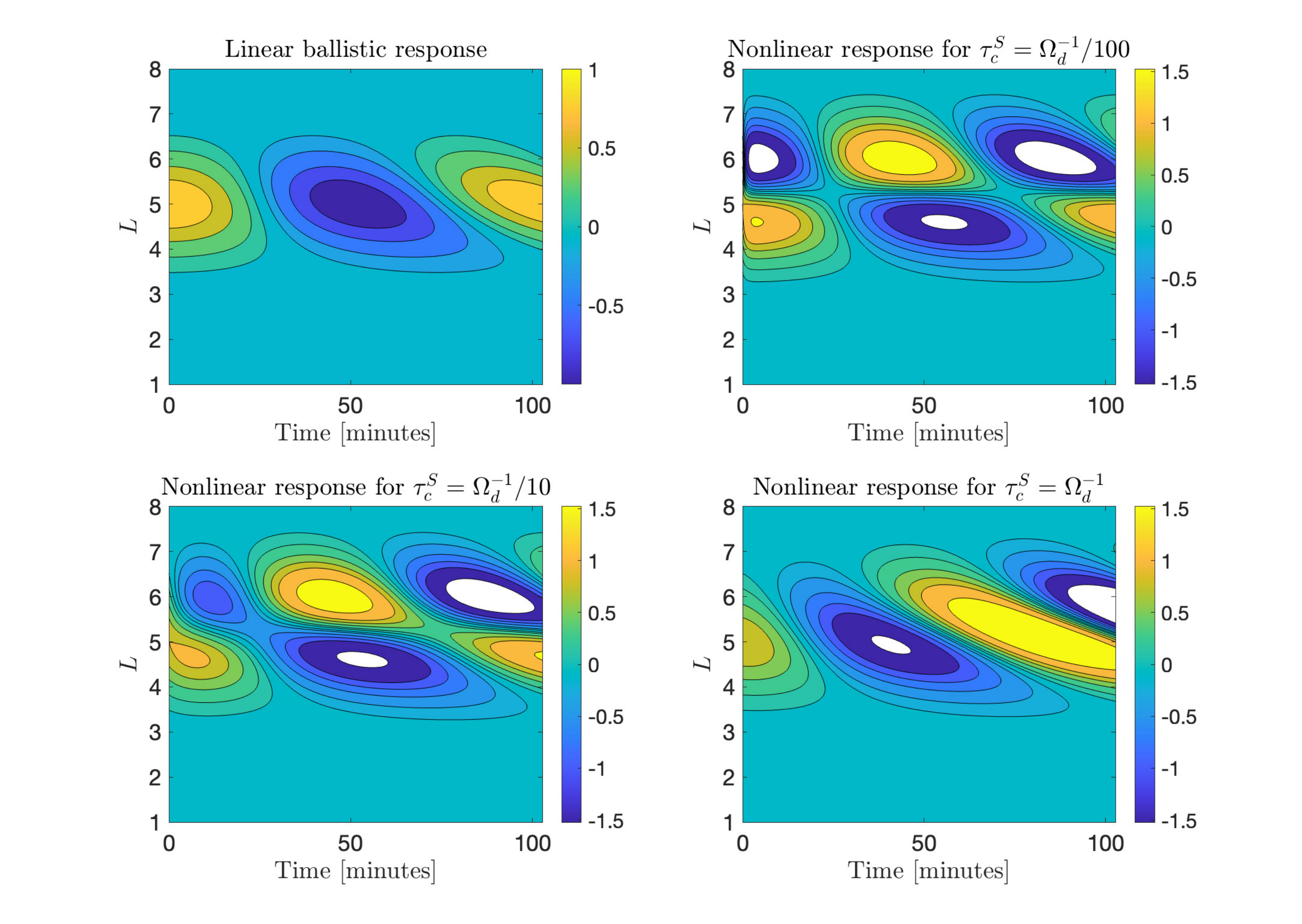}
\caption{Same as Figure \ref{Figure_4} but with symmetric perturbation of amplitude $\delta b=0.62 B_0$ at $L=5$.}
\label{Figure_6}
\end{figure}

\noindent The physical process responsible for this mechanism is illustrated in Figure (\ref{schematic2}). A symmetric ULF compression with amplitude $S(t)$ results in an $E\times B$ differential gradient that is larger in amplitude at higher than lower drift shells. Drift shells with negative (positive) gradients result in particles being driven inward (outward). If the ULF compression is adiabatic particles phase-mix along $L$, but if the compression is non-adiabatic and the $E\times B$ drift decays or grow too fast (compared to the azimuthal drift period) for phase-mixing to occur, the net radial drift is inward. This net motion of particles inward is shown in Figures (\ref{Figure_5}) and (\ref{Figure_6}), for a ULF symmetric amplitude corresponding to 25\% and 62\% of the background field at $L=5$. The inward moving particles increase in energy in order to conserve the first adiabatic invariant whereas the outward moving particles loose energy. This process can result in the fast and irreversible acceleration of particles as well as losses associated with shadowing even though there is no drift-resonance with the ULF modes. These results demonstrate that the inclusion of higher order effects can lead to non-diffusive and irreversible radial transport on fast timescales. Such a process cannot be modeled with quasi-linear radial diffusion.

%for an injection of particles centred at $L_c=5$ and with a spread $\Delta L=1$. The ratio of the perturbation $\delta b $ is given by $0.0008 B_E$, making $S/B_0 \ll1$ for $L\leq 8$. The amplitude of the nonlinear perturbation remains much smaller than linear one as required by the perturbation analysis leading to Equation (\ref{fast_2}). The time evolution is plotted for $t=[0, 30]$ minutes and corresponds to the evolution of the distribution for a fraction of a single drift period. The top left panel shows the linear solution and the remaining five panels the nonlinear perturbation for decaying times of the symmetric ULF perturbation $\tau_c^S=[\frac{1}{2}, 2, 5, 10, 20]$ minutes. \\

%\left[\left( i\frac{m'\mu A_{m-m'}}{qB_0\gamma r} -\frac{m'r\dot{A}_{m-m'}}{7B_0(m-m')}\right)\delta f_{m'}  -\left(\frac{S}{B_0}\delta_{m,m'}+\frac{A_{m-m'} r}{B_0} \right)\frac{\partial \delta f_{m'}}{\partial t}\right] \nonumber\\ 
%&+& \left( \frac{8r^2\dot{A}_{m-m'}}{21 B_0}-i(m-m') \frac{\mu A_{m-m'}}{qB_0\gamma }+\frac{r \dot{S}}{2 B_0}\delta_{m,m'}  \right)\frac{\partial \delta f_{m'}}{\partial r}.

\begin{figure}[ht!]
\plotone{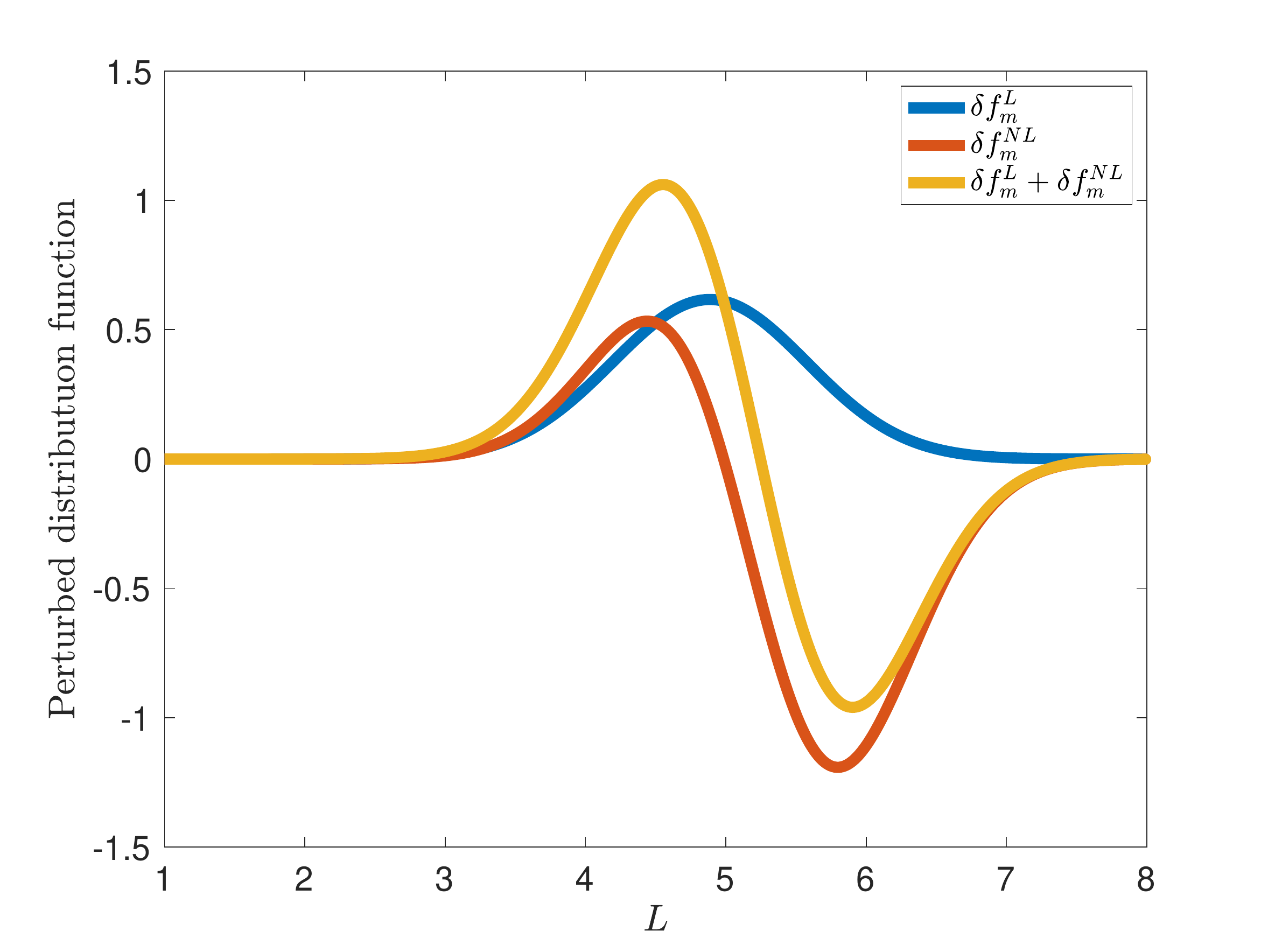}
\caption{Cut of the linear and nonlinear perturbed distribution function at $\varphi=0$ in Figure \ref{Figure_6}. The non-adiabatic symmetric perturbation splits the distribution functions by pushing particles inward and outward.}
\label{Figure_7}
\end{figure}

\begin{figure}[ht!]
\epsscale{1.0}
\plotone{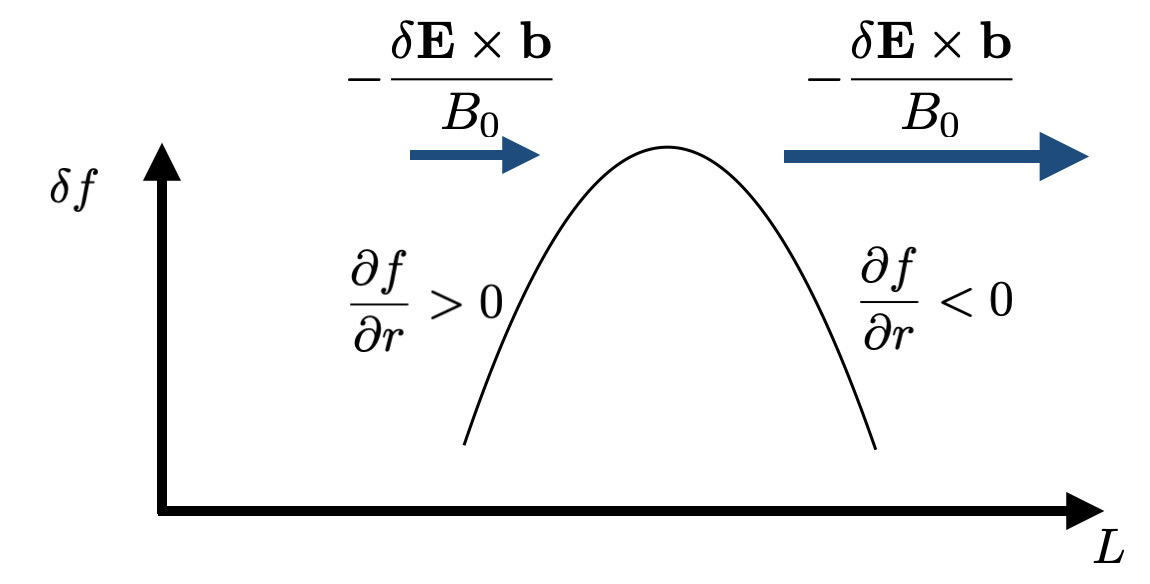}
\caption{Explanation for the nonlinear mechanism presented in Section \ref{fast_symmetric}. A symmetric ULF compression with amplitude $S(t)$ results in an $E\times B$ differential gradient that is larger in amplitude at higher than lower drift shells. Drift shells with negative (positive) gradients result in particles being driven inward (outward). If the ULF compression is adiabatic particles phase-mix along $L$, but if the compression is non-adiabatic and the $E\times B$ drift decays too quickly for phase-mixing to occur, the net drift is inward.}
\label{schematic2}
\end{figure}

\section{Discussion}
\subsection{When can we use quasi-linear radial diffusion?}
\noindent A drift kinetic description of ULF wave interaction with energetic particles is a convenient methodology to define the regime of validity of quasi-linear radial diffusion problems. In comparison, the derivation in terms of the particle's trajectories \citep{Falthammar65, Elkington, Lejosne19} is mathematically more transparent than the one provided in Section \ref{section_diffusion} but since it does not require computation of the perturbed orbits, it does not distinguish explicitly between the fast perturbed part and the slow background part of the distribution function.\\

\noindent The procedure to derive the radial diffusion coefficient is identical to the one pursued for other quasi-linear theories in laboratory and astrophysical plasmas \citep{Kennel66, Diamond10, Schekochihin_notes}. Quasi-linear theories require temporal and spatial scale separation of the distribution function in terms of a slow ensemble-averaged background component and a fast perturbed component. The fast component can evolve on timescales comparable to the periods of electromagnetic fluctuations responsible for the wave-particle interactions. For instance, for seed electrons of 10-100 keV interacting with high frequency whistler waves, the quasi-linear theory of \cite{Kennel66} is explicitly clear that the perturbed component evolves on timescales of the order of the whistler period, and thus the Larmor period as well, since $\omega \simeq \Omega_s$. The diffusive evolution of the distribution function requires a large number of interactions with whistler waves and is therefore computed on timescales that have been averaged over a large number of whistler wave period. The perturbed part is computed linearly, and thus quasi-linear theory assumes that nonlinear effects such as trapping and mode-mode coupling associated with large-amplitudes can be neglected. \\ 

\noindent For a quasi-linear theory of radial transport to be consistent, one needs to preserve the scale separation defined by Equations (\ref{def_average2})  and (\ref{def_average}). The background distribution function is not only independent of magnetic local time, and thus $\varphi$, it cannot change significantly on timescales comparable to the drift period $\Omega_d$. A radial diffusion coefficient that becomes comparable to the drift period ($D_{LL} \simeq \Omega_d$) indicates that a collection of particles can be carried across one drift shell $(\sqrt{\langle\Delta L^2\rangle}\simeq 1$) during a single drift period. This argument stems from the fact that dimensionsally the radial diffusion coefficient scales as $D_{LL}\simeq \langle \Delta L^2 \rangle/t$, and that the inverse of the diffusion coefficients gives a characteristic time for transport across one drift-shell. Taking into account that the derivation of the quasi-linear diffusion coefficients requires a short decorrelation time of the ULF wave amplitude, and the observational fact that ULF waves are long-lived and coherent \citep{Hartinger13}, it is inconceivable that a diffusive scattering along drift-shells can occur over a single drift period \footnote{This heuristic argument is to some degree arbitrary, but for lack of a better alternative, provides a reasonable and reliable constraint on radial diffusion coefficients.}\\ 

\noindent The determination of accurate radial diffusion coefficients is not merely of academic interest and has important consequences on space weather models and in radiation belts' studies focused on distinguishing between local and global acceleration processes \cite{Green04}. Current global magnetospheric models accounting for radial diffusion rely on $D_{LL}$ coefficients that become comparable, and for large geomagnetic activity larger, than the drift periods of energetic electrons trapped in the Earth's radiation belts \citep{Brautigam, Ozeke14}. For instance, the radial diffusion coefficient of \cite{Ozeke14} can become as large as $10^2$ in units of day$^{-1}$ for Kp>5, which corresponds to drift period of electrons with energies of 1 MeV. Additionally, derived radial diffusion coefficients assume that the ULF wave correlation $\langle \delta \mathbf{B} (t) \delta \mathbf{B} (t+\tau)\rangle$ is time and space homogeneous along the particle's orbits. However, ULF waves are sustained by a wide range of processes that are not co-located, ranging from Kelvin-Helmholtz instabilities \citep{Mills99}, pressure pulses in the solar wind \citep{Takahashi07}, foreshock transients \citep{Hartinger13}, magnetospheric substorms \citep{Volwerk16}, and unstable plasma distributions \citep{Southwood69}. As a consequence, ULF waves are not homogeneously distributed in the magnetosphere \citep{Murphy20}, and unless the ULF waves decay very fast compared to the drift period, quasi-linear radial diffusion coefficients accounting for non-homogeneous statistics have to be derived \footnote{\cite{Osmane21a} show that spatially homogeneous ULF waves with finite correlation time comparable to the drift period result in sub-diffusive radial transport and the slowing down of radial diffusion. The inclusion of non-homogeneous effects in radial diffusion are, to the best our knowledge, currently missing. }. \\

\noindent The above mentioned limitations of quasi-linear radial diffusion do not imply that ULF waves cannot sustain transport on timescales comparable to the drift period. Rather, what is argued is that current quasi-linear radial diffusion models have clear limitations, and should not be used beyond their range of validity. Radial transport coefficients encoding the impact of ULF waves on fast timescales require models that are not quasi-linear. A drift kinetic approach to radial transport is not confined to theoretical or modelling studies. With GPS flux measurements calibrated by Van Allen Probes' instruments, it is now  possible to quantify observationally radial transport on timescales of the order of a single drift period for electrons with energies less than 1 MeV \citep{Morley16, Morley17, Kalliokoski22b}.

\subsection{Fast radial transport}\label{subsection_fast}
\subsubsection{Distinguishing between drift resonant and non resonant interactions}
\noindent The scale separation described in Section (\ref{sec:multiscale}) forms the basis to derive a quasi-linear theory of radial transport, but is also appropriate to quantify the linear and nonlinear response of the distribution function that occurs on fast timescales comparable to a few drift periods. Section (\ref{sec:linear}) described three different types of linear responses associated with a ULF wave of frequency $\omega_m$, growth or damping rate $\gamma_m$, and azimuthal wave number $m$. Three of these responses are non-resonant and one corresponds to drift resonance of particles drifting the Earth's magnetic field with frequency $\Omega_d \simeq \omega_m/m$. The first type of non-resonant response is a modulation of the distribution function with the frequency of the ULF wave $\omega_m$, and the second type of non-resonant response is an oscillation of the distribution function at the drift frequency $\Omega_d$. While both resonant and non-resonant responses to a ULF wave are a function of the local gradient in the background distribution function, the resonant particles are energy dependent and the perturbed distribution is amplified by up to one order of magnitude and is therefore distinguishable from non-resonant responses. \\

\noindent Models of ULF drift resonance predict that satellites should observe the largest modulations in particle flux at energies corresponding to the resonant energy, with smaller modulation at lower/higher energy \citep{Southwood81}. Equation (\ref{linear_solution3}) confirms this signature for resonance but also demonstrates that non-resonant, as well resonant particles, can oscillate at the ULF wave frequency. \textit{In situ} observation of distribution functions or fluxes oscillating at a ULF frequency $\omega_m$ should therefore not be assumed as a signature of drift-resonance unless the response is localised in energy spectrograms. For drift resonance, the timescales associated with the resonant interaction and the width are a function of the growth rate, and we here stress that seeing comparable modulation across multiple energy levels for a monochromatic ULF wave spectrum is an indication that the interaction is non-resonant. \\ 

\noindent In the study of \cite{Claudepierre13} fluxes of energetic electrons between 20 and 500 keV are identified as unambiguous signatures of localised drift-resonant interaction with a ULF wave. However, no analysis is provided to quantify the radial gradient of the distribution function for each respective energy fluxes. As shown in this report, the modulation of particle fluxes in terms of a the ULF wave frequency does not require drift-resonance and can be observed for non resonant particles as well. The difference in amplitude between fluxes can be explained in terms of radial gradient differences between energetic fluxes. The localised modulation in time can be explained by a ULF wave that is being damped at a rate $\gamma_m$, and the spatially localised modulation seen on one Van Allen Probe but missed by the second probe can be an indication that the radial gradient of the distribution function is highly spatially localised. Large and localised radial gradients of the distribution function have been reported for case studies. For instance, \cite{Hartinger20} points out that at $L=4.5$ and $L=6.6$ the reported radial PSD gradients are 30-300 times larger at values corresponding to energies of 200 keV compared to 1 MeV. Consequently, residual flux oscillations in this particular case would be 30-300 larger for electrons with energies of 200 keV rather than 1 MeV.  Thus, characterizing flux oscillations without accounting for radial gradients, known to vary by several order of magnitude, can lead to erroneous interpretation of wave-particle processes.\\

\noindent As shown in Figures (\ref{fig:Resonance_condition_Pc5}) and (\ref{fig:Resonance_condition_Pc4}) ULF waves in the Pc4 and Pc5 range can be resonant with electrons of energies ranging between 100 keV and a few MeV, yet signatures of drift resonances for the most energetic MeV populations is rare. \cite{Hartinger20} addresses this inconsistency between observations and theoretical assumptions. On the basis of the theoretical study of \cite{Southwood81}, in order for drift resonances to be observed, one requires finite radial gradients in the background distribution function. Drift resonant interactions could still occur but would be masked by small radial gradients in the background distribution function. While we are in agreement with the conclusions of \cite{Hartinger20}, that drift resonance requires observable gradients to be detected, our analysis of the resonant response provides one additional constraint. Drift resonant signatures result in an amplification of the particle's response, as shown in Equation (\ref{linear_solution3}), that is localised in time and can be comparable to a single drift period. Moreover, if the ULF wave damps quickly, that is, on timescales comparable to a few drift periods, the resonant exchange could be too weak to be observed or distinguishable from the non-resonant one. Keeping in mind the conclusions of \cite{Hartinger20} regarding the importance of radial gradients, our analysis provides an additional explanation as to why observation of drift resonant signatures have been rare when detected by a few spacecrafts. Drift-resonance is a transient process and detection by one spacecraft can be entirely missed by another spacecraft sampling the same orbit but on timescales larger than a few drift periods.   \\

\subsubsection{Mechanisms for zebra stripes formation}
\noindent Even though phase-space structures in the radiation belts are not necessarily indicative of violation of the third adiabatic invariant, and thus acceleration, their observed signatures can be used to test the validity of radial transport models or be used as diagnostic for electric fields or particle injections. In Section \ref{sec:linear} we showed that phase-mixing of trapped electrons can result in the formation of structures known as zebra stripes. Zebra stripes are transient structured peaks and valleys observed on spectrograms of inner radiation belts' electrons with energies ranging between tens to hundreds of keV. The zebra stripes that are measured \textit{in situ} are also characterised by energy peaks and dips that vary as the inverse of the radial distance, i.e., $E_c\sim 1/L$ \citep{Sauvaud13, Lejosne16zebra, Lejosne20a, Lejosne20b}. Since the zebra stripes can be produced on timescales of the order of a few drift periods, a radial diffusion mechanism should be immediately rejected. Our analysis also shows that zebra stripes can form without drift resonance with ULF waves, and as a result of a phase-mixing process described for non resonant particles. The phase-mixing process described in this report is triggered by particle injection or losses from the the radiation belts, and the requirement for an electric field that sustains drift-resonance, as shown in \cite{Sasha_zebra}, is therefore unnecessary. The requirement for a drift resonant interactions to produce zebra stripes is also more constraining than a non resonant phase-mixing mechanisms, since resonance requires a ULF fluctuations with a narrow set of parameters and finite radial gradient in $f_0$ \footnote{On the basis of Occam's razor argument we would favor a phase-mixing mechanism of non resonant particles to explain zebra stripes formation \citep{Popper}.}.\\

\noindent How can we distinguish between zebra stripes formation mechanisms? We note that the first phase-mixing mechanisms, described in Section \ref{zebra_stripes_section}, requires injection or losses of particles but no electric field. The second type, appearing as the ballistic term in Equation (\ref{linear_solution2}), requires a ULF fluctuations and a finite radial gradient in the distribution function. The third type, described by \cite{Sasha_zebra}, but appearing as the drift-resonant term in Equation (\ref{linear_solution2}), requires a ULF fluctuations that can resonate witha wide rage of energies, and also a finite radial gradient in the distribution function. For all three types the formation and shearing occurs on the same timescales. In order to distinguish both phase-mixing mechanisms one needs to measure radial gradients in the phase-space density and determine if the amplitude of the ULF fluctuations can provide the amplitude of stripes structures observed. If such a test proves inconclusive, the phase-mixing process requiring injection such as the one detailed in \cite{Zhao13}, might be favored. If future observational studies demonstrate that injection or loss of particles in the inner belts correlate with phase-mixed structures, one could use zebra stripes as proxies for injection and losses. Similarly, if the phase-mixing process is primarily driven by ULF fluctuations, appearance of zebra stripes could be used as proxies to extract properties of electric fields in the inner belts. 

\subsection{Nonlinear Parker mechanism}

\noindent The first radial transport model resulting in irreversible acceleration of particles was presented by \cite{Parker60} and did not require drift resonance processes. In the \cite{Parker60} scenario, magnetically confined particles experience non-adiabatic transport as a result of an asymmetric magnetic field perturbations. Since particles at different MLT sectors of the same drift shells sensed a different perturbation, they collectively experienced a net radial transport. The mechanism presented in Section \ref{fast_symmetric} is a higher order generalisation of the \cite{Parker60} mechanism in that it does not require drift-resonance with ULF waves. This nonlinear mechanism is also the product of non-adiabatic perturbations but does not require asymmetric magnetic fluctuations. Rather the only two ingredients required for this nonlinear process to result in irreversible radial transport is 
\begin{enumerate}
    \item Large amplitude symmetric perturbation $\delta B/B_0 \simeq 10\%$ decaying or growing non-adiabatically, and
    \item Opposite radial gradients in the distribution function, or put differently, a localised minimum or maximum of the distribution function along the radial distance.
\end{enumerate}
 While particles on the same drift-shells sense the same electromagnetic field and radial drift speeds, particles on different drift shells drift at a different speed, and the combined inward and outward transport in the presence of opposite gradients results in irreversible acceleration as more particles are pushed inward than outward. If the waves decorrelate very slowly (adiabatically) compared to the drift period, particles will phase-mix radially and instead of a net injection inward, a plateau along the radial distance will form.\\

\noindent Are symmetric ULF fluctuations observed in the Earth's radiation belts? In a recent observational study, \cite{Takahashi22} provides the first description of symmetric compressional ULF fluctuations with magnetic field amplitudes comparable to the background magnetic field. The symmetric ULF waves are excited outside of the plasmasphere, and localised in MLT and radial distance. The large-amplitude ($\delta B/B_0 \geq 0.1$) and compressional nature of the fluctuation described by \cite{Takahashi22} are consistent with the one used for the acceleration processe presented in Section \ref{nonlinear section}. Moreover the waves are observed in association with injection of particles, and thus symmetric fluctuations are associated with local radial enhancements of particles. Even though it is too speculative at this point to determine whether this mechanism is commonly occurring in the radiation belts, we want to stress that the two required ingredients for the occurrence of this nonlinear mechanism have been observed in the radiation belts. Unlike radial diffusion, which operates on long timescales and requires a large number of drift-resonant interactions, fast and nonlinear acceleration mechanisms can be both seldom and more efficient. 

\section{Conclusion}

\noindent In this report, we have presented a drift kinetic description of ULF radial transport for the Earth's radiation belts. The use of a drift kinetic formalism is particularly convenient to distinguish quasi-linear diffusion occurring on slow timescales, with fast wave-particle interactions associated with linear or nonlinear processes. Theoretically, current global models of the Earth's magnetosphere account for ULF radial transport solely in terms of quasi-linear diffusion models. Our analysis demonstrates that linear and nonlinear processes occurring on timescales of the order of the drift period and with a spatial dependence on magnetic local time cannot be modelled in terms of quasi-linear diffusion. Observationally, fast and localised radial transport have been known for decades, but have been limited to extreme driving events or serendipitous satellite measurements \citep{Li93, Kanekal16}.  In the recent years, calibration of GPS electron flux measurements with Van Allen Probes' instruments are offering for the first time unprecedented spatial and temporal coverage of the Earth's radiation belts on timescales comparable to the azimuthal drift period \citep{Morley16, Morley17, Kalliokoski22b}. Thus, a modelling framework that distinguishes between fast and slow radial transport is not only of theoretical interest, but can also be tested for the first time with \textit{in situ} measurements for a wide-range of geomagnetic driving conditions. \\

\noindent In the last two decades, dominant acceleration processes in the Earth's radiation belts have been categorised as belonging to local wave-particle interactions or global ULF radial diffusion. The observational signature of local wave-particle processes in the phase-space density consists in localised enhancements, whereas ULF radial diffusion results in the flattening of the phase-space density along the radial distance \citep{Green04, Reeves13}. When including higher order terms in the radial transport equation, we found that seed electrons with $50-100$ keV injected in the outer belts can experience additional betatron acceleration in the presence of symmetric ULF wave amplitudes with amplitudes comparable to the one reported by \cite{Takahashi22}. This impulsive nonlinear process requires no drift resonance yet results in a localised enhancement of the phase-space density on timescales that are much shorter than the drift period. This theoretical result is therefore of particular interest to observational studies radiation belts since ULF waves are also able to produce localised signatures attributed to small-scale wave-particle interactions. With growing satellite coverage and the capacity to measure electron fluxes on timescales comparable to the drift period the binary quasi-linear framework developped in the past decades need to be revisited. \\

\noindent The main focus of this paper has been on radial transport of energetic electrons in the Earth's radiation belts. However, a drift kinetic description based on the work of \citep{Hazeltine73} can also be used to describe energetic ring current protons ($>100 $ keV) with Larmor frequencies $\Omega_p\sim 1-10$ Hz responding to ULF fluctuations $\omega\sim 1 $ mHz  \citep{Murphy14} and energetic electrons in a wide-range of planetary environments, such as those of Jupiter or Saturn \citep{Lejosne20}\footnote{It should, however, be noted that ions of energy less than 10 keV can sustain fluctuations that violate the drift kinetic scale separation with $k_\perp \rho \simeq 1$ \citep{Crabtree04}.}. The main limitations of our paper is that it focused solely on equatorially trapped particles, and it ignored boundary effects that are known observationally as a sink for energetic electron fluxes \citep{Millan07}. A growing number of \textit{in situ} experiments are showing that energetic electrons can be depleted on timescales comparable to a few drift periods \citep{Turner12, Jaynes18, Olifer18}. While such sudden losses can in theory be explained by local wave-particle interactions \citep{Zhang22}, in some events the small scale waves appear insufficient to account for the losses \citep{Albert_dropout}. Since ULF waves can effectively transport energetic electrons on fast timescales, it is worth investigating the net impact that they have when it comes to particle losses. The nonlinear Parker scenario described in Section \ref{fast_symmetric}, and in association with a sudden and symmetric reduction of the magnetic field, will result in the outward transport of particles to the outer magnetopause boundary. Future studies will quantify the role of radial transport in losses occurring on timescales comparable to the drift period and therefore too fast to be explained by radial diffusion. \\

\begin{acknowledgments}
\textbf{Acknowledgments}\\
Support for AO was provided by the Academy of Finland profiling action Matter and Materials (grant \# 318913). OA gratefully acknowledges financial support from the University of Exeter, the University of Birmingham, and also from the United Kingdom Research and Innovation (UKRI) Natural Environment Research Council (NERC) Independent Research Fellowship NE/V013963/1. We are grateful to Yohei Kawazura, Sol\`{e}ne Lejosne, Lucile Turc, Jay Albert, Richard Horne, Leonid Olifer, Hannu Koskinen and Jacob Bortnik for discussions of this work. 
\end{acknowledgments}
\newpage
  
\appendix

\section{Derivation of the Quasi-Linear Equation}\label{AppendixA}
\noindent In this Appendix we provide a detailed derivation of the quasi-linear equation (\ref{kinetic4}). For equatorial particles with a conserved first adiabatic invariant $\mu$ interacting with a Mead field, the kinetic equation takes the form: 
%%%%%%%%
\begin{eqnarray}
g B_0 \frac{\partial f}{\partial t} + \frac{3\mu B_0}{q\gamma r^2} \frac{\partial f}{\partial \varphi} &+& \sum_me^{im\varphi}\left[\frac{\mu A_m}{q\gamma r}+i \frac{r\dot{A}_m}{7m}\right]\frac{\partial f}{\partial \varphi}=-\left[\frac{r \dot{S}}{2} +\sum_me^{im\varphi}\left(\frac{8r^2\dot{A}_m}{21}-im \frac{\mu A_m}{q\gamma }\right)\right]\frac{\partial f}{\partial r}  \nonumber
\end{eqnarray}
%%%%%%%%
with the function $g(r, \varphi, t)= 1- S(t)/B_0-\sum_m e^{im\varphi}r A_m(t)/B_0$.  After decomposing the perturbed fluctuations along the azimuthal angle in Fourier space $f(r, \varphi, t)=f_0(r, t)+\sum_me^{i m\varphi} \delta f_m(r,t)$, the kinetic equation takes the following form: 
%%%%%%%%
\begin{eqnarray}
\label{Kinetic1A}
g\frac{\partial f_0}{\partial t}&=&-\sum_m e^{im\varphi} \left[ \frac{\partial \delta f_m}{\partial t}+i m\Omega_d \delta f_m +\left( \frac{8r^2\dot{A}_m}{21 B_0}-im \frac{\mu A_m}{qB_0\gamma }+\frac{r \dot{S}}{2 B_0}\delta_{m0} \right)\frac{\partial f_0}{\partial r} \right] \nonumber \\
&-&\sum_m\sum_ne^{i(m+n)\varphi}\left[\left( i\frac{m\mu A_n}{qB_0\gamma r} -\frac{mr\dot{A}_n}{7nB_0}\right)\delta f_m +\left( \frac{8r^2\dot{A}_n}{21 B_0}-in \frac{\mu A_n}{qB_0\gamma }+\frac{r \dot{S}}{2 B_0}\delta_{n0}  \right)\frac{\partial \delta f_m}{\partial r} \right] \nonumber \\ 
&+&\sum_m\sum_ne^{i(m+n)\varphi}\left(\frac{S}{B_0}\delta_{n0}+\frac{A_n r}{B_0} \right)\frac{\partial \delta f_m}{\partial t}, 
\end{eqnarray}
%%%%%%%%
with the Kroenecker delta, 
\begin{equation}
\delta_{mn} =
    \begin{cases}
            1, &         \text{if } m=n,\\
            0, &         \text{if } m\neq n.
    \end{cases}
\end{equation}
The first term in bracket of Equation (\ref{Kinetic1A}) contains the linear term, and the following two brackets with the double sums contain the nonlinear terms. We solve this equation with the aid of the Fourier Convolution theorem:
%%%%%%%%
\begin{eqnarray}
\frac{1}{2\pi}\int_{-\pi}^{\pi} \mathcal{F}(\varphi)\mathcal{G}(\varphi) e^{-ip\varphi} \ d \varphi &=& \frac{1}{2\pi}\int_{-\pi}^{\pi} \left(\sum_m \mathcal{F}_m e^{im\varphi} \right) \left(\sum_n \mathcal{G}_n e^{in\varphi}\right) e^{-ip\varphi} \ d \varphi \nonumber \\
&=& \sum_m  \sum_n \mathcal{F}_m \mathcal{G}_n \frac{1}{2\pi}\int_{-\pi}^{\pi} d\varphi  e^{i(m+n-p)\varphi}  \nonumber \\
&=& \sum_m  \sum_n \mathcal{F}_m \mathcal{G}_n \delta_{m+n, p} \nonumber \\
&=& \sum_m \mathcal{F}_m  \mathcal{G}_{p-m},
\end{eqnarray}
%%%%%%%%
which gives us
%%%%%%%%
\begin{eqnarray}
\label{Kinetic2A}
\left(1-\frac{S}{B_0}-\frac{A_p r}{B_0}\right)\frac{\partial f_0}{\partial t}&=&- \left[ \frac{\partial \delta f_p}{\partial t}+i p\Omega_d \delta f_p +\left( \frac{8r^2\dot{A}_p}{21 B_0}-ip \frac{\mu A_p}{qB_0\gamma }+\frac{r \dot{S}}{2 B_0}\delta_{p0} \right)\frac{\partial f_0}{\partial r} \right] \nonumber\\
&-&\sum_m\left[\left( i\frac{m\mu A_{p-m}}{qB_0\gamma r} -\frac{mr\dot{A}_{p-m}}{7B_0(p-m)}\right)\delta f_m  -\left(\frac{S}{B_0}\delta_{p-m,0}+\frac{A_{p-m} r}{B_0} \right)\frac{\partial \delta f_m}{\partial t}\right] \nonumber\\ 
&-&\sum_m \left( \frac{8r^2\dot{A}_{p-m}}{21 B_0}-i(p-m) \frac{\mu A_{p-m}}{qB_0\gamma }+\frac{r \dot{S}}{2 B_0}\delta_{p-m,0}  \right)\frac{\partial \delta f_m}{\partial r}.
\end{eqnarray}
%%%%%%%%
In order to obtain the quasi-linear equation we first set $p=0$ which corresponds to the spatial average of the kinetic equation,
%%%%%%%%
\begin{eqnarray}
\label{Kinetic3A}
\left(1-\frac{S}{B_0}-\frac{A_0 r}{B_0}\right)\frac{\partial f_0}{\partial t}&=&- \left[ \frac{\partial \delta f_{p=0}}{\partial t} +\left( \frac{8r^2\dot{A}_0}{21 B_0}+\frac{r \dot{S}}{2 B_0} \right)\frac{\partial f_0}{\partial r} \right] \nonumber\\
&-&\sum_m\left[\left( i\frac{m\mu A_{-m}}{qB_0\gamma r} +\frac{r\dot{A}_{-m}}{7B_0}\right)\delta f_m  -\left(\frac{S}{B_0}\delta_{m,0}+\frac{A_{-m} r}{B_0} \right)\frac{\partial \delta f_m}{\partial t}\right] \nonumber\\ 
&-&\sum_m \left( \frac{8r^2\dot{A}_{-m}}{21 B_0}+im\frac{\mu A_{-m}}{qB_0\gamma }+\frac{r \dot{S}}{2 B_0}\delta_{m,0}  \right)\frac{\partial \delta f_m}{\partial r}.
\end{eqnarray}
%%%%%%%%
and then perform the time average defined in Equation (\ref{def_average}) to find Equation (\ref{kinetic4}),
%%%%%%%% 
\begin{eqnarray}
\label{Kinetic4A}
\frac{\partial f_0}{\partial t}&=&-\sum_m\left( \frac{i m\mu}{qB_0\gamma r}\langle A^*_{m}\delta f_m\rangle +\frac{r}{7B_0}\langle \dot{A}^*_{m}\delta f_m\rangle-\frac{r}{B_0}\bigg{\langle} A^*_{m}\frac{\partial \delta f_m}{\partial t}\bigg{\rangle}+\frac{8r^2}{21 B_0}\frac{\partial }{\partial r}\langle\dot{A}^*_{m}\delta f_m\rangle +\frac{im \mu}{qB_0\gamma } \frac{\partial}{\partial r} \langle A_m^* \delta f_m\rangle \right) \nonumber \\
&=&-\sum_m \left[ \frac{i m\mu}{qB_0\gamma r}\frac{\partial}{\partial r} \left(r\langle A^*_{m}\delta f_m\rangle\right) +\frac{8}{21}\frac{1}{rB_0}\frac{\partial }{\partial r}\langle r^3\dot{A}^*_{m}\delta f_m\rangle -\frac{r}{B_0}\langle \frac{\partial}{\partial t}({A}^*_{m}\delta f_m)\rangle \right] \end{eqnarray}
%%%%%%%%
The right-hand side of (\ref{Kinetic4A}) describes the slow evolution of the background distribution due to the effect of fluctuations. \\

\section{Derivation of the nonlinear perturbed equation (\ref{QLT2})}\label{AppendixB}
In order to obtain an equation for perturbed part of the distribution function for Fourier modes $m\neq 0$, we substract Equation (\ref{Kinetic3A}) from  (\ref{Kinetic2A}), which results in Equation (\ref{QLT2})
%%%%%%%%
\begin{eqnarray}
\label{QLT2A} 
\frac{\partial \delta f_m}{\partial t}+{i m\Omega_d \delta f_m} =\frac{A_m r}{B_0}\frac{\partial f_0}{\partial t}{-\left( \frac{8r^2\dot{A}_m}{21 B_0}-im \frac{\mu A_m}{qB_0\gamma } \right)\frac{\partial f_0}{\partial r}}{-\sum_{m'} \mathcal{Q}[A_{m-m'}; \delta f_{m'}].} 
\end{eqnarray}
%%%%%%%%
with the nonlinear term given by
\begin{eqnarray}
\label{nonlinear_term}
\mathcal{Q}[S; A_{m-m'}; \delta f_{m'}]&=&\left[\left( i\frac{m'\mu A_{m-m'}}{qB_0\gamma r} -\frac{m'r\dot{A}_{m-m'}}{7B_0(m-m')}\right)\delta f_{m'}  -\left(\frac{S}{B_0}\delta_{m,m'}+\frac{A_{m-m'} r}{B_0} \right)\frac{\partial \delta f_{m'}}{\partial t}\right] \nonumber\\ 
&+& \left( \frac{8r^2\dot{A}_{m-m'}}{21 B_0}-i(m-m') \frac{\mu A_{m-m'}}{qB_0\gamma }+\frac{r \dot{S}}{2 B_0}\delta_{m,m'}  \right)\frac{\partial \delta f_{m'}}{\partial r}.
\end{eqnarray}
The linear wave-particle interaction only depends on the anti-symmetric magnetic field fluctuation $A_m$. However, the nonlinear perturbations is also a function of the symmetric magnetic field fluctuations, i.e. $S(t)$. The traditional quasilinear assumption consists in ignoring the nonlinearities by setting $\mathcal{Q} =0$, and thus compute the fast linear response due to anti-symmetric ULF waves. It is however possible, as shown in Sections  (\ref{fast_symmetric}), to derive the fast nonlinear response on timescales less than a drift period, and a nonlinear quasi-linear theory for long timescales, by accounting for the nonlinear terms associated with the symmetric ULF perturbations. \\

\section{Justification for neglecting the temporal variation of the background distribution in the linear response (\ref{QLT2})}\label{AppendixC}
\noindent We note that the linear equation in (\ref{QLT2A}) contains a term proportional to $\frac{A_m r}{B_0} \frac{\partial f_0} {\partial t}$. In the quasi-linear limit of short autocorrelation this term will introduce an additional term in the diffusion equation that results in the following correction
\begin{equation}
\left(1+\sum_m\frac{8}{3}\frac{r^2|A_m|^2}{B_0^2}\right) \frac{\partial f_0}{\partial t} =L^2 \frac{\partial}{\partial L}\left(\frac{D_LL}{L^2}\frac{\partial f_0}{\partial L}\right)
\end{equation}
and thus in the limit of small ULF wave amplitude given by the Mead field $|\delta B_m|^2 = r^2|A_m^2| \ll B_0^2$ and the correction reduces the diffusion by a factor much less than one. \\

\noindent One can also give a dimensional argument to neglect the first term on the right-hand side of Equation (\ref{QLT2A}) to compute the linear response under quasi-linear assumptions. The diffusion coefficient $D_{LL}$ has units of one over time, and is bounded by the drift period $\Omega_d$ of a particle. With $D_{LL}\ll \Omega_d$ and thus $D_{LL}\ll1$, the diffusion equation requires that $\frac{\partial f_0}{\partial t} \simeq D_{LL} \frac{\partial^2 f_0}{\partial L^2}$. If the time and spatial variations of the background distribution are slow and determined by non-dimensional small parameters $\varepsilon_t\ll 1$ and $\varepsilon_L\ll1$, respectively, then $f_0=f_0(\varepsilon_t t, \varepsilon_L L)$ inserted into the diffusion equation gives the following scaling: $\varepsilon_t \simeq D_{LL} \varepsilon_L^2$. Therefore,  the time variation of the background is smaller than the radial gradient of the background distribution in the linear response by a factor of $\frac{\partial f_0/\partial t}{{\partial f_0}/{\partial L}}\simeq D_{LL} \varepsilon_L \ll 1$. We note that even if the diffusion coefficient is artificially increased to values comparable to the drift period, thereby implying that transport across one drift shell is possible for one single drift period, the short autocorrelation time limit $\Omega_d \tau_c<1$ would nonetheless hold the above dimensional analysis and justify the neglect of the partial time variation of $f_0$ in the linearised Equation (\ref{QLT2A}).

\newpage

\section{List of symbols}

\begin{table}[htbp]
\begin{tabular}{ll}
$a_m$ & Wave mode amplitude \\
$A_m(t)$ & ULF asymmetric fluctuation amplitude \\
$\textbf{B}$ & Magnetic field \\
$B_0$ & Earth's magnetic field dipole magnitude \\
$B_E$ & Earth's magnetic field dipole moment \\
$B^{EK}$ & Asymmetric magnetic field model of \cite{Elkington} \\
$c$ & Speed of light \\
$C_i$ & Correlator \\
$D$ & Root mean square of antisymmetric field perturbation amplitude $A_m$ \\
$D_{LL}$ & Quasi-linear radial diffusion coefficient \\
$\delta \textbf{B}$ & Magnetic field perturbation \\
$\delta \textbf{E}$ & Electric field perturbation \\
$\textbf{E}$ & Electric field \\
$E_c$ & Relativistic kinetic energy \\
$f$ & Distribution function \\
$\langle f \rangle$ & Gyro-averaged distribution function \\
$\delta f_m^{L}$ & Linear perturbation of the distribution function \\
$\delta f_m^{NL}$ & Non-linear perturbation of the distribution function \\
$f_0$ & Background distribution function \\
$\mathcal{J}$ & Second adiabatic invariant\\
$I_1 \&$  $I_2$ & Nonlinear criteria associated with the symmetric ULF perturbations \\
$I_3\&$ $I_4$  & Nonlinear criteria associated with the anti-symmetric ULF perturbations \\
$l$ & Characteristic scale size \\
$L$ & Normalised radial distance from the Earth's midplane\\
$L^*$ & Magnetic drift shell and third adiabatic invariant \\
$m_s$ & Rest mass of particle species 's'\\
$m$ & Wave number \\
$\mathcal{Q}$ & Nonlinear term \\
$\mathbf{p_\parallel}$ & Relativistic momentum along the local magnetic field direction \\
$\mathbf{p_\perp}$ & Relativistic momentum perpendicular to the local magnetic field direction \\
$q_s$ & Charge of particle specie $s$\\
$\textbf{r}$ & Position \\
$R_E$ & Earth's radius \\
$S(t)$ & Azimuthally symmetric fluctuation amplitude \\
$s$ & Label for particle specie $s=i,e$ \\
$t$ & Time \\
$t_{eq}$ & Timescale to reach stationary state\\
$v$ & Characteristic speed \\
\end{tabular}
\end{table}
\begin{table}[htbp]
\begin{tabular}{ll}
$\alpha$ & Pitch angle \\
$\epsilon$ & Nondimensional small parameter \\
$\gamma$ & Lorentz factor \\
$\gamma_m$ & Wave mode growth rate \\
$\mu$ & First adiabatic invariant \\
$\tilde{\mu}$ & First adiabatic invariant correction\\
$\varphi$ & Azimuthal angle \\
$\rho$ & Larmor radius \\
$\tau_C$ & Correlation/decay time for the anti-symmetric ULF perturbation \\
$\tau_C^s$ & Correlation/decay time for the symmetric ULF perturbation\\
$\tau_D$ & Drift period \\
$\theta$ & Polar angle \\
$\theta_g$ & Gyrophase \\
$\chi$ & Gaussian white noise \\
$\Phi$ & Third adiabatic invariant and magnetic flux \\
$\omega$ & Frequency \\
$\omega_m$ & ULF wave mode frequency \\
$\Omega_d$ & Azimuthal drift frequency \\
$\Omega_s$ & Larmor frequency for specie $s$ \\
\end{tabular}
\end{table}

\onecolumngrid
\newpage
\hspace{2mm}

\newpage

%\bibliography{Radial_diffusion.bib}
%\bibliography{owpi.bib}

\begin{thebibliography}{}
\expandafter\ifx\csname natexlab\endcsname\relax\def\natexlab#1{#1}\fi
\providecommand{\url}[1]{\href{#1}{#1}}
\providecommand{\dodoi}[1]{doi:~\href{http://doi.org/#1}{\nolinkurl{#1}}}
\providecommand{\doeprint}[1]{\href{http://ascl.net/#1}{\nolinkurl{http://ascl.net/#1}}}
\providecommand{\doarXiv}[1]{\href{https://arxiv.org/abs/#1}{\nolinkurl{https://arxiv.org/abs/#1}}}

\bibitem[{Adkins \& Schekochihin(2018)}]{Adkins18}
Adkins, T., \& Schekochihin, A. 2018, J. Plasma Phys., 84,
  \dodoi{https://doi.org/10.1017/S0022377818000089}

\bibitem[{{Agapitov et al.}(2015)}]{Agapitov15}
{Agapitov et al.}, O. 2015, Geophys. Res. Lett., 42, 3715,
  \dodoi{https://doi.org/10.1002/2015GL064145}

\bibitem[{Albert(2014)}]{Albert_dropout}
Albert, J. 2014, Ann. Geophys., 32, 10.5194/angeo

\bibitem[{Albert {et~al.}(2012)Albert, Tao, \& Bortnik}]{Albert12}
Albert, J.~M., Tao, X., \& Bortnik, J. 2012, Aspects of Nonlinear Wave-Particle
  Interactions (American Geophysical Union (AGU)), 255--264,
  \dodoi{https://doi.org/10.1029/2012GM001324}

\bibitem[{Allanson {et~al.}(2022)Allanson, Elsden, Watt, \&
  Neukirch}]{Allanson22}
Allanson, O., Elsden, T., Watt, C., \& Neukirch, T. 2022, Front. in Astron, and
  Space Sci., \dodoi{https://doi.org/10.3389/fspas.2021.805699}

\bibitem[{{Artemyev {et al}.}(2012)}]{Art13}
{Artemyev {et al}.}, A.~V. 2012, Phys. Plasmas, 19, 122901,
  \dodoi{10.1063/1.4769726}

\bibitem[{{Artemyev {et al}.}(2015)}]{Art15}
---. 2015, Nat. Comm., 6, \dodoi{https://doi.org/10.1038/ncomms8143}

\bibitem[{Aryan {et~al.}(2020)Aryan, Agapitov, Artemyev, Mourenas, Balikhin,
  Boynton, \& Bortnik}]{Aryan20}
Aryan, H., Agapitov, O., Artemyev, A., {et~al.} 2020, J. Geophys. Res.: Space
  Physics, 125, e2020JA028018, \dodoi{https://doi.org/10.1029/2020ja028018}

\bibitem[{Barani {et~al.}(2019)Barani, Tu, Sarris, Pham, \& Redmon}]{Barani19}
Barani, M., Tu, W., Sarris, T., Pham, K., \& Redmon, R.~J. 2019, Journal of
  Geophysical Research: Space Physics, 124, 5009,
  \dodoi{https://doi.org/10.1029/2019JA026927}

\bibitem[{{Bernstein} {et~al.}(1957){Bernstein}, {Greene}, \& {Kruskal}}]{BGK}
{Bernstein}, I.~B., {Greene}, J.~M., \& {Kruskal}, M.~D. 1957, Physical Review,
  108, 546, \dodoi{10.1103/PhysRev.108.546}

\bibitem[{Bortnik {et~al.}(2008)Bortnik, Thorne, \& Inan}]{Bortnik08}
Bortnik, J., Thorne, R.~M., \& Inan, U.~S. 2008, Geophys. Res. Lett., 35,
  \dodoi{https://doi.org/10.1029/2008GL035500}

\bibitem[{{Bortnik et al.}(2022)}]{Bortnik22}
{Bortnik et al.}, J. 2022, Geophys. Res. Lett., 49,
  \dodoi{https://doi.org/10.1029/2022GL098365}

\bibitem[{Brautigam \& Albert(2000)}]{Brautigam}
Brautigam, D., \& Albert, J. 2000, J. Geophys. Res. Space Physics, 105, 291,
  \dodoi{doi:10.1029/1999JA900344.}

\bibitem[{Brizard \& Chan(2022)}]{Brizard22}
Brizard, A., \& Chan, A. 2022, Front. Astron, and Space Sciences, 9,
  \dodoi{10.3389/fspas.2022.1010133}

\bibitem[{{Cannon}(2013)}]{Cannon13}
{Cannon}, P.~S. 2013, Space Weather, 11, 138, \dodoi{10.1002/swe.20032}

\bibitem[{Cary \& Brizard(2009)}]{Cary09}
Cary, J.~R., \& Brizard, A.~J. 2009, Rev. Mod. Phys., 81, 693

\bibitem[{{Case}(1959)}]{Case}
{Case}, K.~M. 1959, Annals of Physics, 7, 349,
  \dodoi{10.1016/0003-4916(59)90029-6}

\bibitem[{{Cattell et al.}(2008)}]{Cattell08}
{Cattell et al.}, C. 2008, Geophys. Res. Lett., 35, L01105,
  \dodoi{10.1029/2007GL032009}

\bibitem[{Chen {et~al.}(2007)Chen, Reeves, \& Friedel}]{Chen07}
Chen, Y., Reeves, G., \& Friedel, R. 2007, Nature Physics, 3, 614,
  \dodoi{https://doi.org/10.1038/nphys655}

\bibitem[{Claudepierre~et al.(2013)}]{Claudepierre13}
Claudepierre~et al., S.~G. 2013, Geophys. Res. Lett., 40, 4491,
  \dodoi{https://doi.org/10.1002/grl.50901}

\bibitem[{Crabtree \& Chen(2004)}]{Crabtree04}
Crabtree, C., \& Chen, L. 2004, Geophys. Res. Lett., 31,
  \dodoi{https://doi.org/10.1029/2004GL020660}

\bibitem[{Cronin(1999)}]{Cronin99}
Cronin, J. 1999, Rev. Mod. Phys., 71, S165, \dodoi{10.1103/RevModPhys.71.S165}

\bibitem[{{Cully} {et~al.}(2008){Cully}, {Bonnell}, \& {Ergun}}]{Cully08}
{Cully}, C.~M., {Bonnell}, J.~W., \& {Ergun}, R.~E. 2008, Geophys. Res. Lett.,
  35, 17, \dodoi{10.1029/2008GL033643}

\bibitem[{Cunningham(2016)}]{Cunningham16}
Cunningham, G. 2016, J. Geophys. Res.: Space Physics, 121, 5149,
  \dodoi{https://doi.org/10.1002/2015JA021981}

\bibitem[{Datlowe {et~al.}(1985)Datlowe, Imhof, Gaines, \& Voss}]{Datlowe85}
Datlowe, D., Imhof, W., Gaines, E., \& Voss, H. 1985, J. Geophys. Res.: Space
  Physics, 90, 8333, \dodoi{https://doi.org/10.1029/JA090iA09p08333}

\bibitem[{Davidson(2012)}]{Davidson}
Davidson, R. 2012, Methods in nonlinear plasma theory (Elsevier)

\bibitem[{Degeling~et al.(2008)}]{Degeling08}
Degeling~et al., A. 2008, J. Geophys. Res.: Space Physics, 113,
  \dodoi{doi:10.1029/2007JA012411}

\bibitem[{{Desai et al.}(2021)}]{Desai21}
{Desai et al.}, R.~T. 2021, J. Geophys. Res.: Space Physics, 126,
  e2021JA029802, \dodoi{https://doi.org/10.1029/2021JA029802}

\bibitem[{{Diamond} {et~al.}(2010){Diamond}, {Itoh}, \& {Itoh}}]{Diamond10}
{Diamond}, P.~H., {Itoh}, S.-I., \& {Itoh}, K. 2010, {Modern Plasma Physics :
  Physical kinetics of turbulent plasmas} (Cambridge University Press,
  Cambridge, England), \dodoi{https://doi.org/10.1017/CBO9780511780875}

\bibitem[{{Dupree}(1966)}]{Dupree66}
{Dupree}, T.~H. 1966, Physics of Fluids, 9, 1773, \dodoi{10.1063/1.1761932}

\bibitem[{Dupree(1972)}]{Dupree72}
Dupree, T.~H. 1972, Phys. Fluids, 15, 334,
  \dodoi{https://doi.org/10.1063/1.1693911}

\bibitem[{Elkington {et~al.}(1999)Elkington, Hudson, \& Chan}]{Elkington}
Elkington, S.~R., Hudson, M.~K., \& Chan, A.~A. 1999, Geophys. Res. Lett., 26,
  3273, \dodoi{https://doi.org/10.1029/1999GL003659}

\bibitem[{Elkington {et~al.}(2003)Elkington, Hudson, \& Chan}]{Elkington03}
---. 2003, J. Geophys. Res.: Space Physics, 108,
  \dodoi{https://doi.org/10.1029/2001JA009202}

\bibitem[{F{\"a}lthammar(1965)}]{Falthammar65}
F{\"a}lthammar, C.-G. 1965, J. Geophy. Res., 70, 2503,
  \dodoi{https://doi.org/10.1029/JZ070i011p02503}

\bibitem[{Fei {et~al.}(2006)Fei, Chan, Elkington, \& Wiltberger}]{Fei06}
Fei, Y., Chan, A.~A., Elkington, S.~R., \& Wiltberger, M.~J. 2006, J. Geophys.
  Res.: Space Physics, 111, \dodoi{https://doi.org/10.1029/2005JA011211}

\bibitem[{George~et al.(2022)}]{George22b}
George~et al., H. 2022, J. Geophys. Res.: Space Physics, 127,
  \dodoi{https://doi.org/10.1029/2022JA030751}

\bibitem[{{Goldreich} \& {Sridhar}(1995)}]{GS95}
{Goldreich}, P., \& {Sridhar}, S. 1995, Astrophys. J., 438, 763,
  \dodoi{https://doi.org/10.1086/304442}

\bibitem[{Goldston \& Rutherford(1995)}]{Goldston95}
Goldston, R.~J., \& Rutherford, P.~H. 1995, Introduction to plasma physics (CRC
  Press)

\bibitem[{Grach~et al.(2022)}]{Grach22}
Grach~et al., V.~S. 2022, Geophys. Res. Lett., 49, e2022GL099994,
  \dodoi{https://doi.org/10.1029/2022GL099994}

\bibitem[{Green \& Kivelson(2004)}]{Green04}
Green, J., \& Kivelson, M. 2004, J. Geophys. Res., 109, A03213,
  doi:10.1029/2003JA010153, \dodoi{https://doi.org/10.1029/2003JA010153}

\bibitem[{{Hands et al.}(2018)}]{Hands18}
{Hands et al.}, A. D.~P. 2018, Space Weather, 16, 1216,
  \dodoi{https://doi.org/10.1029/2018SW001913}

\bibitem[{{Hartinger et al.}(2013)}]{Hartinger13}
{Hartinger et al.}, M. 2013, J. Geophys. Res.: Space Physics, 118,
  doi:10.1029/2012JA018349, \dodoi{https://doi.org/10.1029/2012JA018349}

\bibitem[{Hartinger~et al.(2020)}]{Hartinger20}
Hartinger~et al., M. 2020, J. Geophys. Res.: Space Physics, 125, e2020JA027924,
  \dodoi{https://doi.org/10.1029/2020JA027924}

\bibitem[{Hazeltine(1973)}]{Hazeltine73}
Hazeltine, R. 1973, Plasma Physics, 15, 77, \dodoi{10.1088/0032-1028/15/1/009}

\bibitem[{Hazeltine(2018)}]{Hazeltine18}
---. 2018, The framework of plasma physics (CRC Press),
  \dodoi{https://doi.org/10.1201/9780429502804}

\bibitem[{Hazeltine \& Meiss(2013)}]{Hazeltine13}
Hazeltine, R., \& Meiss, J. 2013, Plasma Confinement (Courier Corporation)

\bibitem[{Hendry~et al.(2019)}]{Hendry19}
Hendry~et al., A.~T. 2019, Geophys. Res. Lett., 46, 7248,
  \dodoi{https://doi.org/10.1029/2019GL082401}

\bibitem[{{Horne et al.}(2018)}]{Horne18}
{Horne et al.}, R. 2018, Space Weather, 16, 1202,
  \dodoi{https://doi.org/10.1029/2018SW001948}

\bibitem[{Imhof \& Smith(1965)}]{Imhof65}
Imhof, W., \& Smith, R. 1965, Phys. Rev. Lett., 14, 885,
  \dodoi{https://doi.org/10.1103/PhysRevLett.14.885}

\bibitem[{Jaynes~et al.(2018)}]{Jaynes18}
Jaynes~et al., A. 2018, Geophys. Res. Lett., 45, 10,
  \dodoi{https://doi.org/10.1029/2018GL079786}

\bibitem[{Jaynes~et al.(2015)}]{Jaynes15}
Jaynes~et al., A.~N. 2015, J. Geophys. Res.: Space Physics, 120, 7240,
  \dodoi{https://doi.org/10.1002/2015JA021234}

\bibitem[{{Kalliokoski et al.}(2023)}]{Kalliokoski22b}
{Kalliokoski et al.} 2023, J. Geophys. Res. (Space Physics), accepted,
  \dodoi{10.1029/2022JA030708}

\bibitem[{{Kanekal et al.}(2016)}]{Kanekal16}
{Kanekal et al.} 2016, J. Geophys. Res. (Space Physics), 121, 7622,
  \dodoi{https://doi.org/10.1002/2016JA022596}

\bibitem[{Kawazura {et~al.}(2019)Kawazura, Barnes, \&
  Schekochihin}]{kawazura19}
Kawazura, Y., Barnes, M., \& Schekochihin, A.~A. 2019, Proc. Nation. Acad.
  Sciences, 116, 771, \dodoi{https://doi.org/10.1073/pnas.181249111}

\bibitem[{{Kennel} \& {Engelmann}(1966)}]{Kennel66}
{Kennel}, C.~F., \& {Engelmann}, F. 1966, Phys. of Fluids, 9, 2377,
  \dodoi{10.1063/1.1761629}

\bibitem[{Kokubun~et al.(1977)}]{Kokobun77}
Kokubun~et al., S. 1977, J. Geophys. Res, 82, 2774,
  \dodoi{https://doi.org/10.1029/JA082i019p02774}

\bibitem[{Kulsrud(2005)}]{Kulsrud}
Kulsrud, R. 2005, Plasma physics for astrophysics (New Jersey: Princeton
  University Press)

\bibitem[{Kunz {et~al.}(2018)Kunz, Abel, Klein, \& Schekochihin}]{Kunz18}
Kunz, M.~W., Abel, I.~G., Klein, K.~G., \& Schekochihin, A.~A. 2018, J. Plasma
  Phys., 84, \dodoi{https://doi.org/10.1017/S0022377818000296}

\bibitem[{Landau(1946)}]{Landau}
Landau, L. 1946, Zh. Eksp. Teor. Fiz., 10, 25

\bibitem[{Lanzerotti {et~al.}(1967)Lanzerotti, Roberts, \&
  Brown}]{Lanzerotti67}
Lanzerotti, L., Roberts, C., \& Brown, W. 1967, J. Geophys. Res., 72, 5893,
  \dodoi{https://doi.org/10.1029/JZ072i023p05893}

\bibitem[{Lejosne(2019)}]{Lejosne19}
Lejosne, S. 2019, J. Geophys. Res: Space Physics, 124, 4278,
  \dodoi{https://doi.org/10.1029/2019JA026786}

\bibitem[{Lejosne {et~al.}(2013)Lejosne, Boscher, Maget, \&
  Rolland}]{Lejosne13}
Lejosne, S., Boscher, D., Maget, V., \& Rolland, G. 2013, J. Geophys. Res.:
  Space Physics, 118, 3147, \dodoi{https://doi.org/10.1029/2012JA018011}

\bibitem[{Lejosne {et~al.}(2022)Lejosne, Fejer, Maruyama, \&
  Scherliess}]{Lejosne22}
Lejosne, S., Fejer, B., Maruyama, N., \& Scherliess, L. 2022, Front. Astron.
  and Space Scien, 9, 823695, \dodoi{doi:10.2289/fspas.2022.823695}

\bibitem[{Lejosne \& Kollmann(2020)}]{Lejosne20}
Lejosne, S., \& Kollmann, P. 2020, Space Sci. Rev., 216, 1,
  \dodoi{https://doi.org/10.1007/s11214-020-0642-6}

\bibitem[{Lejosne \& Mozer(2020{\natexlab{a}})}]{Lejosne20a}
Lejosne, S., \& Mozer, F. 2020{\natexlab{a}}, J. Geophys. Res.: Space Physics,
  125, e2020JA027889, \dodoi{https://doi.org/10.1029/2020JA027889}

\bibitem[{Lejosne \& Mozer(2020{\natexlab{b}})}]{Lejosne20b}
---. 2020{\natexlab{b}}, Geophys. Res. Lett., 47, e2020GL088564,
  \dodoi{https://doi.org/10.1029/2020GL088564}

\bibitem[{Lejosne \& Roederer(2016)}]{Lejosne16zebra}
Lejosne, S., \& Roederer, J. 2016, J. Geophys. Res.: Space Physics, 121, 507,
  \dodoi{10.1002/2015JA02192}

\bibitem[{Li {et~al.}(2018)Li, Zhou, Omura, Wang, Zong, Liu, Hao, Fu, Kivelson,
  Rankin, Claudepierre, \& Wygant}]{Li18}
Li, L., Zhou, X., Omura, Y., {et~al.} 2018, Geophys. Res. Lett., 45, 8773,
  \dodoi{https://doi.org/10.1029/2018GL079038}

\bibitem[{Li {et~al.}(2017)Li, Selesnick, Schiller, Zhang, Zhao, Baker, \&
  Temerin}]{Li2017}
Li, X., Selesnick, R., Schiller, Q., {et~al.} 2017, Nat., 552, 382,
  \dodoi{https://doi.org/10.1038/nature24642}

\bibitem[{{Li et al.}(1993)}]{Li93}
{Li et al.} 1993, Geophys. Res. Lett., 20, 2423,
  \dodoi{https://doi.org/10.1029/93GL02701}

\bibitem[{{Lichtenberg} \& {Lieberman}(1983)}]{LichLieb}
{Lichtenberg}, A.~J., \& {Lieberman}, M.~A. 1983, {Regular and stochastic
  motion} (Springer-Verlag)

\bibitem[{Malaspina {et~al.}(2014)Malaspina, Andersson, Ergun, Wygant, Bonnell,
  Kletzing, Reeves, Skoug, \& Larsen}]{Malaspina14}
Malaspina, D.~M., Andersson, L., Ergun, R.~E., {et~al.} 2014, Geophys. Res.
  Lett., 41, 5693, \dodoi{10.1002/2014GL061109}

\bibitem[{Mead(1964)}]{Mead}
Mead, G.~D. 1964, J. Geophys. Res., 69, 1181,
  \dodoi{https://doi.org/10.1029/JZ069i007p01181}

\bibitem[{Meyrand {et~al.}(2019)Meyrand, Kanekar, Dorland, \&
  Schekochihin}]{Meyrand2019}
Meyrand, R., Kanekar, A., Dorland, W., \& Schekochihin, A. 2019, Proc. Nat.
  Acad. Sci., 116, 1185, \dodoi{https://doi.org/10.1073/pnas.181391311}

\bibitem[{Millan \& Thorne(2007)}]{Millan07}
Millan, R., \& Thorne, R. 2007, J Atmos. and Sol.-Terr. Phys., 69, 362,
  \dodoi{https://doi.org/10.1016/j.jastp.2006.06.019}

\bibitem[{{Mills} \& {Wright}(1999)}]{Mills99}
{Mills}, K.~J., \& {Wright}, A.~N. 1999, J. Geophys. Res., 104, 22667,
  \dodoi{10.1029/1999JA900280}

\bibitem[{{Morley et al.}(2016)}]{Morley16}
{Morley et al.}, S.~K. 2016, Space Weather, 14, 76,
  \dodoi{10.1002/2015SW001339}

\bibitem[{Morley~et al.(2017)}]{Morley17}
Morley~et al., S.~K. 2017, Space Weather, 15, 283,
  \dodoi{https://doi.org/10.1002/2017SW001604}

\bibitem[{Mozer~et al. {et~al.}(2013)Mozer~et al., Bale, Bonnell, Chaston,
  Roth, \& Wygant}]{Mozer13}
Mozer~et al., F., Bale, S., Bonnell, J., {et~al.} 2013, Phys. Rev. Lett., 111,
  235002, \dodoi{10.1103/PhysRevLett.111.235002}

\bibitem[{Murphy {et~al.}(2014)Murphy, Mann, \& Ozeke}]{Murphy14}
Murphy, K.~R., Mann, I.~R., \& Ozeke, L.~G. 2014, Geophysical Research Letters,
  41, 6595, \dodoi{https://doi.org/10.1002/2014GL061253}

\bibitem[{{Murphy et al.}(2020)}]{Murphy20}
{Murphy et al.}, K.~R. 2020, J. Geophys. Res.: Space Physics, 125, e27887,
  \dodoi{10.1029/2020JA027887}

\bibitem[{O'Brien(2014)}]{Obrien14a}
O'Brien, T.~P. 2014, Geophys. Res. Lett., 41, 216,
  \dodoi{https://doi.org/10.1002/2013GL058712}

\bibitem[{Olifer~et al.(2018)}]{Olifer18}
Olifer~et al., L. 2018, J. Geophys. Res: Space Physics, 123, 3692,
  \dodoi{https://doi.org/10.1029/2018JA025190}

\bibitem[{Omura(2021)}]{Omura21}
Omura, Y. 2021, Earth, Planets and Space, 73, 1,
  \dodoi{https://doi.org/10.1186/s40623-021-01380-w}

\bibitem[{Orszag \& Kraichnan(1967)}]{Orszag67}
Orszag, S.~A., \& Kraichnan, R.~H. 1967, The Physics of Fluids, 10, 1720,
  \dodoi{https://doi.org/10.1063/1.1762351}

\bibitem[{Osmane \& Lejosne(2021)}]{Osmane21a}
Osmane, A., \& Lejosne, S. 2021, Astrophys. J., 912, 142,
  \dodoi{10.3847/1538-4357/abf04b}

\bibitem[{Osmane {et~al.}(2022)Osmane, Savola, Kilpua, Koskinen, Borovsky, \&
  Kalliokoski}]{Osmane22}
Osmane, A., Savola, M., Kilpua, E., {et~al.} 2022, Ann. Geophys., 40, 37,
  \dodoi{10.5194/angeo-40-37-2022}

\bibitem[{{Osmane et al.}(2016)}]{Osmane16}
{Osmane et al.}, A. 2016, Astrophys. J., 816, 51,
  \dodoi{10.3847/0004-637X/816/2/51}

\bibitem[{{Osmane et al.}(2017)}]{Osmane17}
---. 2017, Astrophys. J., 846, 83, \dodoi{10.3847/1538-4357/aa8367}

\bibitem[{Ozeke~et al.(2014)}]{Ozeke14}
Ozeke~et al., L. 2014, J. Geophys. Res: Space Physics, 119, 1587,
  \dodoi{https://doi.org/10.1002/2013JA019204}

\bibitem[{Papoulis(1991)}]{Papoulis}
Papoulis, A. 1991, Probability, Random Variables, and Stochastic Processes,
  Electrical Engineering Series (McGraw-Hill)

\bibitem[{Parker(1960)}]{Parker60}
Parker, E. 1960, J. Geophys. Res., 65, 3117,
  \dodoi{https://doi.org/10.1029/JZ065i010p03117}

\bibitem[{Parra(2019)}]{Parra}
Parra, F. 2019, {Collisionless Plasma Physics. Lecture Notes for an Oxford
  MMathPhys course},
  \url{http://www-thphys.physics.ox.ac.uk/people/FelixParra/CollisionlessPlasmaPhysics/CollisionlessPlasmaPhysics.html}

\bibitem[{Parra \& Catto(2008)}]{Parra08}
Parra, F., \& Catto, P. 2008, Plasma Phys. and Control. Fusion, 50, 065014,
  \dodoi{10.1088/0741-3335/50/6/065014}

\bibitem[{Popper(2005)}]{Popper}
Popper, K. 2005, The logic of scientific discovery (Routledge)

\bibitem[{Quataert \& Gruzinov(1999)}]{Eliot99}
Quataert, E., \& Gruzinov, A. 1999, Astrophys. J., 520, 248,
  \dodoi{10.1086/307423}

\bibitem[{{Reeves {et al}.}(2013)}]{Reeves13}
{Reeves {et al}.}, G.~D. 2013, Science, 341, 991,
  \dodoi{10.1126/science.1237743}

\bibitem[{Roederer \& Zhang(2014)}]{Roederer}
Roederer, J.~G., \& Zhang, H. 2014, Dynamics of Magnetically Trapped Particles
  Foundations of the Physics of Radiation Belts and Space Plasmas (Springer),
  \dodoi{https://doi.org/10.1007/978-3-642-41530-2}

\bibitem[{Sandhu~et al.(2021)}]{Sandhu21}
Sandhu~et al., J.~K. 2021, J. Geophys. Res.: Space Physics, 126, e2020JA029024,
  \dodoi{https://doi.org/10.1029/2020JA029024}

\bibitem[{Santolík~et al.(2014)}]{Santolik14}
Santolík~et al., O. 2014, Geophys. Res. Lett., 41, 293,
  \dodoi{10.1002/2013GL058889}

\bibitem[{Sarma~et al.(2020)}]{Sarma20}
Sarma~et al., R. 2020, J. Geophys. Res: Space Physics, 125, e2019JA027618,
  \dodoi{https://doi.org/10.1029/2019JA027618}

\bibitem[{Sarris(2014)}]{Sarris14}
Sarris, T.~E. 2014, J. Geophys. Res: Space Physics, 119, 5539,
  \dodoi{https://doi.org/10.1002/2013JA019238}

\bibitem[{{Sarris et al}(2021)}]{Sarris21}
{Sarris et al}, T.~E. 2021, J. Geophys. Res.: Space Physics, 126,
  e2020JA028891, \dodoi{https://doi.org/10.1029/2020JA028891}

\bibitem[{{Sarris et al.}(2022)}]{Sarris22}
{Sarris et al.}, T.~E. 2022, J. of Geophys. Res.: Space Physics, e2022JA030469,
  \dodoi{https://doi.org/10.1029/2022JA030469}

\bibitem[{Sauvaud {et~al.}(2013)Sauvaud, Walt, Delcourt, Benoist, Penou, Chen,
  \& Russell}]{Sauvaud13}
Sauvaud, J.-A., Walt, M., Delcourt, D., {et~al.} 2013, J. Geophys. Res: Space
  Physics, 118, 1723, \dodoi{https://doi.org/10.1002/jgra.50125}

\bibitem[{Schekochihin(2017)}]{Schekochihin_notes}
Schekochihin, A. 2017, Lecture notes on kinetic theory and magnetohydrodynamics
  of plasmas,
  \url{http://www-thphys.physics.ox.ac.uk/people/AlexanderSchekochihin/KT/2015/KTLectureNotes.pdf},
  Oxford University

\bibitem[{Schekochihin~et al.(2016)}]{Scheko2016}
Schekochihin~et al., A. 2016, J. Plasma Phys., 82,
  \dodoi{https://doi.org/10.1017/S0022377816000374}

\bibitem[{{Schekochihin et al.}(2008)}]{Scheko08}
{Schekochihin et al.}, A.~A. 2008, Plasma Phys. and Contr. Fusion, 50, 124024,
  \dodoi{10.1088/0741-3335/50/12/124024}

\bibitem[{Schulz \& Eviatar(1969)}]{Schulz69}
Schulz, M., \& Eviatar, A. 1969, J. Geophys. Res., 74, 2182,
  \dodoi{https://doi.org/10.1029/JA074i009p02182}

\bibitem[{Schulz \& Lanzerotti(1974)}]{Schulz74}
Schulz, M., \& Lanzerotti, L. 1974, Physics and Chemistry in Space,
  \dodoi{https://doi.org/10.1007/978-3-642-65675-0}

\bibitem[{{Servidio et al.}(2017)}]{Servidio2017}
{Servidio et al.}, S. 2017, Phys. Rev. Lett., 119, 205101,
  \dodoi{https://doi.org/10.1103/PhysRevLett.119.205101}

\bibitem[{{Shprits et al.}(2006)}]{Shprits06}
{Shprits et al.}, Y. 2006, J. Geophys. Res.: Space Physics, 111,
  \dodoi{https://doi.org/10.1029/2006JA011725}

\bibitem[{Simms~et al.(2018)}]{Simms18}
Simms~et al., L.~E. 2018, J. Geophys. Res.: Space Physics, 123, 4755,
  \dodoi{https://doi.org/10.1029/2017JA025003}

\bibitem[{{Sironi} \& {Narayan}(2015)}]{Sironi15}
{Sironi}, L., \& {Narayan}, R. 2015, \apj, 800, 88,
  \dodoi{10.1088/0004-637X/800/2/88}

\bibitem[{{Southwood} {et~al.}(1969){Southwood}, {Dungey}, \&
  {Etherington}}]{Southwood69}
{Southwood}, D.~J., {Dungey}, J.~W., \& {Etherington}, R.~J. 1969, \planss, 17,
  349, \dodoi{10.1016/0032-0633(69)90068-3}

\bibitem[{{Southwood} \& {Kivelson}(1981)}]{Southwood81}
{Southwood}, D.~J., \& {Kivelson}, M.~G. 1981, J. Geophys. Res: Space Physics,
  86, 5643, \dodoi{10.1029/JA086iA07p05643}

\bibitem[{Summers(2005)}]{Summers05}
Summers, D. 2005, Journal of Geophysical Research: Space Physics, 110,
  \dodoi{https://doi.org/10.1029/2005JA011159}

\bibitem[{{Takahashi} \& {Ukhorskiy}(2007)}]{Takahashi07}
{Takahashi}, K., \& {Ukhorskiy}, A. 2007, J. Geophys. Res. (Space Physics),
  112, A11205, \dodoi{10.1029/2007JA012483}

\bibitem[{Takahashi~et al.(2022)}]{Takahashi22}
Takahashi~et al., K. 2022, J. Geophys. Res.: Space Physics, 127, e2021JA030115,
  \dodoi{https://doi.org/10.1029/2021JA030115}

\bibitem[{Tao {et~al.}(2020)Tao, Zonca, Chen, \& Wu}]{Tao20}
Tao, X., Zonca, F., Chen, L., \& Wu, Y. 2020, Sci. China Earth Sciences, 63,
  \dodoi{https://doi.org/ 10.1007/s11430-019-9384-6}

\bibitem[{{Thorne}(2010)}]{Thorne10}
{Thorne}, R.~M. 2010, Geophys. Res. Lett., 37, L22107,
  \dodoi{10.1029/2010GL044990}

\bibitem[{{Thorne et al.}(2013)}]{Thorne13}
{Thorne et al.}, R.~M. 2013, Nature, 504, 411, \dodoi{doi:10.1038/nature12889}

\bibitem[{{Tu et al}(2012)}]{Weichao12}
{Tu et al}, W. 2012, J. Geophys. Res.: Space Physics, 117,
  \dodoi{https://doi.org/10.1029/2012JA017901}

\bibitem[{{Turner et al.}(2012)}]{Turner12}
{Turner et al.}, D.~L. 2012, Geophys. Res. Lett., 39, 9101,
  \dodoi{10.1029/2012GL051722}

\bibitem[{Turner~et al.(2017)}]{Turner17}
Turner~et al., D.~L. 2017, J. Geophys. Res.: Space Physics, 122, 11,481,
  \dodoi{https://doi.org/10.1002/2017JA024554}

\bibitem[{{Turunen {et al}.}(2009)}]{Turunen}
{Turunen {et al}.}, E. 2009, J. Atm. Sol.-Terr. Phys., 71, 1176,
  \dodoi{https://doi.org/10.1016/j.jastp.2008.07.005}

\bibitem[{Ukhorskiy \& Sitnov(2012)}]{sasha_review}
Ukhorskiy, A., \& Sitnov, M. 2012, Dynamics of radiation belt particles
  (Springer), 545--578, \dodoi{10.1007/s11214-012-9938-5}

\bibitem[{Ukhorskiy {et~al.}(2014)Ukhorskiy, Sitnov, Mitchell, Takahashi,
  Lanzerotti, \& Mauk}]{Sasha_zebra}
Ukhorskiy, A., Sitnov, M., Mitchell, D., {et~al.} 2014, Nature, 507, 338,
  \dodoi{10.1038/nature13046}

\bibitem[{{Van Allen et al.}(1958)}]{VanAllen}
{Van Allen et al.}, J. 1958, J. Jet Propulsion, 28, 588,
  \dodoi{https://doi.org/10.2514/8.7396}

\bibitem[{Van~Kampen(1955)}]{Kampen}
Van~Kampen, N.~G. 1955, Physica, 21, 949,
  \dodoi{https://doi.org/10.1016/S0031-8914(55)93068-8}

\bibitem[{Vanden~Eijnden(1997)}]{Eijnden}
Vanden~Eijnden, E. 1997, Phys. Plasmas, 4, 1486,
  \dodoi{https://doi.org/10.1063/1.872548}

\bibitem[{{Volwerk}(2016)}]{Volwerk16}
{Volwerk}, M. 2016, AGU Geophysical Monograph Series, 216, 139,
  \dodoi{10.1002/9781119055006.ch9}

\bibitem[{{Walton et al.}(2022)}]{Walton22}
{Walton et al.}, S. 2022, J. Geophys. Res.: Space Physics, 127, e2021JA030069,
  \dodoi{https://doi.org/10.1029/2021JA030069}

\bibitem[{Wang {et~al.}(2018)Wang, Rankin, Wang, Zong, Zhou, Takahashi,
  Marchand, \& Degeling}]{Wang18}
Wang, C., Rankin, R., Wang, Y., {et~al.} 2018, J. Geophys. Res.: Space Physics,
  123, 4652, \dodoi{https://doi.org/10.1029/2017JA025123}

\bibitem[{{Wu et al.}(2019)}]{Wu19}
{Wu et al.}, H. 2019, Astrophys. J., 870, 106, \dodoi{10.3847/1538-4357/aaef77}

\bibitem[{{Zhang et al.}(2022)}]{Zhang22}
{Zhang et al.}, X. 2022, Nature Communications, 13, 1611,
  \dodoi{https://doi.org/10.1038/s41467-022-29291-8}

\bibitem[{Zhao \& Li(2013)}]{Zhao13}
Zhao, H., \& Li, X. 2013, J. Geophys. Res.: Space Physics, 118, 6936,
  \dodoi{https://doi.org/10.1002/2013JA019240}

\bibitem[{Zhdankin(2022)}]{Zhdankin22}
Zhdankin, V. 2022, Phys. Rev. X, 12, 031011,
  \dodoi{https://doi.org/10.1103/PhysRevX.12.031011}

\bibitem[{Öztürk \& Wolf(2007)}]{Ozturk07}
Öztürk, M.~K., \& Wolf, R. 2007, J. Geophys. Res: Space Physics, 112,
  \dodoi{https://doi.org/10.1029/2006JA012102}

\end{thebibliography}
\bibliographystyle{aasjournal}

\end{document}